\documentclass[a4paper,11pt]{article}
\pdfoutput=1 

\usepackage{jcappub} 

\usepackage[T1]{fontenc} 
\usepackage{times}
\usepackage{graphicx}	
\usepackage{amsmath}	
\usepackage{amssymb}	
\usepackage[british]{babel}             
\usepackage{bm}
\usepackage{physics}
\usepackage{academicons}
\usepackage{xcolor}
\usepackage{booktabs}
\usepackage{natbib}
\usepackage{url}
\usepackage{hyperref}
\usepackage[nolist]{acronym}
\usepackage{accents}
\usepackage{ulem}
\newacro{3D}{three-dimensional}
\newacro{MG}{modified gravity}
\newacro{FFT}{Fast Fourier Transformation}
\newacro{GR}{General Relativity}
\newacro{DGP}{Dvali-Gabadadze-Porrati}
\newacro{RSD}{redshift space distortion}

\newcommand{\Mpch}{\,h^{-1}{\rm Mpc}}
\newcommand{\Gpch}{\,h^{-1}{\rm Gpc}}

\newcommand{\hMpc}{\,h\,{\rm Mpc}^{-1}}

\newcommand{\Msh}{\,h^{-1}M_\odot}
\newcommand{\myvec}[1]{\boldsymbol{#1}}
\newcommand{\rd}{{\rm d}}

\newcommand\ringring[1]{%
  {
   \mathop{\kern0pt #1}\limits^{
     \vbox to-1.85ex{
       \kern-2ex 
       \hbox to 0pt{\hss\normalfont\kern.2em \r{}\kern-.1em \r{}\hss}%
       \vss 
     }
   }
  }
}

\title{Fast full $N$-body simulations of generic modified gravity: derivative coupling models}

\author[a,b]{C\'{e}sar Hern\'{a}ndez-Aguayo,}
\author[c]{Cheng-Zong Ruan,}
\author[c]{Baojiu Li,}
\author[c]{Christian Arnold,}
\author[c]{Carlton M. Baugh,}
\author[d]{Anatoly Klypin,}
\author[e]{and Francisco Prada}

\affiliation[a]{Max-Planck-Institut f\"ur Astrophysik, Karl-Schwarzschild-Str 1, D-85748 Garching, Germany}
\affiliation[b]{Excellence Cluster ORIGINS, Boltzmannstrasse 2, D-85748 Garching, Germany}
\affiliation[c]{Institute for Computational Cosmology, Department of Physics, Durham University, South Road, Durham DH1 3LE, UK}
\affiliation[d]{Astronomy Department, New Mexico State University, Las Cruces, NM 88001, USA}
\affiliation[e]{Instituto de Astrof\'{i}sica de Andaluc\'{i}a (CSIC), Glorieta de la Astronom\'{i}a, E-18080 Granada, Spain}

\emailAdd{cesarhdz@mpa-garching.mpg.de}
\emailAdd{cheng-zong.ruan@durham.ac.uk}
\emailAdd{baojiu.li@durham.ac.uk}
\emailAdd{christian.arnold@durham.ac.uk}
\emailAdd{c.m.baugh@durham.ac.uk}
\emailAdd{aklypin@nmsu.edu}
\emailAdd{f.prada@csic.es}

\abstract{We present {\sc mg-glam}, a code developed for the very fast production of full $N$-body cosmological simulations in modified gravity (MG) models. We describe the implementation, numerical tests and first results of a large suite of cosmological simulations for two broad classes of MG models with derivative coupling terms---the {\it Vainshtein}- and {\it Kmouflage}-type models---which respectively features the Vainshtein and Kmouflage screening mechanism. Derived from the parallel particle-mesh code \textsc{glam}, {\sc mg-glam} incorporates an efficient multigrid relaxation technique to solve the characteristic nonlinear partial differential equations of these models. For Kmouflage, we have proposed a new algorithm for the relaxation solver, and run the first simulations of the model to understand its cosmological behaviour. In a companion paper, we describe versions of this code developed for conformally-coupled MG models, including several variants of $f(R)$ gravity, the symmetron model and coupled quintessence. Altogether, \textsc{mg-glam} has so far implemented the prototypes for most MG models of interest, and is broad and versatile. The code is highly optimised, with a tremendous (over two orders of magnitude) speedup when comparing its running time with earlier $N$-body codes, while still giving accurate predictions of the matter power spectrum and dark matter halo abundance. \textsc{mg-glam} is ideal for the generation of large numbers of MG simulations that can be used in the construction of mock galaxy catalogues and accurate emulators for ongoing and future galaxy surveys.}

\begin{document}
\maketitle
\flushbottom

\section{Introduction}
\label{sec:intro}
The accelerated expansion of our Universe \cite{SupernovaCosmologyProject:1998vns,SupernovaSearchTeam:1998fmf} is one of the most challenging problems in modern physics, and after decades of attempts to find its origin, we are still far from reaching a clear conclusion. 
While the current standard cosmological model --- $\Lambda$ Cold Dark Matter ($\Lambda$CDM), which assumes that this accelerated expansion is caused by the cosmological constant, $\Lambda$ --- is in excellent agreement with most observational data to date, this model suffers from the well-known coincidence and fine-tuning problems. This suggests that a more fundamental theory is yet to be developed which can naturally explain the small observationally inferred value of $\Lambda$. The alternative theoretical models proposed so far can be roughly classified into two categories: one involves some exotic new matter species beyond the standard model of particle physics, the so-called \textit{dark energy} \cite{Copeland:2006wr}, which usually has a non-trivial dynamics; the other involves modifications of Einstein's \ac{GR} on certain (usually cosmic) scales \citep{Clifton:2011jh,2015PhR...568....1J,Koyama:2020zce}, or introduces new fundamental forces between matter particles\footnote{The two classes of models can not always be clearly distinguished, and some of the modified gravity models studied in this work can also considered as coupled dark energy.}. Some leading examples are quintessence \cite{Ratra:1988_quintessence,Wetterich:1988_quintessence,Zlatev:1998tr_quintessence,Steinhardt:1999nw_quintessence}, k-essence \cite{Armendariz-Picon:2000nqq:kessence,Armendariz-Picon:2000ulo_kessense}, coupled quintessence \cite{Amendola:1999er}, $f(R)$ gravity \cite{Sotiriou:2008rp,DeFelice:2010aj} and chameleon model  \cite{Khoury:2003aq,Khoury:2003rn,Mota:2006fz,Brax:2008hh}, symmetron model \cite{Hinterbichler:2010es,Hinterbichler:2011ca,Davis:2011pj}, the Dvali-Gabadadze-Porrati braneworld (DGP) model \cite{Dvali:2000hr}, scalar \cite{Nicolis:2008in,Deffayet:2009wt} and vector \cite{Heisenberg:2014_Proca,Allys:2015sht_Proca,BeltranJimenez:2016rff_Proca} Galileons, Kmouflage \cite{Babichev:2009ee}, massive gravity \cite[e.g.,][]{Hinterbichler:2011tt_massive_gravity}, etc.. 

In \ac{MG} models, in addition to a modified, and accelerated,  expansion rate that could explain observations, often the law of gravity is also different from \ac{GR}, which can further affect the evolution of the large-scale structure (LSS) of the Universe. This suggests that we can use various cosmological observations to constrain and test these models \citep[e.g.,][]{Koyama:2015vza,Ferreira:2019xrr,Baker:2019gxo}.
In this sense, the study of \ac{MG} models can be used as a testbed to verify the validity of \ac{GR} on cosmological scales, hence going beyond the usual small-scale or local tests of \ac{GR} \cite{Will:2014_GR_LRR}.

In the last two decades, there have been substantial progresses in the size and quality of cosmological observations, many of which can be excellent probes of dark energy and modified gravity \citep[e.g.,][]{Albrecht:2006um,Weinberg_2013}. 
Some of the leading probes studied in the literature include cosmic microwave background (CMB) \cite{Hinshaw:WMAP9,Hou:SPT2014,Planck2018,Aiola:ACT2020}, supernovae \cite{SupernovaCosmologyProject:1998vns,SupernovaSearchTeam:1998fmf,Astier:2006_SNLS_SN,Wood-Vasey:2007_ESSENCE_SN,Sullivan:2011_SNLS_SN,Scolnic:2013_Pan_STARRS_SN,Rest:2013_Pan_STARRS_SN,Abbott:2018_DES_SN,Betoule:2014_SDSS_SN,Jones:2018_Pan_STARRS_SN}, galaxy clustering \citep{Percival:2004_2dF_Galaxy_Clustering,Guzzo:2008_Galaxy_Clustering,Blake:2011_wiggleZ_Galaxy_Clustering,Beutler:2012_6dF_Galaxy_Clustering,Pezzotta:2016_VIMOS_Galaxy_Clustering,Alam:2017_BOSS_Galaxy_Clustering,Zarrouk:2018_SDSS_Galaxy_Clustering} and baryonic acoustic oscillations (BAO)  \cite{Cole2005_BAO,Eisenstein2005_BAO,Beutler2011_BAO,Blake2011_BAO,Anderson2012_BAO,eBOSS2020_BAO}, gravitational lensing \cite{Heymans2013:CFHTLenS_WL,Abbott:2020_DES_WL_Clusters,Hamana2020:HSC_WL,Amon2021:DES_WL,Secco2021:DES_WL}, and the properties of galaxy clusters \cite{Vikhlinin:2009_Xray_Cluster_count,Planck:2013_SZ_Cluster,Mantz:2014a_Cluster,Mantz:2014b_Cluster,SPT:2016_SZ_Cluster,SPT:2018_SZ_Cluster,Abbott:2020_DES_WL_Clusters,Giocoli:2021_KiDS_WL_Cluster}.
In the near future, a number of large, Stage-IV, galaxy and cluster surveys, such as DESI \citep{DESI:2016zmz}, Euclid \citep{Laureijs:2011gra,EuclidTheoryWorkingGroup:2012gxx}, Vera Rubin observatory \cite{lsst} and eROSITA \cite{erosita:2012arXiv1209.3114M}, are expected to revolutionise our knowledge about the Universe and our understanding of the cosmic acceleration, by providing cutting-edge observational data with unprecedented volume and much better controlled systematics. Further down the line, experiments such as CMB-S4 \cite{CMBS4} and LISA \cite{LISA} will offer other independent tests of models using by improved CMB observables, such as CMB lensing and the kinetic Sunyaev-Zel'dovich effect, and gravitational waves.

To exploit the next generation of observational data, we need to develop accurate theoretical tools to predict the cosmological implications of various models, in particular their behaviour on small scales which encode a great wealth of information. However, predicting LSS formation on small scales is a non-trivial work because structure evolution has entered the highly non-linear regime here, with a lot of complicated physical processes, such as gravitational collapse and baryonic interactions, being at play. The only tool that could accurately predict structure formation in this regime is cosmological simulations, which follow the evolution of matter through the cosmic time, from some initial, linear, density field all the way down to the highly-clustered matter distribution on small, sub-galactic, scales at late times. Modern cosmological simulation codes, e.g., \textsc{ramses} \cite{Teyssier:2001_RAMSES_code_paper}, \textsc{gadget} \cite{Springel:2005_Gadget_code_paper,Springel:2020plp}, \textsc{arepo} \cite{Springel:2010_AREPO_code_paper}, \textsc{pkdgrav} \cite{Potter:2016_PKDGRAV_code_paper}, \textsc{swift} \cite{Schaller:2016_SWIFT_code_paper}, have been able to employ hundreds of billions or trillions of particles in Giga-parsec volumes \cite[e.g.,][]{Angulo:2012_MXXL_sim_paper,Kim:2015_Horizon4_sim_paper,Potter:2016_PKDGRAV_code_paper}, and are nowdays indispensable in the confrontation of theories with observational data. In particular, to achieve the high level of precision required by galaxy surveys, one can generate hundreds or thousands of independent galaxy mocks that cover the expected survey volume, based on these simulations. However, this has so far been impossible for \ac{MG} models, which usually involve highly non-linear partial differential equations that govern the new physics, solving which has proven to be very expensive even with the latest codes, e.g., \textsc{ecosmog} \cite{Li:2011_ECOSMOG_code_paper,Li:2013_ECOSMOGV_code_paper,2012JCAP...10..002B,Brax:2013mua}, \textsc{mg-gadget} \cite{Puchwein:2013_MGGADGET_code_paper}, \textsc{isis} \cite{Llinares:2013_ISIS_code_paper} and \textsc{mg-arepo} \cite{Arnold:2019_MGAREPO_code_paper,Hernandez-Aguayo:2020_MGAREPO_code_paper} (see \cite{Winther:2015_MG_code_comparison} for a comparison of several MG codes). For example, current \ac{MG} simulations can take between $2$ to $\mathcal{O}(10)$ times longer than standard $\Lambda$CDM simulations of the same specifications. Obviously, to best explore the future observations for testing \ac{MG} models, we need a new simulation code for these models with greatly improved efficiency compared with the current generation of codes.

In this paper, we present such a code, {\sc mg-glam}, which is an extension of the parallel particle-mesh (PPM) $N$-body code {\sc glam}\footnote{{\sc glam} stands for GaLAxy Mocks, which is a pipeline for massive production of galaxy catalogues in the $\Lambda$CDM (GR) model.} \citep{Klypin:2017iwu}, where various important classes of modified gravity models are implemented. Efficiency is the main feature of \textsc{mg-glam}, which is partly thanks to the efficiency and optimisations it inherits from its base code,  \textsc{glam}\footnote{The \textsc{glam} code has been shown to be $1.6$--$4$ times faster than similar codes such as {\sc cola} \cite{Koda_15}, {\sc icecola} \cite{Izard_15} and {\sc fastpm} \cite{Feng_16}, while still achieving high resolution and accuracy.}, partly due to optimised numerical algorithms tailored to solve the nonlinear equations of motion in these modified gravity models, and partly thanks to a careful design of the code and data structures to reduce memory footprint of simulations.

Modified gravity models can be classified according to the fundamental properties of their new dynamical degrees of freedom, and the interactions the latter have. Here, we study two classes of MG models which introduce new scalar degrees of freedom that have derivative-coupling interactions: the normal-branch of the DGP \citep{Dvali:2000hr} braneworld model, which is a representative example of Vainshtein-type gravity models, and the Kmouflage model \citep{Babichev:2009ee}. These models generally introduce a new force (fifth force) between matter particles, but they can both employ screening mechanisms to evade Solar System constraints \citep{Vainshtein:1972sx,Babichev:2009ee} on the fifth force. These two models have been widely studied in recent years and, as we argue below, the implementation of them can lead to prototype MG codes that can be modified to work with minimal effort for other classes of interesting models. 
In a twin paper \citep{Ruan:2021_twin_paper}, we will describe the implementation and analysis of several other classes of MG models, including the coupled quintessence \cite{Amendola:1999er}, chameleon \cite{Khoury:2003aq,Khoury:2003rn} $f(R)$ gravity \citep{Hu:2007nk}, and symmetron models \citep{Hinterbichler:2010es,Hinterbichler:2011ca}, which are examples of conformally coupled scalar fields.

As we will demonstrate below, the inclusion of modified gravity solvers in \textsc{mg-glam} adds an overhead to the computational cost of \textsc{glam}, and for the models considered in this paper and its twin paper \cite{Ruan:2021_twin_paper}, a \textsc{mg-glam} run can take about $2$-$5$ times (depending on the resolution) the computing time of an equivalent $\Lambda$CDM simulation run using default \textsc{glam}. All in all, this makes this new code at least around $100$ times faster than other modified gravity simulation codes such as \textsc{ecosmog} \cite{Li:2011_ECOSMOG_code_paper,Li:2013_ECOSMOGV_code_paper,2012JCAP...10..002B,Brax:2013mua} and \textsc{mg-arepo} \cite{Arnold:2019_MGAREPO_code_paper,Hernandez-Aguayo:2020_MGAREPO_code_paper} for the same simulation boxsize and particle number. In spite of such a massive improvement in speed over those latter codes, it is worthwhile to note that \textsc{mg-glam} is \textit{not} an approximate code: it solves the full Poisson and MG equations, and its accuracy is only limited by the resolution of the PM grid used, which can be specified by users based on their particular scientific objectives. This makes it different from fast approximate simulation codes such as those  \cite{Winther:2017j_MGCOLA_code_paper,Wright:2017_NUCOLA_code_paper,Valogiannis:2016_MGCOLA_code_paper,Fiorini:2021_MGCOLA_application_haloes} based on the COmoving Lagrangian Acceleration method (\textsc{cola}) \cite{Tassev:2013_COLA_code_paper}.

The paper is organised as follows. Section~\ref{sec:models} presents the theoretical aspects of the modified gravity models studied here. In Section~\ref{sec:code} we discuss the numerical implementation of {\sc mg-glam}. The description and results of several code tests are shown in Section~\ref{sec:code_tests} and in Section~\ref{sec:cosmo_sims} we analyse the nonlinear power spectra and halo mass functions of the first derivative coupling models performed with {\sc mg-glam}. Finally, we summarise the main results and give our conclusions in Section~\ref{sec:conclusions}. 

Throughout this paper, we adopt the usual conventions that Greek indices label all space-time coordinates ($\mu,\nu,\cdots=0,1,2,3$), while Latin indices label the space coordinates only ($i,j,k,\cdots=1,2,3$). Our metric signature is $(-,+,+,+)$. We will strive to include the speed of light $c$ explicitly in relevant equations, rather than setting it to $1$, given that in numerical implementations $c$ must be treated carefully. Unless otherwise stated, the symbol $\approx$ means `approximately equal' or `equal under certain approximations as detailed in the text', while the symbol $\simeq$ means that two quantities are of similar order of magnitude. An overdot denotes the derivative with respect to (wrt) the cosmic time $t$, e.g., $\dot{a} \equiv {\dd{a}}/{\dd{t}}$ and the Hubble expansion rate $H(a)$ is defined as $H=\dot{a}/a$, while a prime ($'$) denotes the derivative wrt the conformal time $\tau$, e.g., $a'=\dd{a}/\dd{\tau}$, $\mathcal{H}(a)\equiv{a}'/a=aH(a)$. Unless otherwise stated, we use a subscript $_0$ to denote the present-day value of a physical quantity, an overbar for the background value of a quantity, and a tilde for quantities written in code units.

We note that, since they have a lot in common, including the motivation and the design of code structure and algorithms, this paper has identical or similar texts with its twin paper \cite{Ruan:2021_twin_paper} in the Introduction section, as well as in Sections~\ref{subsect:glam}, \ref{sec:glam_units}, \ref{subsect:extradof} until \ref{subsubsect:relaxation}, \ref{subsubsect:relaxation}, \ref{subsubsect:code_struc}, the last paragraph of \ref{sec:BG_solvers}, and part of \ref{sec:bkg_tests}. 

\section{Modified gravity models with derivative coupling terms}
\label{sec:models}
In this section we briefly introduce the modified gravity models with derivative coupling terms that are implemented in the {\sc mg-glam} code. We start with the general action of scalar field models in the Einstein frame, 
\begin{equation}\label{eq:SF_general}
S = \int{\rm d}^4x\sqrt{-g}\left[\frac{M^2_{\rm Pl}}{2}R + K\qty[(\nabla\phi)^2,(\nabla^2\phi)] - V(\phi)\right] + \int{\rm d}^4x\sqrt{-\hat{g}}\hat{\mathcal{L}}_m\left[\hat{\psi}^{(i)}_m,\hat{g}_{\mu\nu}\right]\,,
\end{equation}
where $g$ is the determinant of the metric tensor $g_{\mu\nu}$, $M_{\rm Pl} (= 1/\sqrt{8\pi G})$ is the reduced Planck mass, $G$ is Newton's constant, $R$ is the Ricci scalar, $K$ is a general kinetic function which contains nonlinear terms of the derivatives of the scalar field, $V(\phi)$ the potential energy of the scalar field $\phi$, $\hat{\psi}^{(i)}_m$ are the matter fields, and $\hat{g}_{\mu\nu}$ is the Jordan-frame metric that couples to them.

The Jordan-frame metric $\hat{g}_{\mu\nu}$ and Einstein-frame metric $g_{\mu\nu}$ are assumed to be related to each other by the following conformal mapping,
\begin{equation}\label{eq:conformal_metric}
    \hat{g}_{\mu\nu} = A^2(\phi)g_{\mu\nu}\,,
\end{equation}
where $A$ is a function of the scalar field $\phi$. Disformal relations between the two metrics are possible, but they are not considered here. 

By varying the action Eq.~\eqref{eq:SF_general} with respect to the scalar field, we obtain the following equation of motion
\begin{equation}\label{eq:sf_eom_general}
    \frac{1}{\sqrt{-g}}\partial_\mu\qty[\sqrt{-g}\partial^\mu\phi K^\prime]  = \frac{{\rm d}\ln A(\phi)}{{\rm d}\phi}\rho_m + \frac{{\rm d}V(\phi)}{{\rm d}\phi},
\end{equation}
where $\rho_m$ is the density of non-relativistic matter. We define the coupling strength $\beta(\phi)$ as a dimensionless function of $\phi$:
\begin{equation}\label{eq:csf_beta_general}
    \beta(\phi) \equiv M_{\rm Pl}\frac{{\rm d}\ln A(\phi)}{{\rm d} \phi}.
\end{equation}
Note the $M_{\rm Pl}$ in this definition, which is because $\phi$ has mass dimension 1. For later convenience, we shall define a dimensionless scalar field as
\begin{equation}
    \varphi \equiv \frac{\phi}{M_{\rm Pl}}. 
\end{equation}

Two classes of models of Eq.~\eqref{eq:SF_general} are of particular interest in the literature. The first is what we call `Vainshtein-type' modified gravity models, which employs the \textit{Vainshtein screening mechanism} \cite{Vainshtein:1972sx} to decouple the scalar field from matter in regions where the second derivatives of the field are large. The second is the `Kmouflage-type' gravity models, which employs the \textit{Kmouflage screening mechanism} \cite{Babichev:2009ee,Brax:2014gra} to hide the effect of the scalar field in regions where the field has a large gradient. In the next subsections we describe the theoretical aspects of both Vainshtein-type and Kmouflage-type gravity models.

\subsection{Vainshtein-type gravity}
\label{sec:vainshtein}
An excellent example of Vainshtein-type models is the Galileon model \cite{Nicolis:2008in} and its covariant extension \cite{Deffayet:2009wt}, which is a generic description of self-interacting scalar field models whose Lagrangian is invariant under the Galilean shift, $\partial_\mu\varphi\rightarrow\partial_\mu\varphi+b_\mu$, with $b_\mu$ being a constant 4-vector. Simulations of these models have been carried out previously, e.g., \cite{Barreira:2013eea,Li:2013tda}, along with other approaches to studying the nonlinear structure formation in these models, e.g., \cite{Barreira:2013xea}. In recent years, the vector Galileon, or generalised Proca, theory has attracted attentions, e.g., \cite{Heisenberg:2014rta,Allys:2015sht,BeltranJimenez:2016rff}. As the Galileon model, these models also employ the Vainshtein screening mechanism to suppress the effect of modified gravity in regions where the second derivative of the field is large. But unlike Galileons, here the dynamical degrees of freedom are the spatial components of some vector field, whose transverse mode plays a negligible role in cosmic structure formation \cite{Becker:2020azq} while the longitudinal mode behaves like the Galileon field $\varphi$ (with the difference that the vector field has no dynamics on the background). Simulations of vector Galileons have been recently carried out in \cite{Becker:2020azq,Becker:2020yim}. These models have rich phenomenology, able to modify the background expansion history as well as the gravitational potential, and hence propagate a modified gravity---or fifth---force between matter particles and affect large-scale structure formation.

In this paper, we consider another class of models that realise the Vainshtein screening mechanism, the Dvali-Gabadadze-Porrati (DGP) \citep{Dvali:2000hr} brane-world model, as our toy Vainshtein-type gravity model. This choice is for a few reasons. First, the DGP model has been very popular in the literature, being widely used as a testbed for the Vainshtein mechanism. Second, it has great flexibility in terms of the background expansion history (although there is a catch as we will see later), and usually one can make the model have an expansion rate identical to that of $\Lambda$CDM, to focus on the anaysis of the effects of the fifth force. Finally and more importantly, owing to its simplicity, this model can be used as a prototype for all Vainshtein-type models, to understand the effects of the screening mechanism; a simulation code model can be easily modified to simulate the Galileon and vector Galileon models, as well as generalised Galileons \cite{DeFelice:2010nf} and kinetic-gravity braiding models \cite{Deffayet:2010qz}, which all share a similar equation of motion for the dynamical field.

In the DGP model, the Universe is a four-dimensional `brane' embedded in a five-dimensional spacetime, or bulk. The total action of the model is written by,
\begin{equation}
S = \int_{\rm brane} {\rm d}^4x \sqrt{-g} \frac{R}{16\pi G} + \int {\rm d}^5x \sqrt{-g^{(5)}} \frac{R^{(5)}}{16\pi G^{(5)}} +\,\, S_{\rm m}(g_{\mu\nu},\psi_i)\,,\label{eq:S_dgp}
\end{equation}
where $g_{\mu\nu}$, $g$, $R$ and $G$ are respectively the metric tensor, the determinant of the metric, the Ricci scalar and the gravitational constant in the 4-D brane, while $g^{(5)}$, $R^{(5)}$ and $G^{(5)}$ are their equivalents in the 5-D bulk, and $S_{\rm m}$ is the action of the matter fields $\psi_i$ which are assumed to be confined on the brane. 

A new parameter can be introduced, which is defined as the ratio of $G^{(5)}$ and $G$ and known as the \textit{crossover scale}, $r_{\rm c}$,
\begin{equation}\label{eq:rc}
r_{\rm c} = \frac{1}{2} \frac{G^{(5)}}{G}\,.
\end{equation}
It has the physical meaning of being roughly the scale at which the behaviour of gravity transitions from 4-D standard Einsteinian ($r\ll r_{\rm c}$) to 5-D ($r\gg r_{\rm c}$), where gravitons could leak into the fifth dimension.

Here we study the normal-branch (nDGP) model, where the variation of the action, Eq.~\eqref{eq:S_dgp}, yields  the modified Friedmann equation 
\begin{equation}\label{eq:H_ndgp}
\frac{H(a)}{H_0} = \sqrt{\Omega_{\rm m}a^{-3} + \Omega_{\rm DE}(a) + \Omega_{\rm rc}} - \sqrt{\Omega_{\rm rc}},
\end{equation}
in a homogeneous and isotropic universe with $\Omega_{\rm rc} \equiv c^2/(4H^{2}_{0}r^2_{\rm c})$ where $c$ is the speed of light, $\Omega_{\rm m}$ is the present-day value of the matter density parameter, the dark energy density parameter $\Omega_{\rm DE}(a)$ is defined as $\Omega_{\rm DE}(a) \equiv 8\pi G\rho_{\rm DE}(a)/3H^2(a)$, $a$ is the scale factor and $H_0$ is the present-day value of the Hubble parameter. The nDGP model on its own cannot lead to an accelerated Hubble expansion, which is why an extra dark energy component has to be added to match observational data: because there is not much \textit{a priori} requirement on this dark energy component, it is often assumed to have such an equation of state that the overall effect of Eq.~\eqref{eq:H_ndgp} is to give a $\Lambda$CDM expansion history (note that this is not possible if this dark energy component is assumed to be a cosmological constant); also, the dark energy component is assumed to be non-clustering so that its effect is only on the background expansion. In this model, deviations from GR can be characterised in terms of the parameter $H_0 r_{\rm c} / c$. As we can see from Eq.~\eqref{eq:H_ndgp} if $H_0 r_{\rm c} / c \rightarrow \infty$ then the equation of state of the dark energy component approaches $-1$ in order to produce a $\Lambda$CDM expansion history. 

The structure formation in the nDGP model is governed by the Poisson and scalar equations in the quasi-static and weak-field limits: \citep{Koyama:2007ih},
\begin{equation}\label{eq:poisson_nDGP}
\nabla^2 \Phi = 4\pi G a^2 \delta \rho_{\rm m} + \frac{1}{2}\nabla^2\varphi\,,
\end{equation}
\begin{equation}\label{eq:phi_dgp}
\nabla^2 \varphi + \frac{r_{\rm c}^2}{3\beta_{\rm DGP}(a)a^2c^2} \left[ (\nabla^2\varphi)^2 - \nabla_i\nabla_j\varphi\nabla^i\nabla^j\varphi \right] = \frac{8\pi\,G\,a^2}{3\beta_{\rm DGP}(a)} \delta\rho_{m}\, ,
\end{equation}
where $\varphi$ is a scalar degree of freedom related to the bending modes of the brane (which describes the position of the brane in the fifth dimension), the total modified gravitational potential $\Phi$ is given by $\Phi=\Phi_{\rm N}+\frac{1}{2}\varphi$ with $\Phi_{\rm N}$ being the standard Newtonian potential, $\delta\rho_{\rm m} = \rho_{\rm m} - \bar{\rho}_{\rm m}$ is the perturbation of non-relativistic matter density, and
\begin{equation}\label{eq:beta_dgp}
\beta_{\rm DGP}(a) = 1 + 2 H\, r_{\rm c} \left ( 1 + \frac{\dot H}{3 H^2} \right ) = 1 + \frac{\Omega_{\rm m}a^{-3} + 2\Omega_\Lambda}{2\sqrt{\Omega_{\rm rc}(\Omega_{\rm m}a^{-3} + \Omega_{\Lambda})}}\,.
\end{equation}
In the last expression we have used the above assumption that the nDGP model has the same expansion history as the $\Lambda$CDM model, i.e., the Hubble parameter is written as
\begin{equation}\label{eq:H_lcdm}
H(a) = H_0\sqrt{\Omega_{\rm m}a^{-3} + \Omega_{\Lambda}}\,,
\end{equation}
where $\Omega_\Lambda$ is the contribution of $\Lambda$ in the $\Lambda$CDM model, defined as $\Omega_\Lambda \equiv 1 - \Omega_{\rm m}$. Note that throughout this paper we assume that the Universe is spatially flat, and neglect the contribution by radiation unless otherwise stated. 

From Eq.~\eqref{eq:poisson_nDGP}, it is straightforward to identify the modified gravity contribution to the gravitational acceleration,
\begin{equation}
\boldsymbol{a}_{\rm MG} = -\frac{1}{2} \myvec{\nabla} \varphi.
\label{eq:dgp_accel}
\end{equation}

If we linearise Eq.~\eqref{eq:phi_dgp}, the two nonlinear terms in the squared brackets vanish and the modified Poisson equation, Eq.~\eqref{eq:poisson_nDGP}, can be re-expressed as 
\begin{equation}\label{eq:dgp_linear}
 \nabla^2 \Phi = 4\pi G a^2 \left( 1 + \frac{1}{3\beta_{\rm DGP}} \right) \delta\rho_{\rm m},
\end{equation}
which represents a time-dependent and scale-independent rescaling of Newton's constant. Since $\beta_{\rm DGP}$ is always positive, the formation of structure is enhanced in this model with respect to $\Lambda$CDM.

The linear growth for the matter fluctuations in the nDGP model can be obtained by solving the equation of the linear growth factor, $D$,
\begin{equation}\label{eq:Dp}
\frac{{\rm d}^2D}{{\rm d}N^2} + \qty[2 - \frac{3}{2}\Omega_{\rm m}(a)]\frac{{\rm d}D}{{\rm d}N} - \frac{3}{2}\Omega_{\rm m}(a)\qty[1 + \frac{1}{3\beta_{\rm DGP}(a)}] D = 0\,,
\end{equation}
where $N=\ln(a)$, and $1/3\beta_{\rm DGP}$ is the ratio between the strengths of the fifth and standard Newtonian forces in the linear regime, which is scale independent (see derivation below). 

\subsubsection{Vainshtein screening mechanism}
\label{sec:Vainshtein_screening}
As mentioned above, the nDGP model is a representative class of modified gravity models that feature the Vainshtein screening mechanism \citep{Vainshtein:1972sx}. To illustrate how the Vainshtein mechanism works, let us for simplicity consider solutions in spherical symmetry, where Eq.~\eqref{eq:phi_dgp} can be written in the following form
\begin{equation}\label{eq:dgp}
\frac{2r_{\rm c}^2}{3\beta_{\rm DGP} c^2 a^2}\frac{1}{r^2}\frac{\rd}{\rd r}\left[r\left(\frac{\rd\varphi}{\rd r}\right)^2\right] + \frac{1}{r^2}\frac{\rd}{\rd r}\left[r^2\frac{\rd\varphi}{\rd r}\right]
= \frac{8\pi G}{3\beta_{\rm DGP}}\delta\rho_{\rm m}a^2\,.
\end{equation}
Defining the excess mass enclosed in radius $r$ as
\begin{equation}\label{eq:Mass_r}
M(r) \equiv 4\pi\int^r_0\delta\rho_{\rm m}(r')r'^2{\rd}r',
\end{equation}
we can rewrite Eq.~(\ref{eq:dgp}) as
\begin{equation}\label{eq:dgp_eqn}
\frac{2r_{\rm c}^2}{3\beta_{\rm DGP} c^2}\frac{1}{r}\left(\frac{\rd\varphi}{\rd r}\right)^2 + \frac{\rd\varphi}{\rd r} = \frac{2}{3\beta_{\rm DGP}}\frac{GM(r)}{r^2}\ \equiv\ \frac{2}{3\beta_{\rm DGP}}g_{\rm N}(r),
\end{equation}
in which for simplicity we have set $a=1$, and $g_{\rm N}$ is the Newtonian acceleration caused by the mass $M(r)$ at distance $r$ from the centre, Eq.~\eqref{eq:Mass_r}. 

If we further assume that $\delta\rho_{\rm m}$ is a constant within a radius $R$ and zero outside,  then Eq.~\eqref{eq:dgp_eqn} has the physical solution
\begin{equation}\label{eq:dvarphidr_in}
\frac{\rd\varphi}{\rd r} = \frac{4}{3\beta_{\rm DGP}}\frac{r^3}{r_{\rm V}^3}\left[\sqrt{1+\frac{r_{\rm V}^3}{r^3}}-1\right]g_{\rm N}(r),
\end{equation}
for $r\geq R$ and
\begin{equation}\label{eq:dvarphidr_out}
\frac{\rd\varphi}{\rd r} = \frac{4}{3\beta_{\rm DGP}}\frac{R^3}{r_V^3}\left[\sqrt{1+\frac{r_V^3}{R^3}}-1\right]g_{\rm N}(r),
\end{equation}
for $r\leq R$. In these expressions $r_{\rm V}$ is the \textit{Vainshtein radius} which is defined as
\begin{equation}\label{eq:r_V}
r_{\rm V} \equiv \left[\frac{8r^2_{\rm c} r_{\rm S}}{9\beta_{\rm DGP}^2}\right]^{1/3} = \left[\frac{4GM(R)}{9\beta_{\rm DGP}^2H^2_0\Omega_{\rm rc}}\right]^{1/3}\,,
\end{equation}
where $r_{\rm S} \equiv2GM(R)/c^2$ is the Schwarzschild radius and $M(R) \equiv 4\pi\int^R_0\delta\rho_{\rm m}(r')r'^2{\rd}r'$ is the total mass of the spherical object.

According to Eq.~\eqref{eq:poisson_nDGP}, the fifth force is given by $\frac{1}{2}\rd\varphi/\rd r$. Therefore at $r\gg r_{\rm V}$ we have
\begin{equation}\label{eq:fifth-force-ratio-asymptotic}
\frac{1}{2}\frac{\rd\varphi}{\rd r} \rightarrow \frac{1}{3\beta_{\rm DGP}}g_N(r),
\end{equation}
indicating that on scales larger than the Vainshtein radius gravity is enhanced (because $\beta_{\rm DGP}>0$ for the nDGP model) by a scale-independent factor $1/3\beta_{\rm DGP}$. On the other hand, for $r,R\ll r_{\rm V}$ we have
\begin{equation}\label{eq:V_screening}
\frac{1}{2}\frac{\rd\varphi}{\rd r} \rightarrow \frac{2}{3\beta_{\rm DGP}}\left[\frac{R}{r_{\rm V}}\right]^{3/2}g_{\rm N}(r) \ll g_{\rm N}(r),
\end{equation}
indicating that the fifth force is suppressed (or screened), relative to the Newtonian force, well within the Vainshtein radius.

\subsection{Kmouflage-type gravity}
\label{sec:kmouflage}
The Kmouflage model \citep{Babichev:2009ee} is another class of screened modified gravity models, in which $V(\phi) = 0$ and the scalar field satisfies an equation of motion, Eq.~\eqref{eq:sf_eom_general}, that takes the following form \citep{Brax:2014wla,Brax:2014yla}:
\begin{equation}\label{eq:kmouflage_eom}
    \nabla^i\left[K_X\left(X\right)\nabla_i\varphi\right] = {8\pi G}\frac{{\rm d}\ln A(\varphi)}{{\rm d}\varphi}a^2\delta\rho_m.
\end{equation}
where $K(X)$ is the kinetic function in Eq.~\eqref{eq:SF_general} which needs to be specified for a given model, which has mass dimension four, $A(\varphi)$ is the coupling function between the scalar field and matter, which in this work we assume to take the exponential form:
\begin{equation}
    A(\varphi) = \exp\left(\beta_{\rm Kmo}\varphi\right) = \exp\left(\beta_{\rm Kmo}\frac{\phi}{M_{\rm Pl}}\right),
\end{equation}
$\beta_{\rm Kmo}$ is a constant model parameter, $K_X = {\rm d}K(X)/{\rm d}X$ for a given function $K(X)$. For convenience, from here on we specify to the dimensionless versions of $K(X)$ and $X$---which for simplicity are still denoted by the same notations---where the dimensionless $K$ will be defined the dimensional kinetic function $K$ in Eq.~\eqref{eq:SF_general} divided by $\Lambda^4$, and
\begin{equation}
    X \equiv -\frac{M_{\rm Pl}^2}{2\Lambda^4}\nabla^\mu\varphi\nabla_\mu\varphi = \frac{M_{\rm Pl}^2}{2\Lambda^4}\dot{\bar{\varphi}}^2 - \frac{M_{\rm Pl}^2}{2\Lambda^4}a^{-2}\nabla^i\varphi\nabla_i\varphi, 
\end{equation}
is a dimensionless quantity and $\Lambda$ is a model parameter of mass dimension 1 related to dark energy. $\bar{\varphi}$ is the background value of the scalar field $\varphi$, $\nabla^i$ is raised by the metric $\delta^{ij}$, and the $a^{-2}$ is because by default $X$ should use the physical derivatives while here we have written things using the comoving derivatives. 

In addition to featuring a qualitatively different---and less explored---screening mechanism, the Kmouflage model can also be considered as a natural generalisation of the well-known \textit{k-essence} model \cite{Armendariz-Picon:2000nqq,Armendariz-Picon:2000ulo} by allowing a direct coupling of the k-essence scalar field with matter via the coupling function $A(\varphi)$. Furthermore, the equation of motion in the Kmouflage model, Eq.~\eqref{eq:kmouflage_eom}, is featured in other models, such as the charged dark matter model proposed in \cite{Jimenez:2020bgw} and the covariant models of MOdified Newtonian Dynamics (MOND; e.g., \cite{Milgrom:1983ApJ...270..365M,Bekenstein:1984ApJ...286....7B}). Thus, a simulation code for Kmouflage can be a prototype for simulating these other models. There has been very little work on the simulations of Kmouflage models so far, and in this work we will develop a code to do this\footnote{We note there have been codes to simulate MOND, e.g., \cite{Llinares:2008ce}, though our algorithm in this work will be different.}.

For convenience, we define a dimensionless parameter $\lambda$ so that
\begin{equation}
    \frac{\Lambda^4}{M^2_{\rm Pl}} \equiv H_0^2\lambda^2,
\end{equation}
and $X$ can be rewritten more as
\begin{equation}
    X = \frac{1}{2H_0^2\lambda^2}\dot{\bar{\varphi}}^2 - \frac{c^2}{2a^2H_0^2\lambda^2}\nabla^i\varphi\nabla_i\varphi,
\end{equation}
where we have explicitly included a factor containing the speed of light $c$. Note that the parameter $\lambda$ satisfies $\lambda\sim\mathcal{O}(1)$, because the model parameter $\Lambda$ is chosen such that it plays the role of accelerating the cosmic expansion at late times, meaning that at low $z$ we have $\Lambda^4/M^2_{\rm Pl}\sim8\pi{G}\rho_{\rm DE}/3\sim{H_0^2}\Omega_{\rm DE}$. We will describe how to determine the numerical value of $\lambda$ in the \textsc{mg}-\textsc{glam} code later.

A possible choice of the function $K(X)$ that has been studied previously \citep{Brax:2014wla,Brax:2014yla,Barreira:2014gwa,Barreira:2015aea} is
\begin{equation}\label{eq:kmf_model_K_function}
    K(X) = -1 + X + K_0X^n,
\end{equation}
where the integer $n$ satisfies $n\geq2$ and $K_0$ is a dimensionless model parameter. In this model, the modified Poisson equation is given by,
\begin{equation}\label{eq:poisson_Kmo}
\nabla^2 \Phi = 4\pi G a^2 A(\varphi) \delta \rho_{\rm m}\,,
\end{equation}
and the total force on matter particles is given by
\begin{equation}\label{eq:kmo_total_force}
\frac{{\rm d}^2\boldsymbol{r}}{{\rm d}t^2} = -\boldsymbol{\nabla}\Phi - c^2\beta_{\rm Kmo}\boldsymbol{\nabla}\varphi - \beta_{\rm Kmo}\dot{\varphi}\frac{{\rm d}\boldsymbol{r}}{{\rm d}t}\,,
\end{equation}
where $\boldsymbol{r}$ is the particle coordinate, $t$ is the physical time, and ${\rm d}\boldsymbol{r}/{\rm d}t$ is the peculiar velocity and 
\begin{equation}
    \beta_{\rm Kmo}(\varphi) \equiv \frac{{\rm d}\ln A(\varphi)}{{\rm d}\varphi} = \beta_{\rm Kmo}.
\end{equation}
The force equation can be rewritten as
\begin{equation}\label{eq:force_Kmo}
    \frac{{\rm d}\boldsymbol{p}}{{\rm d}t} = -\boldsymbol{\nabla}\Phi_{\rm N} - c^2\beta_{\rm Kmo}\boldsymbol{\nabla}\varphi - a^2\beta_{\rm Kmo}\dot{\bar{\varphi}}\frac{{\rm d}\boldsymbol{x}}{{\rm d}t},
\end{equation}
where $\boldsymbol{x}$ is the comoving coordinate and the $\boldsymbol{\nabla}$ symbol denotes the \textit{comoving} gradient, with $\boldsymbol{p}\equiv{a}^2\dot{\boldsymbol{x}}$.

The linearised version of the full Kmouflage equation of motion, Eq.~\eqref{eq:kmouflage_eom}, is
\begin{equation}\label{eq:linearised_Kmo}
\nabla^2 \varphi = 8\pi G a^2 \qty[1+nK_0\left(\frac{\bar{\varphi}'^2}{2\lambda^2a^2H_0^2}\right)^{n-1}]^{-1}\beta^2_{\rm Kmo}e^{\beta_{\rm Kmo}\bar{\varphi}} \delta \rho_{\rm m}\,.     
\end{equation}

For completeness, here is the linear growth equation for matter density contrast $\delta$ (or the linear growth factor itself) in the Kmouflage model:
\begin{equation}\label{eq:kmo_lin_growth}
    \delta'' + \left[\frac{a'}{a}+\frac{{\rm d}\ln{A}(\varphi)}{{\rm d}\varphi}\varphi'\right]\delta' - 4\pi{G}\bar{\rho}_{\rm m}(a){a}^2A(\bar{\varphi})\left[1+\frac{2\beta_{\rm Kmo}^2}{K_X(\bar{X})}\right]\delta = 0,
\end{equation}
where $'$ denotes the derivative with respect to the conformal time $\tau$, and $K_X={\rm d}K/{\rm d}X$ as above---in our case
\begin{equation}\label{eq:K_Xbar}
    K_X(\bar{X}) = 1+nK_0\bar{X}^{n-1} = 1+nK_0\left(\frac{\bar{\varphi}'^2}{2\lambda^2a^2H_0^2}\right)^{n-1}.
\end{equation}
Therefore, we can already observe four effects the Kmouflage scalar field has on structure formation: (\textit{i}) the modified expansion history, cf.~$a'/a$; (\textit{ii}) a fifth force which can (but may not) be screened by the Kmouflage mechanism, described by $2\beta^2_{\rm Kmo}/K_X$; (\textit{iii}) a rescaling of the matter density field by $A(\varphi)\neq1$ in the Poisson equation, implying that the matter particle mass is effectively modified; and (\textit{iv}) a velocity-dependent force\footnote{This force is similar to the `frictional' force on particles caused by the cosmic expansion, but we refrain from using the word `frictional' because, as we will see below, in our Kmouflage model it points to the same, rather the opposite, direction of the particle velocity.} described by the term involving $\left({\rm d}\ln A/{\rm d}\varphi\right)\varphi'\delta'$. The fifth force has a ratio of $2\beta^2_{\rm Kmo}/K_X$ to the Newtonian force, and this will be derived explicitly shortly. 

\subsubsection{The  Kmouflage screening mechanism}
\label{sec:Kmo_screening}
Similarly to the Vainshtein screening mechanism, let us consider the static and spherically symmetric form of the Kmouflage equation of motion, Eq.~\eqref{eq:kmouflage_eom},
\begin{equation}\label{eq:Kmo_eom_spherical}
\frac{1}{r^2}\frac{\rd}{\rd r}\qty[r^2 K_X\frac{\rd\varphi(r)}{\rd r}] = 8\pi G\beta_{\rm Kmo} \delta\rho_m\,,
\end{equation}
which can be integrated once to give,
\begin{equation}\label{eq:Kmo_eom_spherical1}
K_X\frac{\rd\varphi(r)}{\rd r} = 2\beta_{\rm Kmo} \frac{GM(r)}{r^2}\ \equiv\ 2\beta_{\rm Kmo}g_{\rm N}(r)\,,
\end{equation}
in which for simplicity we have set $a=1$, and $g_{\rm N}$ is the Newtonian acceleration caused by the mass $M(r)$ at distance $r$ from the centre, Eq.~\eqref{eq:Mass_r}. For a spherical symmetric object of radius $R$, we can define with mass $M(R)$, we can define the so-called Kmouflage radius
\begin{equation}\label{eq:r_K}
r_{\rm K} = \frac{2\beta_{\rm Kmo}GM(R)M_{\rm pl}}{c\Lambda^2}\,.
\end{equation}
From Eq.~\eqref{eq:Kmo_eom_spherical1} we can see that the fifth force, $F_{\rm 5th} \propto \beta_{\rm Kmo}\rd\varphi/\rd r$ (cf.~Eq.~\eqref{eq:kmo_total_force}), is suppressed for $r<r_{\rm K}$ where $\abs{X}$ is large if $K_X(X)\gg1$,
\begin{equation}
 \frac{\rd\varphi}{\rd r} = \frac{2\beta_{\rm Kmo}}{K_X}g_{\rm N}(r) \ll g_{\rm N}(r)\,.
\end{equation}
The condition for screening, $r<r_{\rm K}$, can be written as
\begin{equation}
\frac{GM}{r^2} > \frac{\Lambda^2}{2\beta_{\rm Kmo} M_{\rm pl}}\,.
\end{equation}

In the linear perturbation regime, we can neglect the contribution to $X$ by the spatial derivatives and therefore $K$, $K_X$ become purely time-dependent quantities, leading to a constant ratio, 
\begin{equation}
    \frac{\beta_{\rm Kmo}{\rm d}\varphi/{\rm d}r}{g_{\rm N}(r)} = \frac{2\beta^2_{\rm Kmo}}{K_X(\bar{X})}, 
\end{equation}
between the strengths of the fifth and standard Newtonian forces. This is what appears in Eq.~\eqref{eq:kmo_lin_growth}.

\section{Numerical Implementation}
\label{sec:code}
This section is the core part of this paper, where we will describe in detail how the different theoretical models of \S\ref{sec:models} can be incorporated in a numerical simulation code, so that the scalar degree of freedom can be solved at any given time with any given matter density field. This way, the various effects of the scalar field on cosmic structure formation can be accurately predicted and implemented. 

\subsection{The {\sc glam} code}
\label{subsect:glam}
The \textsc{glam} code is presented in \cite{Klypin:2017iwu}, and is a promising tool to quickly generate $N$-body simulations with reasonable speed and acceptable resolution, which are suitable for the massive production of galaxy survey mocks. 

As a PM code, \textsc{glam} solves the Poisson equation for the gravitational potential in a periodic cube using fast Fourier Transformation (FFT). The code uses a 3D mesh for density and potential estimates, and only one mesh is needed for the calculation: the density mesh is replaced with the potential. The gravity solver uses FFT to solve the discrete analogue of the Poisson equation, by applying it first in $x$- and then to $y$-direction, and finally transposing the matrix to improve data locality before applying FFT in the third ($z$-)direction. After multiplying this data matrix by the Green's function, an inverse FFT is applied, performing one matrix transposition and three FFTs, to compute the Newtonian potential field on the mesh. The potential is then differentiated using a standard three-point finite difference scheme to obtain the $x,y$ and $z$ force components at the centres of the mesh cells. These force components are next interpolated to the locations of simulation particles, which are displaced using a leapfrog scheme. A standard Cloud-in-Cell (CIC) interpolation scheme is used for both the assignment of particles to calculate the density values in the mesh cells and the interpolation of the forces.  

A combination of parameters that define the resolution and speed of the \textsc{glam} code are carefully selected. For example, it uses the \textsc{FFT}5 code (the Fortran 90 version of \textsc{FFTpack}5.1) because it has an option of real-to-real FFT that uses only half of the memory as compared to \textsc{FFTW}. It typically uses $1/2$--$1/3$ of  the number of particles (in 1D) as compared with the mesh size---given that the code is limited by available RAM, this is a better combination than using the same number of particles and mesh points.

\textsc{glam} uses \textsc{openmp} directives to parallelise the solver. Overall, the code scales nearly perfectly, as has been demonstrated by tests run with different mesh sizes and on different processors (later in the paper we will present some actual scaling test of \textsc{mg-glam} as well, which again is nearly perfect). \textsc{mpi} parallelisation is used only to run many realisations on different supercomputer nodes with very little inter-node communications. Load balance is excellent since theoretically every realisation requires the same number of CPUs. 

Initial conditions are generated on spot by \textsc{glam}, using the standard Zel'dovich approximation \cite{Zeldovich:1970A&A.....5...84Z,Efstathiou:1985ApJS...57..241E} from a user-provided linear matter power spectrum $P(k)$ at $z=0$. The code backscales this $P(k)$ to the initial redshift $z_{\rm ini}$ using the linear growth factor for $\Lambda$CDM with the specified cosmological parameters. Since the Zel'dovich approximation is less accurate at low redshifts \cite{Crocce:2006ve}, the simulation is typically started at an initial redshift $z_{\rm ini}\geq100$.

\textsc{glam} uses a fixed number of time steps, but this number can be specified by the user. The standard choice is about $150$--$200$. In this work, we have compared the model difference of the matter power spectra between modified gravity \textsc{mg-glam} and $\Lambda$CDM \textsc{glam} simulations and found that the result is converged with $160$ time steps. Doubling the number of steps from $160$ to $320$ makes negligible difference. 

The code generates the density field, including peculiar velocities, for a particular cosmological model. Nonlinear matter power spectra and halo catalogues at user-specified output redshifts (snapshots) are measured on the fly. For the latter, \textsc{glam} employs the Bound Density Maximum (BDM; \cite{Klypin:1997sk,Riebe:2011arXiv1109.0003R}) algorithm to get around the usual limitations placed on the completeness of low-mass haloes by the lack of force resolution in PM simulations. Here we briefly describe the idea behind the BDM halo finder, and further details can be found in \cite{Riebe:2011arXiv1109.0003R,Knebe:2011MNRAS.415.2293K}. The code starts by calculating a local density at the positions of individual particles, using a spherical tophat filter containing a constant number $N_{\rm filter}$ (typically 20) of particles. It then gathers all the density maxima and, for each maximum, finds a sphere that contains a mass $M_\Delta = \frac{4}{3}\pi\Delta\rho_{\rm crit}(z)R_\Delta^3$, where $\rho_{\rm crit}(z)$ is the critical density at the halo redshift $z$, and $\Delta$ is the overdensity within the halo radius $R_\Delta$. Throughout this work we will use the virial density definition for $\Delta$ given by \cite{Bryan:1997dn}
\begin{equation}
    \Delta_{\rm vir}(z) = 18\pi^2 + 82\left[\Omega_{\rm m}(z)-1\right] - 39\left[\Omega_{\rm m}(z)-1\right]^2,
\end{equation}
where $\Omega_{\rm m}(z)$ is the matter density parameter at $z$. To find distinct haloes, the BDM halo finder still needs to deal with overlapping spheres. To this end, it treats the density maxima as halo centres and finds the one sphere, amongst a group of overlapping ones, with the deepest Newtonian potential. This is treated as a distinct, central, halo. The radii and masses of the haloes which correspond to the other (overlapping) spheres are then found by a procedure that guarantees a smooth transition of the properties of small haloes when they fall into the larger halo to become subhaloes of the latter. The latter is done by defining the radius of the infalling halo as $\max(R_1, R_2)$, where $R_1$ is its distance to the surface of the larger, soon-to-be host, central halo, and $R_2$ is its distance to the nearest density maximum in the spherical shell  $[\min(R_\Delta,R_1),\max(R_\Delta,R_1)]$ centred around it (if no density maximum exists in this shell, $R_2=R_\Delta$). The BDM halo finder was compared against a range of other halo finders in \cite{Knebe:2011MNRAS.415.2293K}, where good agreement was found. 

\textsc{mg-glam} extends \textsc{glam} to a general class of modified gravity theories by adding extra modules for solving MG scalar field equations, which will be introduced in the following subsection.

\subsubsection{The \textsc{glam} code units}
\label{sec:glam_units}
Like most other $N$-body codes, \textsc{glam} uses its own internal unit system. The code units are designed such that the physical equations can be cast in dimensionless form, which is more convenient for numerical solutions.

Let the box size of simulations be $L$ and the number of grid points in one dimension be $N_{\rm g}$. We can introduce dimensionless coordinates $\tilde{\boldsymbol{x}}$, momenta $\tilde{\boldsymbol{p}}$ and potentials $\tilde{\Phi}$ using the following relations \citep{Klypin:2017iwu} 
\begin{equation}\label{eq:code_units}
\tilde{\boldsymbol{x}} = \left( \frac{N_{\rm g}}{L}\right) {\boldsymbol{x}} \,, \qquad
\tilde{\boldsymbol{p}} = \left( \frac{N_{\rm g}}{H_0 L}\right) {\boldsymbol{p}}\,, \qquad
\tilde\Phi = \left( \frac{N_{\rm g}}{H_0 L}\right)^2\Phi\,.
\end{equation}
Having the dimensionless momenta, we can find the peculiar velocity,
\begin{equation}
{\boldsymbol{v}}_{\rm pec} = 100 \left(\frac{L}{N_{\rm g}}\right)\left(\frac{\tilde{\boldsymbol{p}}}{a}\right)\,{\rm km}~{\rm s}^{-1}\,,
\end{equation}
where we assumed that box size $L$ is given in units of $\Mpch$.
Using these notations, we write the particle equations of motion and the Poisson equation as
\begin{align}
\frac{{\rm d}\tilde{\boldsymbol{p}}}{{\rm d} a} &= -\left(\frac{H_0}{\dot{a}}\right)\tilde{\boldsymbol{\nabla}}\tilde{\Phi}\,,\\
\frac{{\rm d}\tilde{\boldsymbol{x}}}{{\rm d} a} &= -\left(\frac{H_0}{\dot{a}}\right)\frac{\tilde{\boldsymbol{p}}}{a^2}\,,\\
\label{eq:GR_poisson_codeunit}\tilde{\nabla}^2\tilde{\Phi} &= \frac{3}{2}\Omega_{\rm m} a^{-1} \tilde{\delta},
\end{align}
where $\tilde{\delta}$ is the code unit expression of the density contrast $\delta$.

From Eqs.~\eqref{eq:code_units} we can derive the following units,
\begin{equation}\label{eq:code_units2}
\tilde{\boldsymbol{\nabla}} = \left(\frac{L}{N_{\rm g}}\right) \boldsymbol{\nabla}\,, \quad {\rm d} \tilde t = H_0 {\rm d} {t}\,, \quad \tilde{\rho}_{\rm m} = \left( \frac{a^3}{\rho_{\rm crit, 0}\Omega_{\rm m}}\right) \rho_{\rm m}\,, \quad \tilde{\delta} = \delta\,.
\end{equation}
In what follows, we will also use the following definition
\begin{equation}
    \tilde{c} = \left(\frac{N_{\rm g}}{H_0 L}\right) c
\end{equation}
for the code-unit expression of the speed of light, $c$.

{\sc glam} uses a regularly spaced three-dimensional mesh of size $N_{\rm g}^3$ that covers the cubic domain $L^3$ of a simulation box. The size of a cell, $\Delta x =L/N_{\rm g}$, and the mass of each particle, $m_{\rm p}$, define the force and mass resolution respectively:
\begin{eqnarray}
m_{\rm p} &=& \Omega_{\rm m} \, \rho_{\rm crit,0}\left[\frac{L}{N_{\rm p}}\right]^3 = 8.517\times 10^{10}\left[\frac{\Omega_{\rm m}}{0.30}\right]
\left[\frac{L/\Gpch}{N_{\rm p}/1000}\right]^3h^{-1}M_\odot, \label{eqn:mass_resolution_def}\\
\Delta x &=& \left[\frac{L/\Gpch}{N_{\rm g}/1000}\right]\Mpch, \label{eqn:force_resolution_def}
\end{eqnarray}
where $N_{\rm p}^3$ is the number of particles and $\rho_{\rm crit,0}$ is the critical density of the universe at present.

\subsection{Solvers for the extra degrees of freedom}
\label{subsect:extradof}
We have seen in \S\ref{sec:models} that in modified gravity models we usually need to solve a new, dynamical, degree of freedom, which is governed by some nonlinear, elliptical type, partial differential equation (PDE). Being a nonlinear PDE, unlike the linear Poisson equation solved in default {\sc glam}, the equation can not be solved by a one-step fast Fourier transform\footnote{This does not mean that FFT cannot be used under any circumstances. For example, Ref.~\cite{Chan:2009ew} used a FFT-relaxation method to solve nonlinear PDEs iteratively. In each iteration, the equation is treated as if it were linear (by treating the nonlinear terms as a `source') and solved using FFT, but the solution in the previous step is used to update the `source', for the PDE to be solved again to get a more accurate solution, until some convergence is reached.} but requires a \textit{multigrid relaxation} scheme to obtain a solution.

For completeness, we will first give a concise summary of the relaxation method and its multigrid implementation (\S\ref{subsubsect:relaxation}). Next, we will specify the practical side, discussing how to efficiently arrange the memory in the computer, to allow the same memory space to be used for different quantities at different stages of the calculation, therefore minimising the overall memory requirement (\S\ref{subsubsect:code_struc}), and also saving the time for frequently allocating and deallocating operations. 
After that, in \S\ref{sec:Vainshtein_solvers}--\S\ref{sec:Kmouflage_solvers}, we will respectively discuss how the nonlinear PDEs in Vainshtein- and Kmouflage-type gravity models can be solved most efficiently.
In \S\ref{sec:BG_solvers}, we will present how to solve the evolution of the cosmic background in the Kmouflage model.
Much effort will be devoted to replacing the common Newton-Gauss-Seidel relaxation method by a nonlinear Gauss-Seidel, which has been found to lead to substantial speedup of simulations \cite{Bose:2016wms} (but we will generalise this to more models than focused on in Ref.~\cite{Bose:2016wms}). For the coupled quintessence model, we will also briefly describe how the background evolution of the scalar field is numerically solved as an integral part of {\sc mg}-{\sc glam}, to further increase its flexibility. 

\subsubsection{Multigrid Gauss-Seidel relaxation}
\label{subsubsect:relaxation}
Let the partial differential equation (PDE) to be solved take the following form:
\begin{equation}
    \mathcal{L}(u) = 0,
\end{equation}
where $u$ is the scalar field and $\mathcal{L}$ is the PDE operator. To solve this equation numerically, we use finite difference to get a discrete version of it on a mesh\footnote{In this paper we consider the simplest case of cubic cells.}. Since {\sc mg}-{\sc glam} is a particle-mesh (PM) code, it has a uniform mesh resolution and does not use adaptive mesh refinement (AMR). When discretised on a uniform mesh with cell size $h$, the above equation can be denoted as
\begin{equation}\label{eq:general_pde}
    \mathcal{L}^h({u}^h) = {f}^h,
\end{equation}
where we have added a nonzero right-hand side, $f^h$, for generality (while $f^h=0$ on the mesh with cell size $h$, later when we discrete it on coarser meshes needed for the multigrid implementation, $f$ is no longer necessarily zero). Both $u^h$ and $f^h$ are evaluated at the cell centres of the given mesh.

The solution we obtain numerically, $\hat{u}$, is unlikely to be the true solution $u^h$ to the discrete equation, and applying the PDE operator on the former gives the following, slightly different, equation:
\begin{equation}\label{eq:general_pde_numerical}
    \mathcal{L}^h(\hat{u}^h) = \hat{f}^h.
\end{equation}
Taking the difference between the above two equations, we get
\begin{equation}\label{eq:numerical_error}
    \mathcal{L}^h({u}^h) - \mathcal{L}^h(\hat{u}^h) = f^h-\hat{f}^h = -d^h,
\end{equation}
where 
\begin{equation}\label{eq:residual_d}
    d^h \equiv \hat{f}^h-f^h,
\end{equation}
is the {\it local residual}, which characterises the inaccuracy of the solution $\hat{u}^h$ (this is because if $\hat{u}^h=u^h$, we would expect $\hat{f}^h=f^h$ and hence there is zero `inaccuracy'). $d^h$ is also evaluated at cell centres. Later, to check if a given set of numerical solution $\hat{u}^h$ is acceptable, we will use a \textit{global residual}, $\epsilon^h$, which is a single number for the given mesh of cell size $h$. In this work we choose to define $\epsilon^h$ as the root-mean-squared of $d^h$ in all mesh cells (although this is by no means the only possible definition). We will call both $d^h$ and $\epsilon^h$ `residual' as the context will make it clear which one is referred to.

Relaxation solves Eq.~\eqref{eq:general_pde} by starting from some approximate trial solution to $u^h$, $\hat{u}^h_{\rm old}$, and check if it satisfies the PDE. If not, this trial solution can be updated using a method that is similar to the Newton-Ralphson iterative method to solve nonlinear algebraic equations
\begin{equation}\label{eq:relaxation_iteration}
    \hat{u}^h_{\rm new} = \hat{u}^h_{\rm old} - \frac{\mathcal{L}^h\left(\hat{u}^h_{\rm old}\right)-\hat{f}^h}{\partial\mathcal{L}^h\left(\hat{u}^h_{\rm old}\right)/\partial\hat{u}^h}.
\end{equation}
This process can be repeated iteratively, until the updated solution satisfies the PDE to an acceptable level, i.e., $\epsilon^h$ becomes small enough. In practice, because we are solving the PDE on a mesh, Eq.~\eqref{eq:relaxation_iteration} should be performed for all mesh cells, which raises the question of how to order this operation for the many cells. We will adopt the Gauss-Seidel `black-red chessboard' approach, where the cells are split into two classes, `black' and `red', such that all the six direct neighbours\footnote{The direct neighbours of a given cell are the six neighbouring cells which share a common face with that cell.} of a `red' cell are black and vice versa. The relaxation operation, Eq.~\eqref{eq:relaxation_iteration}, is performed in two sweeps, the first for `black' cells (i.e., only updating $\hat{u}^h$ in `black' cells while keeping their values in `red' cells untouched), while the second for all the `red' cells. This is a standard method to solve nonlinear elliptical PDEs by using relaxation, known as the \textit{Newton-Gauss-Seidel method}. However, although this method is generic, it is not always efficient, and later we will describe a less generic alternative which is nevertheless more efficient. 

Relaxation iterations are useful at reducing the Fourier modes of the error in the trial solution $\hat{u}^h$, whose wavelengths are comparable to that of the size of the mesh cell $h$. If we do relaxation on a fine mesh, this means that the short-wave modes of the error are quickly reduced, but the long-wave modes are generally much slower to decrease, which can lead to a slow convergence of the relaxation iterations. A useful approach to solve this problem is by using {\it multigrid}: after a few iterations on the fine level, we `move' the equation to a coarser level where the cell size is larger and the longer-wave modes of the error in $\hat{u}^h$ can be more quickly decreased. The discretised PDE on the coarser level is given by
\begin{equation}\label{eq:general_pde_numerical_coarse}
    \mathcal{L}^H(u^H) = \mathcal{L}\left(\mathcal{R}\hat{u}^h\right) - \mathcal{R}d^h \equiv S^H,
\end{equation}
where the superscript $^H$ denotes the coarse level where the cell size is $H$ (in our case $H=2h$), and $\mathcal{R}$ denotes the \textit{restriction} operator which interpolates quantities from the fine level to the coarse level. In our numerical implementation, a coarse (cubic) cell contains 8 fine (cubic) cells of equal volume, and the restriction operation can be conveniently taken as the arithmetic average of the values of the quantity to be interpolated in the 8 fine cells. 

Eq.~\eqref{eq:general_pde_numerical_coarse} can be solved using relaxation similarly to Eq.~\eqref{eq:general_pde_numerical}, for which the numerical solution is denoted as $\hat{u}^H$. This can be used to `correct' and `improve' the approximate solution $\hat{u}^h$ on the fine level, as
\begin{equation}\label{eq:solution_correction}
    \hat{u}^{h,{\rm new}} = \hat{u}^{h,{\rm old}} + \mathcal{P}\left(\hat{u}^H - \mathcal{R}\hat{u}^h\right),
\end{equation}
where $\mathcal{P}$ is the \textit{prolongation} operation which does the interpolation from the coarse to the fine levels. In this work we shall use the following definition of the prolongation operation: for a given fine cell,
\begin{enumerate}
    \item find its parent cell, i.e., the coarser cell that contains the fine cell;
    \item find the seven neighbours of the parent cell, i.e., the coarser cells which share a face (there are 3 of these), an edge (there are 3 of these) or a vertex (just 1) with the above parent coarser cell;
    \item calculate the fine-cell value of the quantity to be interpolated from the coarse to the fine levels, as a weighted average of the corresponding values in the 8 coarse cells mentioned above: $27/64$ for the parent coarse cell, and $9/64$, $3/64$ and $1/64$ respectively for the coarse cells sharing a face, an edge and a vertex with the parent cell. 
\end{enumerate}

The above is a simple illustration of how multigrid works for two levels of mesh resolution, $h$ and $H$. In principle, multigrid can be and is usually implemented using more than two levels. In this paper we will use a hierarchy of increasingly coarser meshes with the coarsest one having $4^3$ cells. 

\begin{figure}
    \centering
    \includegraphics[width=\textwidth]{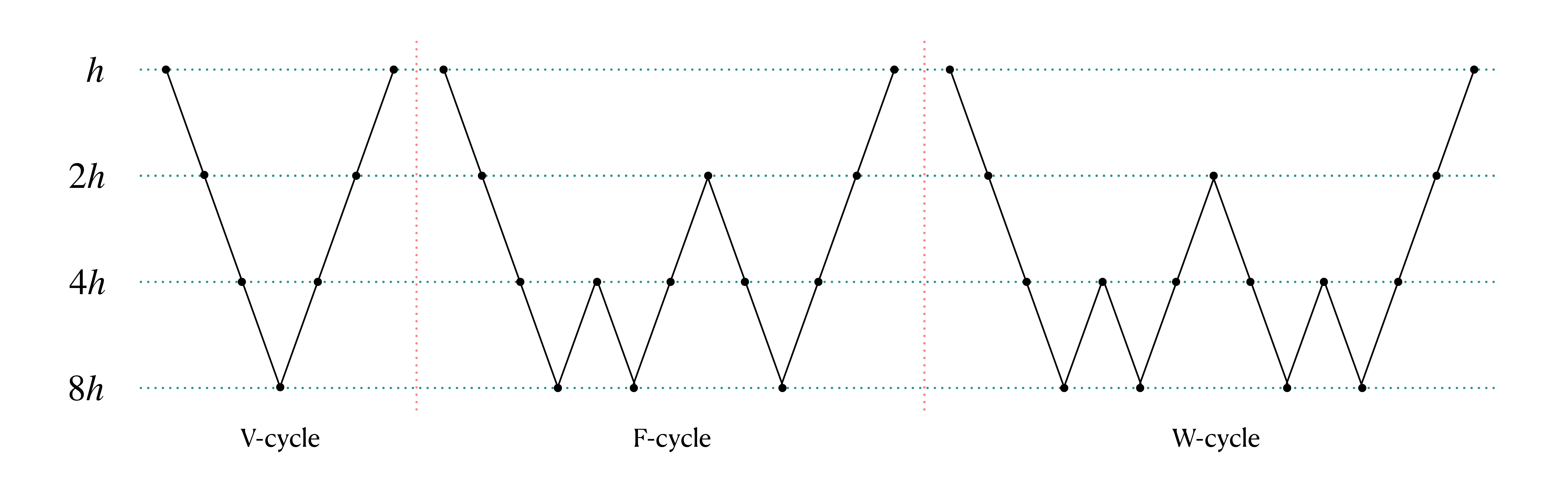}
\caption{An illustration of the three different arrangements of multigrid relaxation method used in this paper: from left to right, V-cycle, F-cycle and W-cycle. The horizontal dotted lines depict 4 multigrid levels of mesh, with the finest mesh (denoted by its cell size $h$) on top, and the coarsest mesh (with cell size $8h$) at the bottom. The relaxation always starts on the finest level, and the solid lines show how the multigrid solver walks through the different levels, performing Gauss-Seidel relaxation iterations at each level (denoted by the circles), called smoothing. Only one single full cycle is shown for each case. The solver walks over the multigrid levels more times in W-cycle than in F-cycle and V-cycle, and thus it requires fewer cycles in the former case to arrive at a converged solution. However, it is also computationally more expensive. 
We will compare the performances of the three different arrangements in real cosmological simulations in \S~\ref{sec:convergence_tests}}
    \label{fig:multigrid_cycles}
\end{figure}

There are flexibilities in how to arrange the relaxations at different levels. The most-commonly used arrangement is the so-called V-cycle, where one starts from the finest level, moves to the coarsest one performing relaxation iterations on each of the intermediate levels (cf.~Eq.~\eqref{eq:general_pde_numerical_coarse}), and then moves straight back to the finest performing corrections using Eq.~\eqref{eq:solution_correction} on each of the intermediate levels. Other arrangements, such as F-cycle and W-cycle (cf.~Fig.~\ref{fig:multigrid_cycles}), are sometimes more efficient in improving the convergence rate of $\hat{u}^h$ to $u^u$, and we have implemented them in {\sc mg}-{\sc glam} as well. 

\subsubsection{Memory usage}
\label{subsubsect:code_struc}
{\sc glam} uses a single array to store mesh quantities, such as the matter density field and the Newtonian potential, because at any given time only one of these is needed. The Newtonian force at cell centres is calculated by finite-differencing the potential and then interpolated to the particle positions. To be memory efficient, \textsc{glam} also opts not to create a separate array to store the forces at the cell centres, but instead directly calculates them at the particle positions immediately before updating the particle velocities.

With the new scalar field to be solved in modified gravity models, we need two additional arrays of size $N_{\rm g}^3$, where $N_{\rm g}^3$ is the number of cells of the PM grid (i.e., there are $N_{\rm g}$ cells in each direction of the cubic simulation box). This leads to three arrays. \texttt{Array 1} is the default array in {\sc glam}, which is used to store the density field $\rho$ and the Newtonian potential $\Phi$ (at different stages of the simulation). Note that the density field is also needed when solving the scalar field equation of motion during the relaxation iterations, and so we cannot use this array to also store the scalar field. On the other hand, we will solve the Newtonian potential after the scalar field, by when it is safe to overwrite this array with $\Phi$. \texttt{Array2} is exclusively used to store the scalar field solution $\hat{u}^h$ on the PM grid, which will be used to calculate the fifth force. \texttt{Array3} is used to store the various intermediate quantities which are created for the implementation of the multigrid relaxation, such as $d^h$, $\hat{u}^H$, $\mathcal{R}\hat{u}^h$, $\mathcal{R}d^h$, $S^H$ and $\rho^H$, the last of which is the density field on the coarser level $^H$, which appears in the coarse-level discrete PDE operator $\mathcal{L}^H$.

To be concrete, we imagine the 3D array (\texttt{Array3}) as a cubic box with $N_{\rm g}^3$ cubic cells of equal size. An array element, denoted by $(i,j,k)$, represents the $i$th cell in the $x$ direction, $j$th cell in the $y$ direction and $k$th cell in the $z$ direction, with $i,j,k=1,\cdots,N_{\rm g}$. We divide this array into 8 sections, each of which can be considered to correspond to one of the 8 octants that equally divide the volume of the cubic box. The range of $(i,j,k)$ of each section and the quantity stored in that section of \texttt{Array3} are summarised in the table below: 
\begin{center}
    \begin{tabular}{|c|p{3cm}|p{3cm}|p{3cm}|p{3cm}|}
    \hline
    Section & $i$ range & $j$ range & $k$ range & Quantity\\ 
    \hline
    \hline
    1 & $1,\cdots,N_{\rm g}/2$ & $1,\cdots,N_{\rm g}/2$ & $1,\cdots,N_{\rm g}/2$ & $d^\ell$, $\mathcal{R}d^\ell$ \\ 
    \hline
    2 & $N_{\rm g}/2+1,\cdots,N_{\rm g}$ & $1,\cdots,N_{\rm g}/2$ & $1,\cdots,N_{\rm g}/2$ & $d^\ell$, $\rho^{\ell-1} = \mathcal{R}\rho^\ell$ \\ 
    \hline
    3 & $1,\cdots,N_{\rm g}/2$ & $N_{\rm g}/2+1,\cdots,N_{\rm g}$ & $1,\cdots,N_{\rm g}/2$ & $d^\ell$, $\mathcal{R}\hat{u}^\ell$ \\ 
    \hline
    4 & $N_{\rm g}/2+1,\cdots,N_{\rm g}$ & $N_{\rm g}/2+1,\cdots,N_{\rm g}$ & $1,\cdots,N_{\rm g}/2$ & $d^\ell$, $\hat{u}^{\ell-1}$ \\ \hline
    5 & $1,\cdots,N_{\rm g}/2$ & $1,\cdots,N_{\rm g}/2$ & $N_{\rm g}/2+1,\cdots,N_{\rm g}$ & $d^\ell$, recursion \\ 
    \hline
    6 & $N_{\rm g}/2+1,\cdots,N_{\rm g}$ & $1,\cdots,N_{\rm g}/2$ & $N_{\rm g}/2+1,\cdots,N_{\rm g}$ & $d^\ell$, $d^{\ell-1}$ \\ \hline
    7 & $1,\cdots,N_{\rm g}/2$ & $N_{\rm g}/2+1,\cdots,N_{\rm g}$ & $N_{\rm g}/2+1,\cdots,N_{\rm g}$ & $d^\ell$, $S^{\ell-1}$ \\ \hline
    8 & $N_{\rm g}/2+1,\cdots,N_{\rm g}$ & $N_{\rm g}/2+1,\cdots,N_{\rm g}$ & $N_{\rm g}/2+1,\cdots,N_{\rm g}$ & $d^\ell$ \\
    \hline
    \end{tabular}
\end{center}

Let us explain this more explicitly. First of all, the whole \texttt{Array3}, of size $N_{\rm g}^3$, will be used to store the residual value $d^h$ on the PM grid (which has $N_{\rm g}^3$ cells). From now on, we label this grid by `level-$\ell$', and use `level-($\ell-m$)' to denote the grid that are $m$ times coarser, i.e., if the cell size of the PM grid is $h$, then the cells in this coarse grid have a size of $2^mh$. 
In the table above we have used $d^\ell$ to denote the $d^h$ on level-$\ell$, and so on. 
Note that we always use $N_{\rm g}=2^\ell$.

The local residual $d^h$ on a fine grid is only needed for two purposes: (1) to calculate the global residual on that grid, $\epsilon^h$, which is needed to decide convergence of the relaxation, and (2) to calculate the coarse-level PDE operator $\mathcal{L}^H$ that is needed for the multigrid acceleration, as per Eq.~\eqref{eq:general_pde_numerical_coarse}. This suggests that $d^h$ does not have to occupy \texttt{Array3} all the time, and so this array can be reused to store other intermediate quantities (see the last column of the above table) \textit{after} we have obtained $\epsilon^h$. 

In our arrangement, Section 1 stores the residual residual $\mathcal{R}d^\ell$, Section 2 stores the restricted density field $\rho^{\ell-1}=\mathcal{R}\rho^\ell$, Sections 3 and 4 store, respectively, the restricted scalar field solution $\mathcal{R}\hat{u}^\ell$ and the coarse-grid scalar field solution $\hat{u}^{\ell-1}$ --- the former is needed to calculate $S^{\ell-1}$ in Eq.~\eqref{eq:general_pde_numerical_coarse} and to correct the fine-grid solution using Eq.~\eqref{eq:solution_correction}, which is fixed after calculation, while the latter is updated during the coarse-grid relaxation sweeps\footnote{We use $\mathcal{R}\hat{u}^\ell$ as the initial guess for $\hat{u}^{\ell-1}$ for the Gauss-Seidel relaxations on the coarse level.}. Section 7 stores the coarse-grid source $S^{\ell-1}$ for the PDE operator $\mathcal{L}^{\ell-1}$ as defined in Eq.~\eqref{eq:general_pde_numerical_coarse}, and finally Section 6 stores the residual on the coarse level, $d^{\ell-1}$. Note that all these quantities are for level-$(\ell-1)$, so that they can be stored in section of \texttt{Array3} of size $\left(N_{\rm g}/2\right)^3$. Section 8 is not used to store anything other than $d^\ell$. 

We have not touched Section 5 so far --- this section is reserved to store the same quantities as above, but for level-$(\ell-2)$, which are needed if we want to use more than two levels of multigrid. It is further divided into 8 section, each of which will play the same roles as detailed in the table above\footnote{The exception is that, as $d^{\ell-1}$ is already stored in Section 6, it does not have to be stored in Section 5 again.}. In particular, the (sub)Section 5 of Section 5 is reserved for quantities on level-$(\ell-3)$, and so on. In this way, there is no need to create separate arrays of various sizes to store the intermediate quantities on different multigrid levels which therefore saves memory. 

There is a small tricky issue here: as we mentioned above, the local residual $d^\ell$ on the PM grid is needed to calculate the coarse-grid source $S^{\ell-1}$ using Eq.~\eqref{eq:general_pde_numerical_coarse}, thus we will be using the quantity $d^\ell$ stored in \texttt{Array3} to calculate $\mathcal{R}d^{\ell}$ and then write it to (part of) the same array, running the risk of overwriting some of the data while it is still needed. To avoid this problem, we refrain from using the $d^\ell$ data already stored in \texttt{Array3}, but instead recalculate it in the subroutine to calculate $\mathcal{R}d^\ell$ (this only needs to be done for level-$\ell$). With a bit of extra computation, this enables use to avoid creating another array of similar size to \texttt{Array3}.

Since \texttt{Array3} stores different quantities in different parts, care must be excised when assessing these data. There is a simple rule for this: suppose that we need to read or write the quantities on the coarse grid of level-$(\ell-m)$ with $m\geq1$. These are 3-dimensional quantities with the three directions labelled by $I,J,K$, which run over $1,\cdots,2^{\ell-m}$, and we have
\begin{align}
    \mathcal{R}(d^{\ell-m+1}) \left[I,J,K\right] &\leftrightarrow \texttt{Array3}[i=I,&j=J\phantom{\ +2^{\ell-m}},k=K + \left(2^{m}-2\right)\cdot2^{\ell-m}],\nonumber\\
    \mathcal{R}(\rho^{\ell-m+1})\left[I,J,K\right] &\leftrightarrow \texttt{Array3}[i=I+2^{\ell-m},&j=J\phantom{\ +2^{\ell-m}},k=K + \left(2^{m}-2\right)\cdot2^{\ell-m}],\nonumber\\
    \mathcal{R}(u^{\ell-m+1})\left[I,J,K\right] &\leftrightarrow\texttt{Array3}[i=I,&j=J+2^{\ell-m},k=K + \left(2^{m}-2\right)\cdot2^{\ell-m}],\nonumber\\
    \hat{u}^{\ell-m}\left[I,J,K\right] &\leftrightarrow \texttt{Array3}[i=I+2^{\ell-m},&j=J+2^{\ell-m},k=K + \left(2^{m}-2\right)\cdot2^{\ell-m}],\nonumber\\
    d^{\ell-m}\left[I,J,K\right] &\leftrightarrow \texttt{Array3}[i=I+2^{\ell-m},&j=J\phantom{\ +2^{\ell-m}},k=K + \left(2^{m}-1\right)\cdot2^{\ell-m}],\nonumber\\
    S^{\ell-m}[I,J,K] &\leftrightarrow \texttt{Array3}[i=I,&j=J+2^{\ell-m},k=K + \left(2^{m}-1\right)\cdot2^{\ell-m}],
\end{align}
where $i,j,k=1,\cdots,N_{\rm g}$ run over the entire \texttt{Array3}. 

We can estimate the required memory for \textsc{mg-glam} simulations as follows. As mentioned above, the code uses a 3D array of single precision to store both the density field and the Newtonian potential, and one set of arrays for particle positions and velocities. In addition, two arrays are added to store the scalar field solution (\texttt{Array2}) and various intermediate quantities in the multigrid relaxation solver (\texttt{Array3}). In the cosmological simulations described in this paper, we have used double precision for the two new arrays, and we have checked that using single precision slightly speeds up the simulation, while agreeing with the double-precision results within $0.001\%$ and $0.5\%$ respectively for the matter power spectrum and halo mass function. Given its fast speed and its shared-memory nature, memory is expected to be the main limiting factor for large \textsc{mg-glam} jobs. For this reason, we assume that all arrays are set to be single precision for future runs, and this leads to the following estimate of the total required memory:
\begin{align}
M_{\rm tot} &= 12N^3_{\rm g} + 24N^3_{\rm p} \, {\rm bytes}\,,\nonumber\\
&= 89.41 \qty(\frac{N_{\rm g}}{2000})^3 + 22.35\qty(\frac{N_{\rm p}}{1000})^3 \, {\rm GB}\,,\nonumber\\
&\approx 112 \, \qty(\frac{N_{\rm p}}{1000})^3 \, {\rm GB}\,,\quad {\rm for} \  N_{\rm g}=2N_{\rm p}\,,
\end{align}
where we have used $1~\mathrm{GB} = 1024^3~\mathrm{bytes}$. This is slightly more than twice the memory requirement of the default \textsc{glam} code, which is $52\left(N_{\rm p}/1000\right)^3$ GB \citep{Klypin:2017iwu}.

\subsubsection{Implementation of Vainshtein-type gravity models}
\label{sec:Vainshtein_solvers}
Having described the code and data structure of \textsc{mg-glam}, we next discuss in greater detail how each of the two classes of models studied in this paper is implemented, starting from Vainshtein-type models. 

Since $\varphi$ plays the role of the conservative potential of the fifth force (\S~\ref{sec:vainshtein}), we can choose the same code unit for it as for the Newtonian potential $\Phi$:
\begin{equation}
    \tilde{\varphi} = \left(\frac{N_{\rm g}}{H_0L}\right)^2\varphi.
\end{equation}
We also introduce the code-unit counterpart of the cross-over scale $r_{\rm c}$ as
\begin{equation}
    \tilde{r}_{\rm c} = \frac{N_{\rm g}}{L}r_{\rm c},
\end{equation}
which is consistent with the code unit for comoving coordinate or length. Using the code unit expression for the speed of light $c$, Eq.~\eqref{eq:code_units2}, it can be shown that 
\begin{equation}
    \frac{r_{\rm c}H_0}{c} = \frac{1}{2\sqrt{\Omega}_{\rm rc}} = \frac{\tilde{r}_{\rm c}}{\tilde{c}} \equiv R_{\rm c},
\end{equation}
where $R_{\rm c}$ is a new dimensionless model parameter and $\Omega_{\rm rc}$ has been introduced above. We can then recast the DGP equation of motion, Eq.~\eqref{eq:phi_dgp}, in code unit as
\begin{equation}\label{eq:dgp_eom_code}
    \tilde{\nabla}^2\tilde{\varphi} + \frac{R_{\rm c}^2}{3\beta_{\rm DGP}(a)a^2}\left[ (\tilde{\nabla}^2\tilde{\varphi})^2 - \tilde{\nabla}_i\tilde{\nabla}_j\tilde{\varphi}\tilde{\nabla}^i\tilde{\nabla}^j\tilde{\varphi}\right] = \frac{1}{\beta_{\rm DGP}(a)} \Omega_{\rm m}a^{-1}\tilde{\delta}. 
\end{equation}
\begin{equation}
 \tilde{\boldsymbol{a}}_{\rm MG} = -\frac{1}{2} \tilde{\boldsymbol{\nabla}} \tilde{\varphi},
\label{eq:dgp_accel}
\end{equation}
where $\tilde{\boldsymbol{a}}_{\rm MG}$ denotes the modified gravity contribution to the gravitational acceleration in code units.

For simplicity, in what follows we neglect the tildes in Eq.~\eqref{eq:dgp_eom_code}. Making the following defining decomposition of the second derivative of the scalar field \cite{Chan:2009ew,Li:2013tda},
\begin{equation}\label{eq:operator_split}
    \nabla_i\nabla_j\varphi = \hat{\nabla}_i\hat{\nabla}_j\varphi + \frac{1}{3}\delta_{ij}\nabla^2\varphi,
\end{equation}
so that $\hat{\nabla}_i\hat{\nabla}_j\varphi$ has zero trace, i.e., $\hat{\nabla}^i\hat{\nabla}_i\varphi=0$, one can show that
\begin{equation}
    \nabla_i\nabla_j\varphi\nabla^i\nabla^j\varphi = \hat{\nabla}_i\hat{\nabla}_j\varphi\hat{\nabla}^i\hat{\nabla}^j\varphi + \frac{1}{3}\left(\nabla^2\varphi\right)^2.
\end{equation}
Eq.~\eqref{eq:dgp_eom_code} can then be rewritten as \cite{Li:2013nua,Hernandez-Aguayo:2020_MGAREPO_code_paper}
\begin{equation}\label{eq:dgp_quadratic_eom}
    \frac{2}{3}\left(\nabla^2\varphi\right)^2 + \alpha\nabla^2\varphi - \Sigma = 0,
\end{equation}
where
\begin{eqnarray}
    \alpha &\equiv& \frac{3\beta_{\rm DGP}(a)a^2}{R_{\rm c}^2},\nonumber\\
    \Sigma &\equiv& \hat{\nabla}_i\hat{\nabla}_j\varphi\hat{\nabla}^i\hat{\nabla}^j\varphi + \frac{\alpha}{\beta_{\rm DGP}}\Omega_{\rm m}a^{-1}\delta\,.
\end{eqnarray}
Eq.~\eqref{eq:dgp_quadratic_eom} has two branches of solutions:
\begin{equation}
    \nabla^2\varphi = \frac{3}{4}\left[-\alpha\pm\sqrt{\alpha^2+\frac{8}{3}\Sigma}\right].
\end{equation}
Which branch is the physical solution depends on the sign of $\alpha$ and hence of the function $\beta_{\rm DGP}(a)$. The requirement is that, as $\delta\rightarrow0$, i.e., for a homogeneous density field, we must have a homogeneous scalar field, and so $\nabla^2\varphi\rightarrow0$. Therefore, the solution can be written as
\begin{equation}\label{eq:DGP}
    \nabla^2\varphi = \frac{3}{4}\left[-\alpha+{\rm sign}(\alpha)\sqrt{\alpha^2+\frac{8}{3}\Sigma}\right],
\end{equation}
with the function ${\rm sign}(x)=1$ for $x\geq0$ and $-1$ for $x<0$.

The solve it on a discrete mesh, the continuous equation, \eqref{eq:DGP}, is first discretised as $\mathcal{L}^h\varphi_{i,j,k}=0$, where the operator $\mathcal{L}^h$ is defined as
\begin{eqnarray}
\mathcal{L}^h\varphi_{i,j,k} &\equiv& \frac{1}{h^2}\Big(\varphi_{i+1,j,k}+\varphi_{i-1,j,k}+\varphi_{i,j+1,k}+\varphi_{i,j-1,k}+\varphi_{i,j,k+1}+\varphi_{i,j,k-1}-6\varphi_{i,j,k}\Big)\nonumber\\
&& - \frac{3}{4}\left[-\alpha+{\rm sign}(\alpha)\sqrt{\alpha^2+\frac{8}{3}\Sigma_{i,j,k}}\right],\ \ \ \
\end{eqnarray}
with
\begin{eqnarray}
\Sigma_{i,j,k} &\equiv&\frac{2}{3h^4}\bigg[\Big(\varphi_{i+1,j,k}+\varphi_{i-1,j,k}-2\varphi_{i,j,k}\Big)^2+\Big(\varphi_{i,j+1,k}+\varphi_{i,j-1,k}-2\varphi_{i,j,k}\Big)^2\nonumber\\
&& +\Big(\varphi_{i,j,k+1} + \varphi_{i,j,k-1}-2\varphi_{i,j,k}\Big)^2\bigg]\nonumber\\
&& - \frac{2}{3h^4}\left(\varphi_{i+1,j,k}+\varphi_{i-1,j,k}-2\varphi_{i,j,k}\right)\left(\varphi_{i,j+1,k}+\varphi_{i,j-1,k}-2\varphi_{i,j,k}\right)\nonumber\\
&& - \frac{2}{3h^4}\left(\varphi_{i+1,j,k}+\varphi_{i-1,j,k}-2\varphi_{i,j,k}\right)\left(\varphi_{i,j,k+1}+\varphi_{i,j,k-1}-2\varphi_{i,j,k}\right)\nonumber\\
&& - \frac{2}{3h^4}\left(\varphi_{i,j+1,k}+\varphi_{i,j-1,k}-2\varphi_{i,j,k}\right)\left(\varphi_{i,j,k+1}+\varphi_{i,j,k-1}-2\varphi_{i,j,k}\right)\nonumber\\
&& + \frac{1}{8h^4}\Big(\varphi_{i+1,j+1,k}+\varphi_{i-1,j-1,k}-\varphi_{i+1,j-1,k}-\varphi_{i-1,j+1,k}\Big)^2\nonumber\\
&& + \frac{1}{8h^4}\Big(\varphi_{i+1,j,k+1}+\varphi_{i-1,j,k-1}-\varphi_{i+1,j,k-1}-\varphi_{i-1,j,k+1}\Big)^2\nonumber\\
&& + \frac{1}{8h^4}\Big(\varphi_{i,j+1,k+1}+\varphi_{i,j-1,k-1}-\varphi_{i,j+1,k-1}-\varphi_{i,j-1,k+1}\Big)^2\nonumber \\ 
&& + \frac{\alpha}{\beta_{\rm DGP}} \Omega_{\rm m}a^{-1}\delta_{i,j,k}\,,
\end{eqnarray}
where $h$ here denotes the simulation mesh cell size in code units, as introduced in \S~\ref{subsubsect:relaxation} (this is the same symbol as used for the dimensionless Hubble constant, but not confusion should arise given the context); $i,j,k$ are the indices of cells in the simulation mesh, with, e.g., $(i+1,j,k)$ denoting the neighbouring cell to the right of cell $(i,j,k)$, with the same $y, z$ coordinates. This discretisation has second-order accuracy, meaning that its deviation from the true value reduces as $\mathcal{O}\left(h^2\right)$. 

This equation can be solved using the multigrid relaxation method described above, for which the code iterates to update the value of $\varphi_{i,j,k}$ in all cells, and at each iteration the field values changes as
\begin{equation}\label{eq:nonlinear_GS_DGP}
\varphi^{h,{\rm new}}_{i,j,k} = \varphi^{h,{\rm old}}_{i,j,k} - \frac{\mathcal{L}^h\left(\varphi^{h,{\rm old}}_{i,j,k}\right)}{\frac{\partial\mathcal{L}^h\left(\varphi^{h,{\rm old}}_{i,j,k}\right)}{\partial\varphi_{i,j,k}^{h,{\rm old}}}},
\end{equation}
where we have
\begin{equation}
\frac{\partial\mathcal{L}^h\left(\varphi^{h,{\rm old}}_{i,j,k}\right)}{\partial\varphi_{i,j,k}^{h,{\rm old}}}
= -\frac{6}{h^2}\,.
\end{equation}
As mentioned in \cite{Li:2013nua}, the operator splitting of Eq.~\eqref{eq:operator_split} and the manipulation of the default discrete DGP equation into the Poisson-equation-like form of Eq.~\eqref{eq:DGP} are critical for obtaining reasonable convergence properties of the relaxation solver. The latter also makes the code more efficient as there is no need for expensive and approximate Newton-Gauss-Seidel iterations\footnote{Eq.~\eqref{eq:nonlinear_GS_DGP} can be considered as the exact solution of a linear equation for $\varphi^{h,{\rm new}}_{i,j,k}$ so that there is no need for the Newton iterations, though we note that this linear equation itself is only an approximation to the full DGP equation, \eqref{eq:DGP}, where $\Sigma$ depends on the field itself. The key point here is that the discretion of $\Sigma$ does {\it not} depend on $\varphi_{i,j,k}$ but only depends on the field values in neighbouring cells to $(i,j,k)$.}. We will follow the same spirit in designing the relaxation algorithm for Kmouflage-type models next.

\subsubsection{Implementation of Kmouflage-type models}
\label{sec:Kmouflage_solvers}
For this model, we define the following code unit for $\varphi$,
\begin{equation}\label{eq:Kmo_unit}
    \tilde{\varphi} = \frac{cN_{\rm g}}{H_0L}\varphi = \tilde{c}\varphi.
\end{equation}
Crucially, we note that this unit only applies to the scalar field when we take the spatial derivatives of it, while the time derivative of the scalar field is treated differently. Alternatively, one can understand the $\varphi$ here as the spatial perturbation of the total Kmouflage field, i.e., $\delta\varphi=\varphi-\bar{\varphi}$. In the quasi-static approximation with which we work in this paper, the equations to be solved contain only the \textit{spatial derivatives} of $\delta\varphi$ and the \textit{field value} or \textit{time derivatives} of $\varphi\approx\bar{\varphi}$ (because $|\delta\varphi|\ll|\bar{\varphi}|$). Therefore, we opt to use $\varphi$ to also denote $\delta\varphi$ for simplicity, and the context should make it clear which quantity is being referred to.

With this, we get the following expression of $\tilde{X}$, which is the code-unit counterpart of $X$,
\begin{equation}
    \tilde{X} = X = \frac{1}{2a^2\lambda^2}\left[\left(\frac{\bar{\varphi}'}{H_0}\right)^2-\left(\tilde{\boldsymbol{\nabla}}\tilde{\varphi}\right)^2\right],
\end{equation}
where, as stated in the introduction, $'$ denotes the derivative with respect to the conformal time $\tau$, so that $\varphi'=a\dot{\varphi}$. Then, in code units, the equation of motion, Eq.~\eqref{eq:kmouflage_eom}, can be recast as
\begin{equation}\label{eq:kmo_eom}
    \tilde{c}\tilde{\nabla}^i\left[\left(1+\gamma\left(\frac{\bar{\varphi}'^2}{H_0^2}-{\tilde{\nabla}^j\tilde{\varphi}\tilde{\nabla}_j\tilde{\varphi}}\right)^{n-1}\right)\tilde{\nabla}_i\tilde{\varphi}\right] = 3\frac{{\rm d}A(\bar{\varphi})}{{\rm d}\varphi}\Omega_{\rm m}a^{-1}\tilde{\delta},
\end{equation}
with
\begin{equation}
    \gamma \equiv nK_0\left(\frac{1}{2a^2\lambda^2}\right)^{(n-1)}.
\end{equation}
Here we have evaluated ${\rm d}A(\varphi)/{\rm d}\varphi$ at $\bar{\varphi}$ because the perturbation to the scalar field is generally much smaller than the background value $\bar{\varphi}$ itself, which is of order $0.1\sim1$ at late times (see \S~\ref{sec:bkg_tests}).

This equation, however, has a potential issue. To see this, let's consider the simple case of a 1D density field, say, which depends only on the $x$ coordinate. Then the equation becomes
\begin{equation}\label{eq:1D_kmo_eqn}
    \tilde{c}\left[1+\gamma\left(\frac{\bar{\varphi}'^2}{H_0^2}-\left(\tilde{\partial}\tilde{\varphi}\right)^2\right)^{n-2}\left(\frac{\bar{\varphi}'^2}{H_0^2}-(2n-1)\left(\tilde{\partial}\tilde{\varphi}\right)^2\right)\right]\tilde{\partial}_x^2\tilde{\varphi} = 3\frac{{\rm d}A(\bar{\varphi})}{{\rm d}\varphi}\Omega_{\rm m}a^{-1}\tilde{\delta}.
\end{equation}
The second term in the square brackets on the left-hand side is negative in the regime of
\begin{equation}
    \frac{1}{2n-1}\frac{\bar{\varphi}'^2}{H_0^2} < \left(\tilde{\partial}_x\tilde{\varphi}\right)^2 < \frac{\bar{\varphi}'^2}{H_0^2}.
\end{equation}
While $|\bar{\varphi}'/H_0|\simeq{\mathcal{O}}\left(10^{-3}\right)$ at late times, at $z>10$ it can be much larger (note that the denominator is $H_0$). For $\gamma>0$, in certain regimes the coefficient of $\tilde{\partial}^2_x\tilde{\varphi}$ can cross 0, which leads to a singularity. Instead of the model being unphysical in these regimes, this is more likely a consequence of deriving the equation in the quasi-static and weak-field approximations, because even when the coefficient of $\tilde{\partial}^2\tilde{\varphi}$ is zero, the left-hand side of Eq.~\eqref{eq:1D_kmo_eqn} should have had terms that involve time derivatives of the field so that the full equation is still physical. As we are mostly interested in the Kmouflage screening mechanism in this work, we circumvent this potential numerical issue by slightly modify Eq.~\eqref{eq:kmo_eom} to the following form:
\begin{equation}\label{eq:kmo_eom1}
    \tilde{c}\tilde{\nabla}^i\left[\left(1+\gamma\left(\frac{\varphi'^2}{H_0^2}\right)^{n-1}+\gamma\left({\tilde{\nabla}^j\tilde{\varphi}\tilde{\nabla}_j\tilde{\varphi}}\right)^{n-1}\right)\tilde{\nabla}_i\tilde{\varphi}\right] = 3\frac{{\rm d}A(\bar{\varphi})}{{\rm d}\varphi}\Omega_{\rm m}a^{-1}\tilde{\delta}.
\end{equation}
This should not affect the Kmouflage screening because it mainly takes effect in the highly nonlinear regime, where the spatial term in $X$ (or $\tilde{X}$) is much larger than the temporary contribution. In the linear regime, when the spatial contribution in $X$ is subdominant, the above equation should also reproduce the perturbation behaviour of the fifth force.

Eq.~\eqref{eq:kmo_eom1} is a nonlinear equation in $\tilde{\varphi}$. As mentioned towards the end of the last subsection, we also apply the operator splitting of Eq.~\eqref{eq:operator_split} to improve the stability and convergence properties of the relaxation solver for the Kmouflage model. After some manipulation, this leads to the following equivalent form of the Kmouflage equation,
\begin{eqnarray}
    &&\tilde{c}\left[1+\gamma\left(\frac{\varphi'^2}{H_0^2}\right)^{n-1}+\frac{(2n+1)\gamma}{3}\left({\tilde{\nabla}^j{\tilde\varphi}\tilde{\nabla}_j\tilde{\varphi}}\right)^{n-1}\right]\tilde{\nabla}^2\tilde{\varphi}\nonumber\\
    &=& 3\frac{{\rm d}A(\bar{\varphi})}{{\rm d}\varphi}\Omega_{\rm m}a^{-1}\tilde{\delta} - 2\tilde{c}(n-1)\gamma\left(\tilde{\nabla}^k\tilde{\varphi}\tilde{\nabla}_k\tilde{\varphi}\right)^{n-2}\hat{\tilde{\nabla}}^i\hat{\tilde{\nabla}}^j\tilde{\varphi}\tilde{\nabla}_i\tilde{\varphi}\tilde{\nabla}_j\tilde{\varphi},
\end{eqnarray}
where we notice that, after discretisation, only the left-hand side contains $\tilde{\varphi}_{i,j,k}$ because $\hat{\tilde{\nabla}}_i\hat{\tilde{\nabla}}_j\tilde{\varphi}$ does not contain $\tilde{\varphi}_{i,j,k}$, and neither does $\tilde{\nabla}_i\tilde{\varphi}$. The latter is because, at second order accuracy, we have the following discrete version of the scalar field gradient:
\begin{equation}
    \nabla_x\varphi = \partial_x\varphi = \frac{1}{2h}\left(\varphi_{i+1,j,k}-\varphi_{i-1,j,k}\right).
\end{equation}
Therefore, the code-unit equation can be written in the following simplified form:
\begin{equation}\label{eq:kmf_eqn_codeunit_final}
    \tilde{c}\tilde{\nabla}^2\tilde{\varphi} = \frac{1}{\Sigma_1}\left(3\frac{{\rm d}A(\bar{\varphi})}{{\rm d}\varphi}\Omega_{\rm m}a^{-1}\tilde{\delta}+\Sigma_2\right),
\end{equation}
where
\begin{eqnarray}
    \Sigma_1 &\equiv& 1+\gamma\left(\frac{\varphi'^2}{H_0^2}\right)^{n-1}+\frac{(2n+1)\gamma}{3}\left({\tilde{\nabla}^j\tilde{\varphi}\tilde{\nabla}_j\tilde{\varphi}}\right)^{n-1},\nonumber\\
    \Sigma_2 &\equiv& -2\tilde{c}(n-1)\gamma\left(\tilde{\nabla}^k\tilde{\varphi}\tilde{\nabla}_k\tilde{\varphi}\right)^{n-2}\hat{\tilde{\nabla}}^i\hat{\tilde{\nabla}}^j\tilde{\varphi}\tilde{\nabla}_i\tilde{\varphi}\tilde{\nabla}_j\tilde{\varphi},
\end{eqnarray}
and $\Sigma_{1,2}$ do not have contribution from the central cell, $\tilde{\varphi}_{i,j,k}$, as described just now. This is therefore essentially a linear equation for $\tilde{\nabla}^2\tilde{\varphi}$. 

The discrete version of $\hat{\nabla}^i\hat{\nabla}^j\varphi\nabla_i\varphi\nabla_j\varphi$ (here we have again neglected the tildes temporarily for simplicity) can be written as
\begin{eqnarray}
    &&\hat{\nabla}^i\hat{\nabla}^j\varphi\nabla_i\varphi\nabla_j\varphi\nonumber\\ 
    &=& \frac{1}{12h^4}\left(\varphi_{i+1,j,k}-\varphi_{i-1,j,k}\right)^2\left(2\varphi_{i+1,j,k}+2\varphi_{i-1,j,k}-\varphi_{i,j+1,k}-\varphi_{i,j-1,k}-\varphi_{i,j,k+1}-\varphi_{i,j,k-1}\right)\nonumber\\
    && + \frac{1}{12h^4}\left(\varphi_{i,j+1,k}-\varphi_{i,j-1,k}\right)^2\left(2\varphi_{i,j+1,k}+2\varphi_{i,j-1,k}-\varphi_{i+1,j,k}-\varphi_{i-1,j,k}-\varphi_{i,j,k+1}-\varphi_{i,j,k-1}\right)\nonumber\\
    && + \frac{1}{12h^4}\left(\varphi_{i,j,k+1}-\varphi_{i,j,k-1}\right)^2\left(2\varphi_{i,j,k+1}+2\varphi_{i,j,k-1}-\varphi_{i+1,j,k}-\varphi_{i-1,j,k}-\varphi_{i,j+1,k}-\varphi_{i,j-1,k}\right)\nonumber\\
    && + \frac{1}{2h^4}\left(\varphi_{i+1,j+1,k}+\varphi_{i-1,j-1,k}-\varphi_{i+1,j-1,k}-\varphi_{i-1,j+1,k}\right)\left(\varphi_{i+1,j,k}-\varphi_{i-1,j,k}\right)\left(\varphi_{i,j+1,k}-\varphi_{i,j-1,k}\right)\nonumber\\
    && + \frac{1}{2h^4}\left(\varphi_{i+1,j,k+1}+\varphi_{i-1,j,k-1}-\varphi_{i+1,j,k-1}-\varphi_{i-1,j,k+1}\right)\left(\varphi_{i+1,j,k}-\varphi_{i-1,j,k}\right)\left(\varphi_{i,j,k+1}-\varphi_{i,j,k-1}\right)\nonumber\\
    && + \frac{1}{2h^4}\left(\varphi_{i,j+1,k+1}+\varphi_{i,j-1,k-1}-\varphi_{i,j+1,k-1}-\varphi_{i,j-1,k+1}\right)\left(\varphi_{i,j+1,k}-\varphi_{i,j-1,k}\right)\left(\varphi_{i,j,k+1}-\varphi_{i,j,k-1}\right).\nonumber
\end{eqnarray}

As mentioned in \S~\eqref{sec:kmouflage}, the Kmouflage field has 4 effects on cosmological structure formation, and thus we also need to write the other effects in code units. Using the code-unit expressions Eqs.~\eqref{eq:code_units}, \eqref{eq:code_units2} and \eqref{eq:Kmo_unit}, we can rewrite the force equation, Eq.~\eqref{eq:force_Kmo}, into
\begin{eqnarray}
    \label{eq:force_Kmo1}\frac{{\rm d}\tilde{\boldsymbol{x}}}{{\rm d}a} &=& \frac{H_0}{a^2\dot{a}}\tilde{\boldsymbol{p}},\\
    \label{eq:force_Kmo2}\frac{{\rm d}\tilde{\boldsymbol{p}}}{{\rm d}a} &=& -\frac{H_0}{\dot{a}}\left[\tilde{\boldsymbol{\nabla}}\tilde{\Phi}_{\rm N}+\beta_{\rm Kmo}\tilde{c}\tilde{\boldsymbol{\nabla}}\tilde{\varphi}\right] - \beta_{\rm Kmo}\frac{{\rm d}\bar{\varphi}}{{\rm d}a}\tilde{\boldsymbol{p}}.
\end{eqnarray}

Consider the linear-theory behaviour of the model, where Eq.~\eqref{eq:kmf_eqn_codeunit_final} can be simplified as
\begin{equation}\label{eq:kmf_eqn_codeunit_lin}
    \tilde{c}\tilde{\boldsymbol{\nabla}}^2\tilde{\varphi} 
    = \left[1+\gamma\left(\frac{\varphi'^2}{H_0^2}\right)^{n-1}\right]^{-1} 3\beta_{\rm Kmo} A(\bar{\varphi}) \Omega_{\rm m}a^{-1}\tilde{\delta}.
\end{equation}
Meanwhile, the Poisson equation is modified to
\begin{equation}\label{eq:kmf_Poission_eqn_codeunit}
    \tilde{\boldsymbol{\nabla}}^2\tilde{\Phi}_{\rm N} = \frac{3}{2} A(\bar{\varphi}) \Omega_{\rm m}a^{-1}\tilde{\delta}.
\end{equation}
This means that as an approximation we have
\begin{eqnarray}
    \frac{\tilde{c}\tilde{\varphi}}{\tilde{\Phi}_{\rm N}} = 2\left[1+\gamma\left(\frac{\bar{\varphi}'^2}{H_0^2}\right)^{n-1}\right]^{-1}\beta_{\rm Kmo},
\end{eqnarray}
and the ratio between the fifth force ($\beta\tilde{c}\tilde{\boldsymbol{\nabla}}\tilde{\varphi}$) and Newtonian gravity ($\tilde{\boldsymbol{\nabla}}\tilde{\Phi}_{\rm N}$) is
\begin{equation}
    \frac{F_5}{F_{\rm N}} = 2\left[1+\gamma\left(\frac{\bar{\varphi}'^2}{H_0^2}\right)^{n-1}\right]^{-1}\beta_{\rm Kmo}^2. 
\end{equation}
Note that here the Newtonian gravity is the force that \textit{already accounts for the particle mass variation}. If $F_{\rm N}$ is the standard Newtonian gravity force (no particle mass variation taken into account yet), the ratio would become 
\begin{equation}
    \frac{F_5}{F_{\rm N}} = 2\left[1+\gamma\left(\frac{\bar{\varphi}'^2}{H_0^2}\right)^{n-1}\right]^{-1}\beta_{\rm Kmo}^2A(\bar{\varphi}). 
\end{equation}
These agree with the fifth-force-to-Newtonian-gravity ratio used in Eq.~\eqref{eq:kmo_lin_growth}, and so it confirms that the code-unit equations are correct and that the modification to Eq.~\eqref{eq:kmo_eom1} indeed does not change the linear theory evolution of the model. 

\subsubsection{Kmouflage background cosmology solver}
\label{sec:BG_solvers}
Because Eqs.~(\ref{eq:kmf_eqn_codeunit_final}, \ref{eq:force_Kmo1}, \ref{eq:force_Kmo2}, \ref{eq:kmf_Poission_eqn_codeunit}) involve various background  quantities such as $\dot{a}$, $\bar{\varphi}$ and ${\rm d}\tilde{\varphi}/{\rm d}a$, for any given Kmouflage model we need to solve its background evolution. This is governed by the following equation \cite{Barreira:2014gwa}, which is the background part of the Kmouflage equation \eqref{eq:kmouflage_eom}:
\begin{equation}\label{eq:kmf_background_eom}
    \left[K_X(\bar{X})+2\bar{X}K_{XX}(\bar{X})\right]\ddot{\bar{\varphi}} + 3HK_X(\bar{X})\dot{\bar{\varphi}} + \frac{{\rm d}A(\bar{\varphi})}{{\rm d}\varphi}8\pi G\bar{\rho}_{\rm m}(a) = 0,
\end{equation}
where $K_{XX}\equiv{{\rm d}^2K}/{{\rm d}X^2}$, along with the modified Friedmann equation (recall that we assume here a flat Universe, $k=0$)
\begin{equation}\label{eq:kmf_background_Friedmann}
    H^2 = \left(\frac{\dot{a}}{a}\right)^2 = \frac{8\pi{G}}{3}\left[\bar{\rho}_{\rm r}(a) + A(\bar{\varphi})\bar{\rho}_{\rm m}(a)\right] + \frac{1}{3}K_X(\bar{X})\dot{\bar{\varphi}}^2 - \frac{1}{3}\lambda^2H_0^2K(\bar{X}),
\end{equation}
and the modified Raychaudhuri equation, 
\begin{equation}\label{eq:kmf_background_Raychaudhuri}
    3\left(\dot{H}+H^2\right) = -4\pi{G}\left[2\bar{\rho}_{\rm r}(a) + A(\bar{\varphi})\bar{\rho}_{\rm m}(a)\right] - K_X(\bar{X})\dot{\bar{\varphi}}^2 - 2\lambda^2H_0^2K(\bar{X}),
\end{equation}
where $\bar{\rho}_{\rm r}$ denotes the background density of radiations (we assume that all three species of neutrinos are massless and thus counted as radiation).

The Friedmann equation \eqref{eq:kmf_background_Friedmann} contains $\dot{\bar{\varphi}}^2$, both explicitly and inside functions of $\bar{X}$, on the right-hand side. Writing 
\begin{equation}
    a\dot{\bar{\varphi}} = {\bar{\varphi}}' = \frac{{\rm d}\bar{\varphi}}{{\rm d}\tau} = \frac{{\rm d}\bar{\varphi}}{{\rm d}N}\frac{{\rm d}N}{{\rm d}\tau} = \frac{a'}{a}\frac{{\rm d}\bar{\varphi}}{{\rm d}N} = \mathcal{H}\frac{{\rm d}\bar{\varphi}}{{\rm d}N} \equiv \mathcal{H}\mathring{\bar{\varphi}},
\end{equation}
where $N\equiv\ln(a)$ and for simplicity we have used an over-circle to denote the derivative with respect to $N$, that equation can be recast, after some manipulation, as
\begin{equation}\label{eq:kmf_background_Friedmann2}
    \frac{\mathcal{H}^2}{H_0^2}\left[1-\frac{1}{6}\mathring{\bar{\varphi}}\right] = \frac{8\pi{G}}{3H_0^2}\left[\bar{\rho}_{\rm r0}a^{-2} + A(\bar{\varphi})\bar{\rho}_{\rm m0}a^{-1}\right] + \frac{1}{3}\lambda^2a^2 + \frac{2n-1}{3}\lambda^2K_0\left[\frac{\mathring{\bar{\varphi}}^2}{2\lambda^2a^2}\right]^n\left[\frac{\mathcal{H}^2}{H_0^2}\right]^n,
\end{equation}
where we have used $\rho_{\rm r}(a)=\rho_{\rm r0}a^{-4}$ and $\rho_{\rm m}(a)=\rho_{\rm m0}a^{-3}$, and have specified to the functional form of $K$ given in Eq.~\eqref{eq:kmf_model_K_function}. Likewise, Eq.~\eqref{eq:kmf_background_Raychaudhuri} can be rewritten as
\begin{align}\label{eq:kmf_background_Raychaudhuri2}
    \frac{\mathcal{H}'}{H_0^2} = &-\frac{4\pi{G}}{3H^2_0}\left[2\bar{\rho}_{\rm r0}a^{-2} + A(\bar{\varphi})\bar{\rho}_{\rm m0}a^{-1}\right] + \frac{2}{3}\lambda^2a^2\nonumber\\ 
    &- \frac{2}{3}\frac{\mathcal{H}^2}{H_0^2}\mathring{\bar{\varphi}}^2 - \frac{1}{3}(n+1)K_0\left[\frac{1}{2\lambda^2a^2}\right]^{n-1}\left[\frac{\mathcal{H}^2}{H_0^2}\right]^n\mathring{\bar{\varphi}}^{2n}.
\end{align}
Finally, using
\begin{equation}
    \bar{\varphi}'' = \mathcal{H}^2\frac{{\rm d}^2\bar{\varphi}}{{\rm d}N^2} + \mathcal{H}'\frac{{\rm d}\bar{\varphi}}{{\rm d}N} = \mathcal{H}^2\ringring{\bar{\varphi}} + \mathcal{H}'\mathring{\bar{\varphi}},\quad \ddot{\bar{\varphi}} = \frac{1}{a^2}\left(\bar{\varphi}''-\mathcal{H}\bar{\varphi}'\right),
\end{equation}
the background Kmouflage field equation, \eqref{eq:kmf_background_eom}, becomes
\begin{equation}\label{eq:kmf_background_eom2}
    \left(K_X+2\bar{X}K_{XX}\right)\left[\frac{\mathcal{H}^2}{H_0^2}\ringring{\bar{\varphi}} + \frac{\mathcal{H}'}{H_0^2}\mathring{\bar{\varphi}}\right] + 2\left(K_X-\bar{X}K_{XX}\right)\frac{\mathcal{H}^2}{H_0^2}\mathring{\bar{\varphi}} + 3\frac{{\rm d}A(\bar{\varphi})}{{\rm d}\varphi}\Omega_{\rm m}\exp(-N) = 0,
\end{equation}
where for simplicity we have not expanded the coefficients of $\ddot{\bar{\varphi}}$ and $\dot{\bar{\varphi}}$. 

Eqs.~(\ref{eq:kmf_background_eom2}, \ref{eq:kmf_background_Friedmann2}, \ref{eq:kmf_background_Raychaudhuri2}) must be solved simultaneously, with Eq.~\eqref{eq:kmf_background_eom2} treated as a differential equation with time variable $N$, and its coefficients depending on Eqs.~\eqref{eq:kmf_background_Friedmann2} and \eqref{eq:kmf_background_Raychaudhuri2}. However, we note that Eqs.~(\ref{eq:kmf_background_Friedmann2}, \ref{eq:kmf_background_Raychaudhuri2}) also both depend on $\mathring{\bar{\varphi}}$, so that these equations are coupled. To solve them, we note that for a given time ($a$ or $N$) and $\mathring{\bar{\varphi}}$, Eq.~\eqref{eq:kmf_background_Friedmann2} can be considered as a quadratic (in case of $n=2$) or cubic (for $n=3$) equation\footnote{Note that in this work we only consider Kmouflage models with $n=2$ or $3$.} of $\mathcal{H}^2/H_0^2$, which can be solved analytically (the expressions of the solutions will not be presented here). This can be substituted into Eq.~\eqref{eq:kmf_background_Raychaudhuri2} to find $\mathcal{H}'/H_0^2$ at the same $a$ (or $N$) and for the same $\mathring{\bar{\varphi}}$. After that, $\mathring{\bar{\varphi}}$, $\mathcal{H}^2/H_0^2$ and $\mathcal{H}'/H_0^2$ at time $a$ or $N$ can be used to calculate $\ringring{\bar{\varphi}}$ using Eq.~\eqref{eq:kmf_background_eom2} and this one we can integrate Eq.~\eqref{eq:kmf_background_eom2} forward in time to obtain the whole evolution of $\bar{\varphi}$ and $\mathcal{H}$. The equation is solved using a fifth-sixth order continuous Runge-Kutta method\footnote{For this numerical integrator we have adapted \texttt{subroutine dverk} from the \href{https://camb.info/}{\textsc{camb} code}, originally developed in Fortran 66 by K.~R.~Jackson.}.

In our calculation we have included both radiation and non-relativistic matter, with `radiation' including CMB photons with a current temperature of $2.7255$ K and $3.046$ flavours of massless neutrinos. We defer the implementation of massive neutrinos which couple to the scalar field in a different way from non-relativistic matter in the Kmouflage model, to future works. 

We remark that $\lambda$ is \textit{not} a free parameter of the model. Rather, once the density parameters $\Omega_{\rm m}$, $\rho_{\rm r0}$ and $H_0$ are specified, $\lambda$, which roughly quantifies the amount of dark energy in this model, must take some certain value in order to ensure consistency --- if $\lambda$ is too large, the predicted $H(a=1)$, by solving Eqs.~(\ref{eq:kmf_background_eom2}, \ref{eq:kmf_background_Friedmann2}, \ref{eq:kmf_background_Raychaudhuri2}) with given initial conditions of $\bar{\varphi}$ and $\mathring{\bar{\varphi}}$, will be larger than the desired (input) value of $H_0$, and vice versa. In practice, \textsc{mg}-\textsc{glam} starts from a trial value of $\lambda=1$, evolves the above equations from some initial redshift ($z_{\rm i}=10^5$) to $z=0$, and checks if the calculated value of $H(a=1)$ is equal to the desired value $H_0$ (within a small relative error of order $\mathcal{O}\left(10^{-6}\right)$) --- if the predicted $H(a=1)$ value overshoots the desired $H_0$, $\lambda$ is decreased, and vice versa. This process is repeated until we have obtained a good approximation to $\lambda$, with the relative error of the predicted $H_0$ less than $10^{-6}$. The initial conditions of $\bar{\varphi}$ and $\mathring{\bar{\varphi}}$ at $z_{\rm i}=10^5$ are not important, as long as their values are sufficiently small (in the \textsc{mg-glam} code we set them to be both $10^{-30}$). Once the value of $\lambda$ has been determined in this way, it is stored to be used in other parts of the code; also stored are a large array for the various background quantities such as $H, \dot{H}, \bar{\varphi}$ and $\dot{\bar{\varphi}}$ --- if needed at any time by the Kmouflage field solver of \textsc{mg}-\textsc{glam}, these quantities will be linearly interpolated in the scale factor $a$ or $N=\ln(a)$.  

\section{Numerical code tests}
\label{sec:code_tests}
We have performed a series of code tests to check that our MG solvers work correctly following the framework of the {\sc ecosmog} and {\sc mg-arepo} codes \citep{Li:2013nua,Hernandez-Aguayo:2020_MGAREPO_code_paper}. To this end, we have run low-resolution simulations with box size $L=256\Mpch$ and $N_{\rm g} = 256$ grid cells in each coordinate direction.

\subsection{Background cosmology tests}
\label{sec:bkg_tests}
\begin{figure*}
    \centering
    \includegraphics[width=0.49\textwidth]{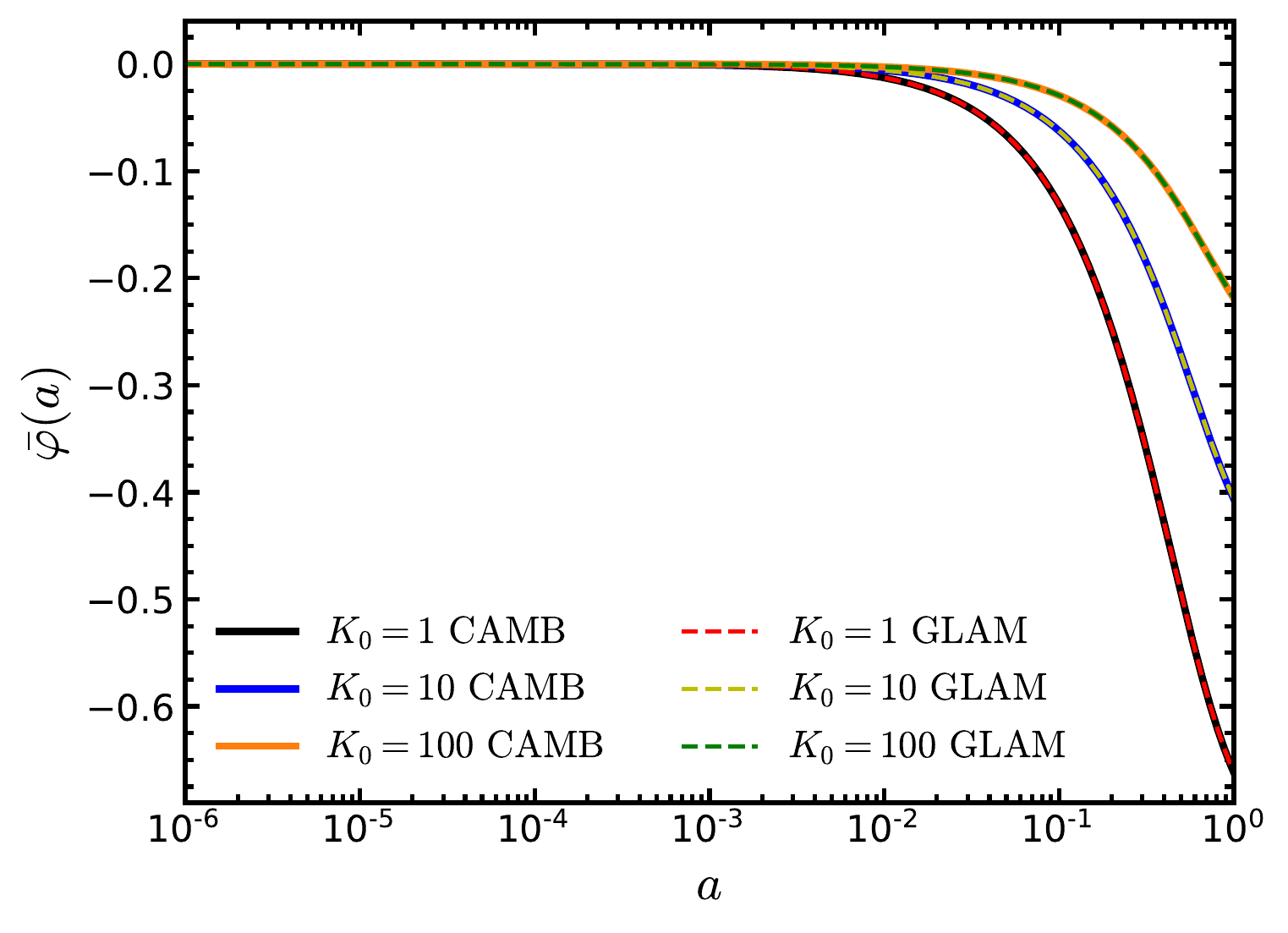}
    \includegraphics[width=0.49\textwidth]{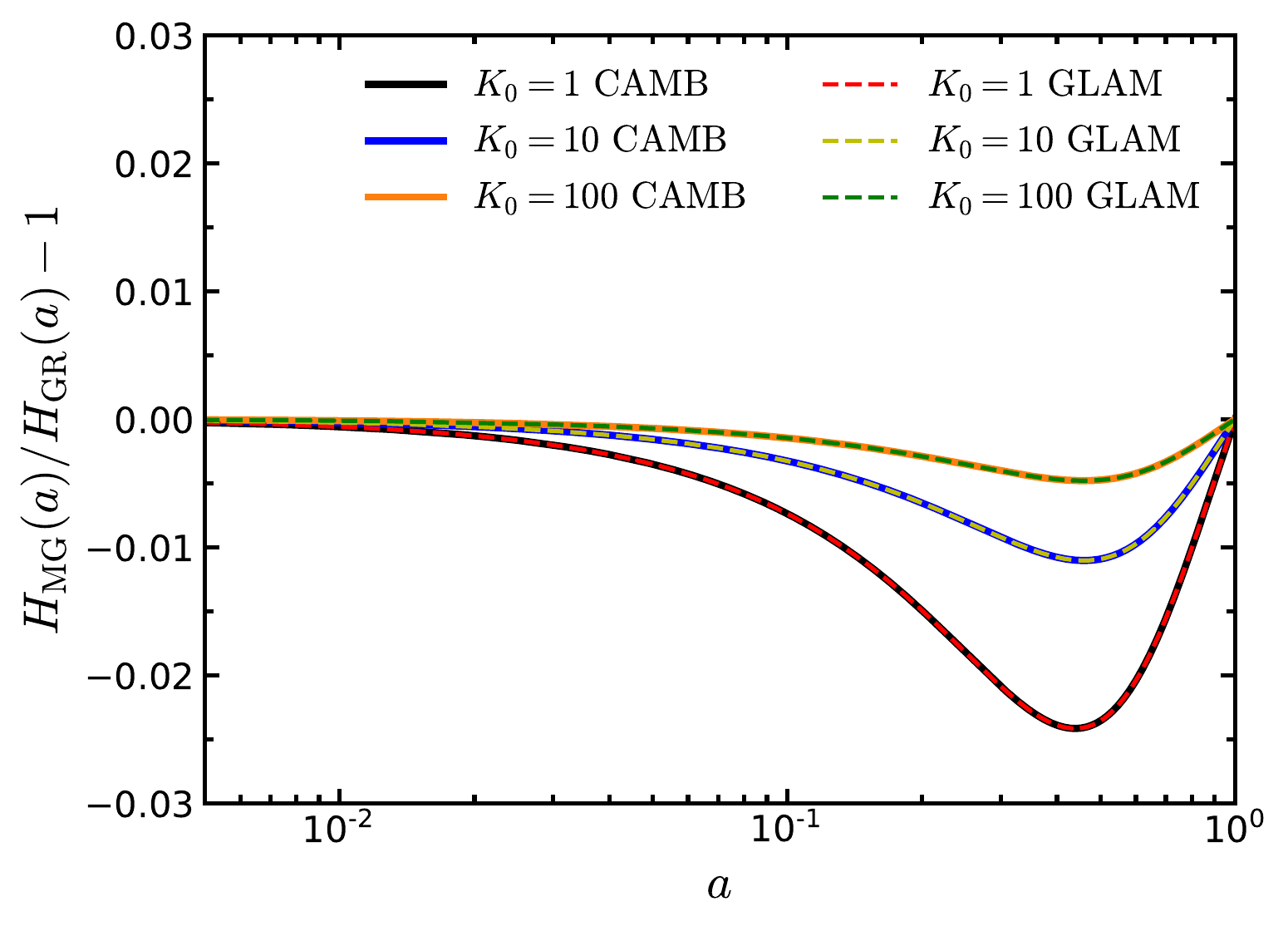}
    \caption{Cosmological background evolution tests. {\it Left panel:} The evolution of the background scalar field in the Kmouflage model predicted by {\sc camb} (solid lines) and {\sc mg-glam} (dashed lines). {\it Right panel:} The relative difference of the Hubble expansion rate between the Kmouflage and the GR models measured from the outputs of a modified \href{https://camb.info/}{{\sc camb}} code (solid lines) and {\sc mg-glam} (dashed lines) codes. Different colours correspond to different values of $K_0$ as shown in the legend. In all cases we have used $n=2$ and $\beta_{\rm Kmo} = 0.2$.}
    \label{fig:bkg_tests}
\end{figure*}

Of the two classes of models considered in this work, the nDGP models have an expansion history identical to that of $\Lambda$CDM by design, but the Kmouflage models can have non-negligible deviations from $\Lambda$CDM in background expansion \citep{Barreira:2014gwa}. Our numerical solver of the background equations have been described in \S~\ref{sec:BG_solvers}, and in this subsection we test the reliability of that implementation. 

To this end, we have compared the predictions by the numerical Kmouflage background solver in \textsc{mg}-\textsc{glam} with the results obtained using a modified version of the \href{https://camb.info/}{\textsc{camb}} code used in \cite{Barreira:2014gwa}. The results are shown in Fig.~\ref{fig:bkg_tests}, where the left panel shows the background Kmouflage field as a function of the scale factor $a$, and the right panel shows the ratio between the modified expansion rate $H_{\rm MG}(a)$ and that of standard $\Lambda$CDM, $H_{\rm GR}(a)$, with the same $\Omega_{\rm m}$ and $H_0$. As we can see, for both quantities and all models tested here, the two codes agree very well. 

In this figure, we have shown the results of fixed $n=2$ and $\beta_{\rm Kmo}=0.2$, but varying values of $K_0$; however, we have checked that the same agreement between the two codes hold for other values of $n$ and $\beta_{\rm Kmo}$. 

We note that in the models studied here, the background scalar field is negative, $\bar{\varphi}<0$, and decays over time. This has two implications: ($i$) the direction-dependent force in Eq.~\eqref{eq:force_Kmo} or Eq.~\eqref{eq:force_Kmo2}, $-\beta_{\rm Kmo}\frac{{\rm d}\bar{\varphi}}{{\rm d}a}\tilde{\boldsymbol{p}}$, points to the direction of the particle's movement, which means that it actually speeds up the particle rather than acting as a `friction' force; ($ii$) given that $\beta_{\rm Kmo}>0$ in the models studied here, we have $A(\bar{\varphi})=\exp\left(\beta_{\rm Kmo}\bar{\varphi}\right)<1$ at late times, which means that the particles contribute less to the Poisson equation, cf.~the discussion below Eq.~\eqref{eq:kmo_lin_growth}; equivalently, we can consider this as a decrease of the effective dark matter particle mass over time.

Therefore, we can have a quick discussion about how the 4 effects of the Kmouflage model in structure formation, discussed below Eq.~\eqref{eq:kmo_lin_growth}, depend on the parameter $K_0$, when $n=2$ and $\beta_{\rm Kmo}$ is fixed. This may also help us appreciate the complexity of this model when discussing its effects on the halo mass function below.
\begin{itemize}
    \item \textit{varying particle mass}: the Kmofulage models have $A(\bar{\varphi})<1$ and the smaller $K_0$ is (we only focus on the cases with $K_0>1$ here), the smaller $A(\bar{\varphi})$ becomes, which reduces the Newtonian force and hence weakens structure formation.
    \item \textit{modified expansion rate}: as shown in the right panel of Fig.~\ref{fig:bkg_tests}, decreasing $K_0$ slows down the expansion rate more, which can enhance structure formation. However, even for $K_0=1$ the expansion rate is only $\approx2\%$ smaller than in $\Lambda$CDM, and so this effect is expected to be small.
    \item \textit{direction-dependent force}: for fixed $\beta_{\rm Kmo}$, the amplitude of this force (for particles moving at the same speed) depends on $|{\rm d}\bar{\varphi}/{\rm d}a|$, which is clearly larger for smaller $K_0$ values.
    \item \textit{the fifth force}: the ratio between the amplitudes of the fifth and Newtonian forces is $2\beta^2_{\rm Kmo}/K_X$, with $K_X(\bar{X})$ given in Eq.~\eqref{eq:K_Xbar}. Neglecting the weak dependence of $\lambda$ on $K_0$, we can see that the size of $K_X$ is a result of the competition between $K_0$ and $|\bar{\varphi}'|$ or equivalently $|{\rm d}\bar{\varphi}/{\rm d}a|$: but from the left panel of Fig.~\ref{fig:bkg_tests} it is evident that $K_0$ varies more than $\left(\bar{\varphi}'\right)^2$, and so $K_X$ decreases with a decreasing $K_0$, making the fifth force force relatively stronger.
\end{itemize}
Therefore, the effect of varying particle mass works against all the remaining three effects, and which side wins the competition of boosting versus weakening structure formation can only be answered by numerical solutions. 

\subsection{Density field tests}
\label{sec:density_tests}
This subsection is devoted to the tests of the multigrid solvers for the nDGP and Kmouflage models, using different density configurations for which the scalar field solution can be solved analytically or using a different numerical code.

\subsubsection{Uniform density field tests}
\label{sec:uniform_tests}
For the first test we consider the case where the solution of the scalar field, $\varphi$, is constant in space. A constant field should be obtained if we choose a homogeneous matter distribution (i.e., the density field is uniform and equal to the cosmological background value). To check this we have set $\tilde{\delta}_{i,j,k} = 0$ and chose a set of random values that follow a uniform distribution in the range $[-0.05,0.05]$ as initial guesses of $\tilde{\varphi}_{i,j,k}$, then we let the code run until the residual is $d^\ell \leq 10^{-8}$. 

The results of this test are shown in the upper left panel of Fig.~\ref{fig:code_tests}, where the orange (blue) dots represent the initial guess, and the orange (blue) solid line is the numerical solution after relaxation, in the nDGP (Kmouflage) case. In both cases a constant solution is obtained by the code, as expected. 

\subsubsection{1D density field tests}
\label{sec:1D_tests}
For our next test, we consider a one-dimensional sine density field (varying in the $x$ direction) given by,
\begin{equation}\label{eq:sine_nDGP}
\tilde{\delta}(\tilde{x}) = -\frac{a\beta_{\rm DGP}}{\Omega_{\rm m}N^2_{\rm g}}\sin\frac{2\pi\tilde{x}}{N_{\rm g}}\,,
\end{equation}
for nDGP and
\begin{equation}\label{eq:sine_Kmo}
\tilde{\delta}(\tilde{x}) = -\frac{a\tilde{c}}{3\beta_{\rm Kmo}\Omega_{\rm m}}\qty{1 + \frac{n(2n+1)K_0}{2a^2\lambda^2}\left(\frac{2\pi{K}}{N_{\rm g}}\right)^2A\sin\frac{2\pi{K}\tilde{x}}{N_{\rm g}} \left[\frac{2\pi{A}}{N_{\rm g}}\cos\frac{2\pi{K}\tilde{x}}{N_{\rm g}}\right]^{2(n-1)}}\,,
\end{equation}
for Kmouflage, where the model parameters are set as $n=2$, $K_0 = 1$, $\beta_{\rm Kmo} = 0.1$, while $A = 0.1$, $K = 4$ are extra parameters describing the specific density field. We have checked other parameter values and found similar agreement, but we only present the results for one set of parameters here, to make the plot easier to read.

The analytical solutions of the nDGP and Kmouflage scalar field equations of motion, Eq.~\eqref{eq:dgp_eom_code} and Eq.~\eqref{eq:kmf_eqn_codeunit_final}, for these density fields are,
\begin{eqnarray}
    \tilde{\varphi}(\tilde{x}) &=& \frac{1}{4\pi^2}\sin\frac{2\pi\tilde{x}}{N_{\rm g}}\,,\\
    \tilde{\varphi}(\tilde{x}) &=& {A}\sin\frac{2\pi{K}\tilde{x}}{N_{\rm g}}\,,
\end{eqnarray}
respectively. 

\begin{figure*}
    \centering
    \includegraphics[width=1\textwidth]{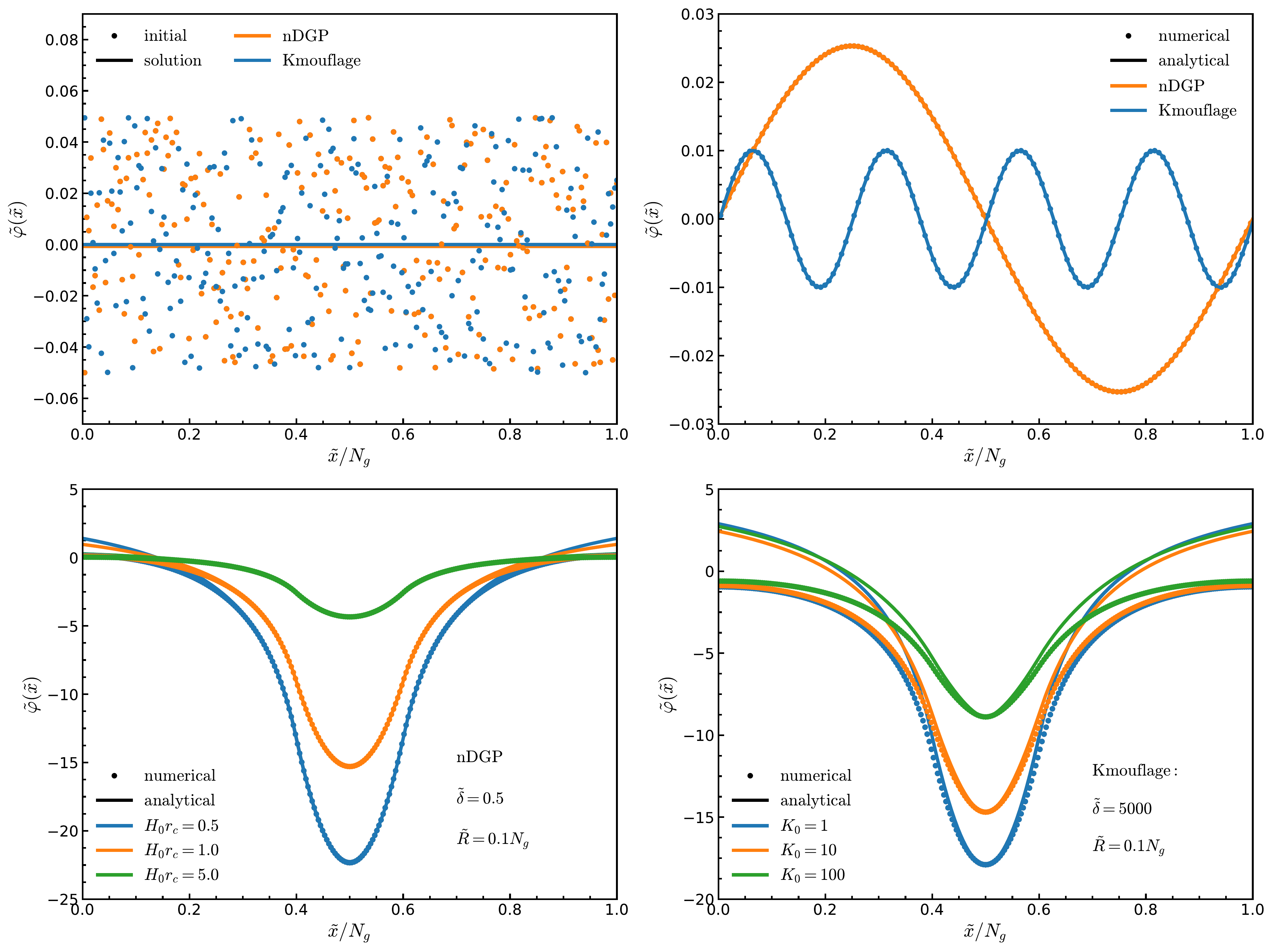}
    \caption{Results of the code tests of the multigrid solver in \textsc{mg-glam}: {\it Upper left panel:} Uniform density test, where the coloured dots represent the random initial guesses of the scalar field uniformly generated in the range $[-0.05,0.05]$ and the solid lines show the final solutions by \textsc{mg-glam}, orange for the nDGP model and blue for Kmouflage. {\it Upper right panel:} The 1D sine density field tests, the solid lines show the analytical solutions and the dots correspond to the numerical results, orange for the nDGP model and blue for Kmouflage. {\it Lower left panel:} Spherical overdensity test using $\tilde{\delta} = 0.5$ and $\tilde{R}=0.1N_{\rm g}$ for three nDGP models with $H_0r_{\rm c} = 0.5$ (blue), $H_0r_{\rm c} = 1$ (orange) and $H_0r_{\rm c} = 5$ (green). The lines represent the analytical solutions, while the dots correspond to \textsc{mg-glam} code test results. {\it Lower right panel:} Spherical overdensity test using $\tilde{\delta} = 5000$ and $\tilde{R}=0.1N_{\rm g}$ for three Kmouflage models, with $K_0 = 1$ (blue), $K_0 = 10$ (orange) and $K_0 = 100$ (green); in all cases we have used $n=2$ and $\beta = 0.2$. The lines represent the analytical solutions, while the dots correspond to the \textsc{mg-glam} results.}
    \label{fig:code_tests}
\end{figure*}

The results of this test are shown in the upper right panel of Fig.~\ref{fig:code_tests}, where the orange (blue) dots correspond to the numerical solution and the orange (blue) solid line represents the analytical solution for the nDGP (Kmouflage) model. The code is able to accurately recover the analytical predictions in both models. 

\subsubsection{3D spherical overdensity field tests}
\label{sec:3D_tests}
The 3D spherical tests help us to check that the code is able to solve the nonlinear terms of the nDGP and Kmouflage equations correctly. For the nDGP spherical test we use the code units and $a=1$, so that Eqs.~\eqref{eq:dvarphidr_in} and \eqref{eq:dvarphidr_out} can be written as
\begin{equation}
\frac{\rd\tilde{\varphi}}{\rd\tilde{r}} = \frac{3 \beta_{\rm DGP}}{4R_{\rm c}} \left[\sqrt{1+\frac{8\Omega_{\rm m} R_{\rm c}}{9\beta_{\rm DGP}^2}\tilde{\delta}}-1\right]\tilde{r},\label{eq:phir_in}
\end{equation}
for $\tilde{r}\leq\tilde{R}$ and
\begin{equation}
\frac{\rd\tilde{\varphi}}{\rd\tilde{r}} = \frac{3 \beta_{\rm DGP}}{4R_{\rm c}} \left[\sqrt{1+\frac{8\Omega_{\rm m} R_{\rm c}}{9\beta_{\rm DGP}^2}\frac{\tilde{R}^3}{\tilde{r}^3}\tilde{\delta}}-1\right]\tilde{r},\label{eq:phir_out}
\end{equation}
for $\tilde{r}\geq\tilde{R}$, where $\tilde{r}$ is the comoving radial distance from the centre of the spherical overdensity, $\tilde{R}$ is the radius of the latter and $\tilde{\delta}$ is the (constant) value of the overdensity inside $\tilde{R}$, all in code units. 

Similarly, the Kmouflage equation, \eqref{eq:Kmo_eom_spherical}, in code units can be solved (for the special case of $n=2$) as
\begin{equation}
\frac{\rd\tilde{\varphi}}{\rd\tilde{r}} = \frac{1}{6\gamma} g^{1/3}(\tilde{r}) - \frac{2}{g^{1/3}(\tilde{r})}\,,    
\end{equation}
where $g(\tilde{r})$ is a function defined as
\begin{equation}
g(\tilde{r}) \equiv \gamma^2\qty[108f(\tilde{r}) +  20.78460969\sqrt{27f^2(\tilde{r}) + \frac{4}{\gamma}}]\,,
\end{equation}
which is obtained by analytically solving a cubic equation satisfied by ${\rm d}\tilde{\varphi}/{\rm d}\tilde{r}$, and the function $f(\tilde{r})$ is defined as
\begin{align}
f(r) \equiv
 \begin{cases}
 \displaystyle \frac{\beta_{\rm Kmo} \Omega_{\rm m}\tilde{r}}{\tilde{c}}\tilde{\delta}\,, & \tilde{r} \leq \tilde{R}\,,\\
 \displaystyle \frac{\beta_{\rm Kmo} \Omega_{\rm m}}{\tilde{c}}\frac{\tilde{R}^3}{\tilde{r}^2}\tilde{\delta}\,, & \tilde{r} > \tilde{R} \,.
 \end{cases}
\end{align}

For these tests, we place the spherical overdensity in the centre of the grid and $\tilde{r}$ is defined as,
\begin{equation}
\tilde{r} \equiv \sqrt{\left(\tilde{x}-N_{\rm g}/2\right)^2 + \left(\tilde{y}-N_{\rm g}/2\right)^2 + \left(\tilde{z}-N_{\rm g}/2\right)^2}\,,    
\end{equation}
where $(\tilde{x}, \tilde{y}, \tilde{z})$ is the coordinate of a mesh cell in code units, with $\tilde{x}, \tilde{y}, \tilde{z}$ running from $0$ to $N_{\rm g}$. For cells with $\tilde{r}\leq\tilde{R}$, we set $\tilde{\delta}$ to a nonzero value; otherwise $\tilde{\delta}=0.0$. We use the values of $\tilde{R}=0.1N_{\rm g}$, $\tilde{\delta} = 0.5$ and $H_0r_{\rm c} = 0.5,\,1,\,5$ for nDGP and $\tilde{R}=0.1N_{\rm g}$, $\tilde{\delta}=5000$ and $K_0 = 1,\,10,\,100$ with $n=2$ and $\beta = 0.2$ for Kmouflage. 

In both models, the above analytical solutions are for ${\rm d}\tilde{\varphi}/{\rm d}\tilde{r}$. We then numerically integrate this quantity to get the radial profiles of $\tilde{\varphi}$. The solutions $\tilde{\varphi}(\tilde{r})$ obtained this way may have a constant shift relative to the numerical solutions obtained by \textsc{mg-glam}, which is because the DGP and Kmouflage equations contain only spatial derivatives of the scalar field\footnote{Recall that for the Kmouflage model what is solved is essentially the spatial perturbation of the scalar field $\delta\varphi=\varphi-\bar{\varphi}$, rather than the total or background scalar field. While the latter does enter the equation, e.g., through $A(\varphi)\approx A(\bar{\varphi})$, what is solved by the relaxation is actually $\delta\varphi$ which does satisfy the shift symmetry, c.f., the discussion below Eq.~\eqref{eq:Kmo_unit}.}, and so any solution to these equations shifted by a constant value everywhere would still be a valid solution. Thus, to compare the analytical and numerical solutions, we shift the former so that it has the same peak value as the latter.

The results from these tests are shown in the lower left and right panels of Fig.~\ref{fig:code_tests} for the nDGP and the Kmouflage models, respectively. The coloured symbols in the different panels represent the numerical solutions from {\sc mg-glam} and the solid lines are the analytical solutions. We can see that the two agree well, especially at small $\tilde{r}$, i.e., close to the centre of the spherical overdensity. Far from the centre, the agreement becomes poorer because the analytical solution does not assume periodicity of the spherical overdensity, while the numerical code uses periodic boundary conditions so that the field sees the overdensities in the replicated boxes as well.  

\begin{figure*}
    \centering
    \includegraphics[width=0.49\textwidth]{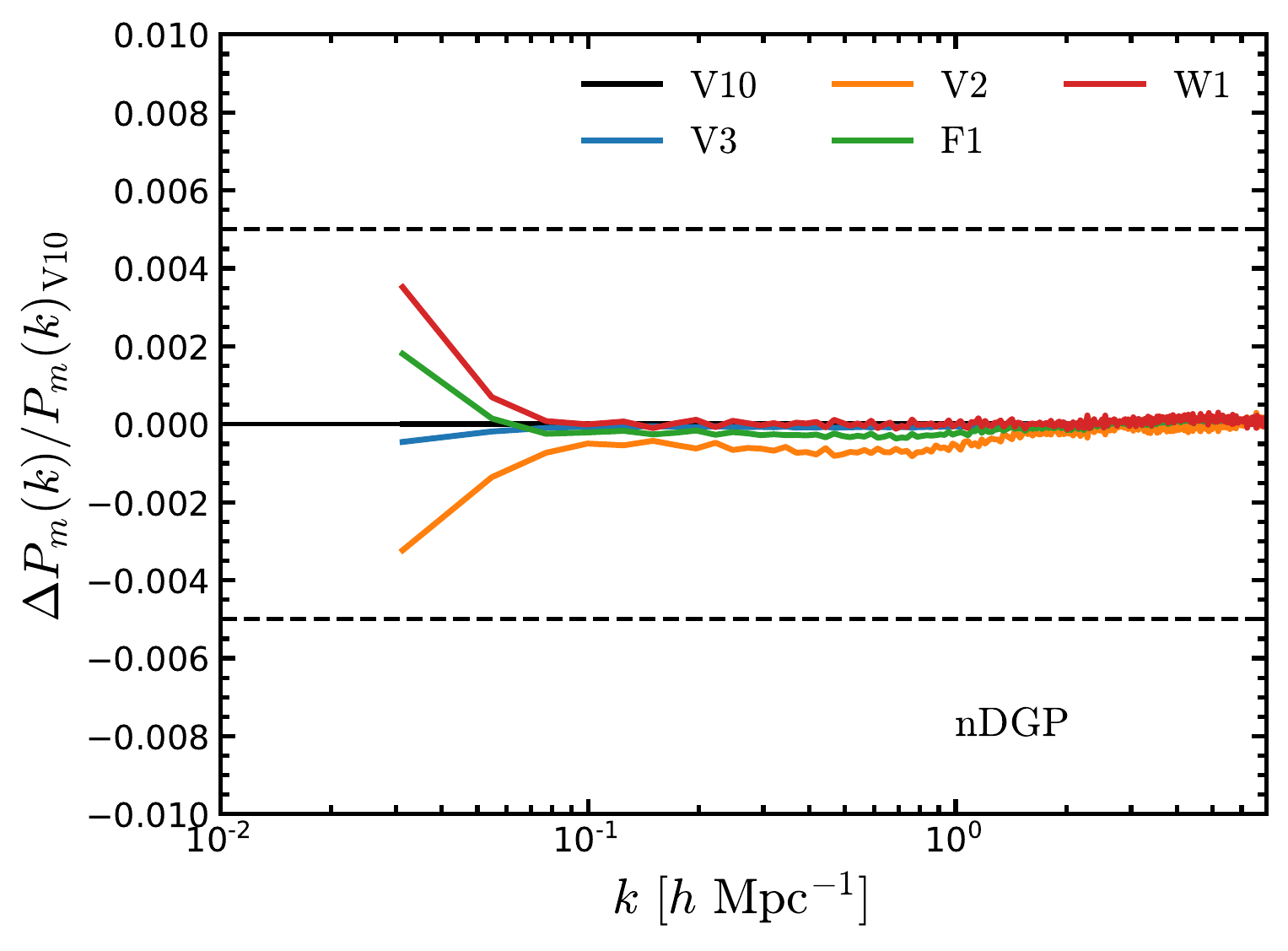}
    \includegraphics[width=0.49\textwidth]{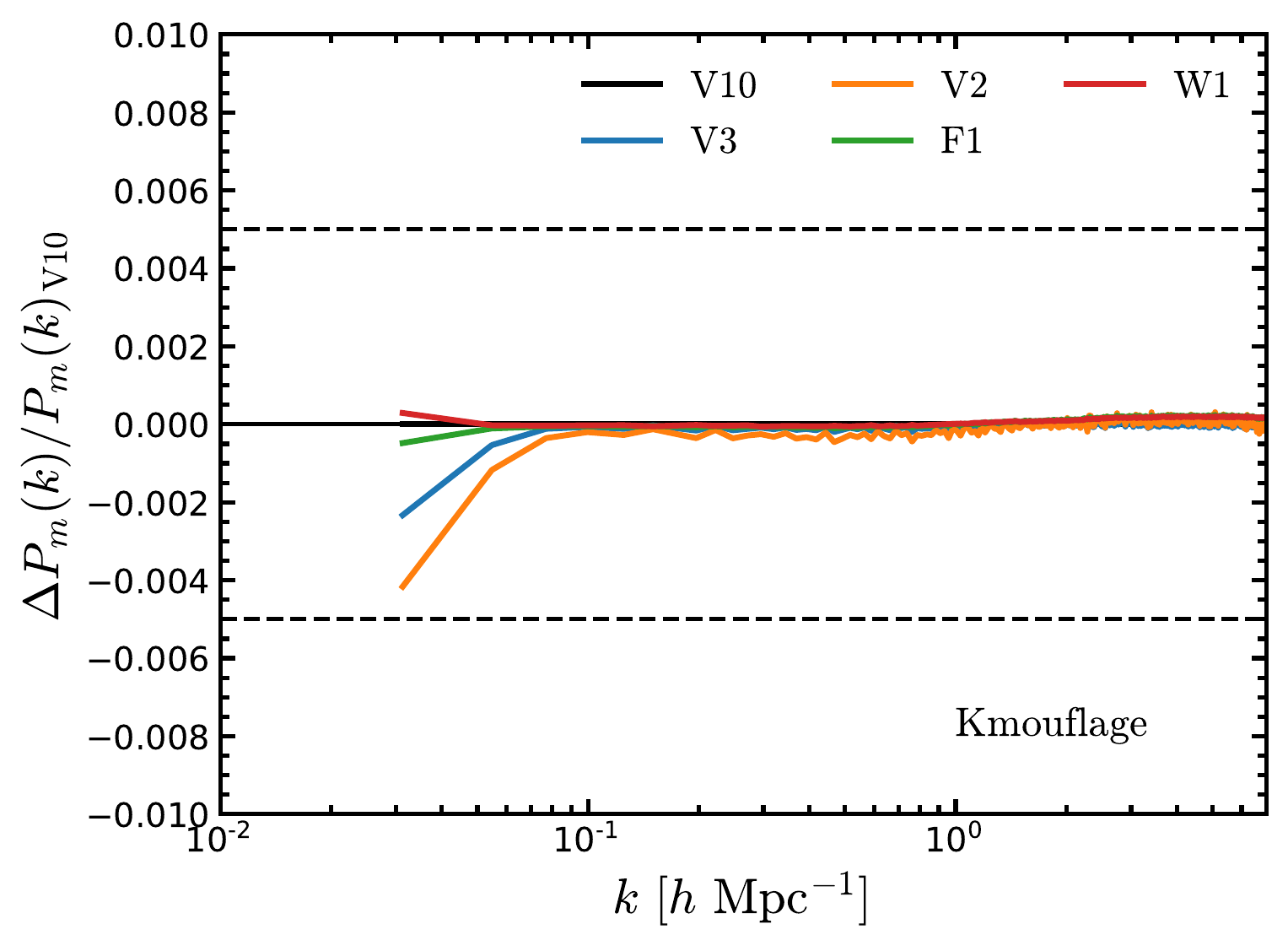}
    \caption{Convergence tests. Comparison of the measured matter power spectrum from \textsc{mg-glam} simulations with $L=256\Mpch$, $N_{\rm p} = 512$ and $N_{\rm g} = 1024$ at $z=0$ for an nDGP model with $H_0r_{\rm c} = 1$ ({\it left panel}) and a Kmouflage model with $n=2$, $\beta=0.2$ and $K_0 = 1$ ({\it right panel}) using multigrid schemes with different number of V-cycles (V10, V3 and V2), F-cycles (F1) and W-cycles (W1). Here `V10' means 10 V-cycles and so on. The coloured lines are the relative differences with respect to V10, and the dashed black lines show the $0.5$ per cent level of difference. Within a given cycle, two Gauss-Seidel sweeps (smoothings) are performed on each multigrid level (which applies to V-cyles, F-cycles and W-cycles). For F-cycles and W-cycles, very good agreement is obtained after just one cycle, and so we have only included results for F1 and W1.}
    \label{fig:cycles_test}
\end{figure*}

\subsection{Convergence tests}
\label{sec:convergence_tests}
As mentioned in \S~\ref{subsubsect:relaxation}, in \textsc{mg-glam} we have implemented three different arrangements of the multigrid solver --- V-cycles, F-cycles and W-cycles. We have compared the accuracy and computational costs of these arrangements. To do so, we have run a series of smaller simulations for the nDGP model with $H_0r_{\rm c} = 1$ and for Kmouflage with $n=2$, $K_0=1$ and $\beta_{\rm Kmo}=0.2$. The simulations follow the evolution of $512^3$ dark-matter particles in a cubic box of length $L=256\Mpch$ with $N_{\rm g}=1024$ grid points in each direction. We use 10, 3 and 2 V-cycles (V10, V3 and V2), one F-cycle (F1) and one W-cycle (W1) to test the convergence of the solution. In all cases, within each cycle the code transverse the mesh twice to perform Gauss-Seidel relaxation.

In Fig.~\ref{fig:cycles_test} we show the relative difference of the nonlinear matter spectrum measured at $z=0$ from our test simulations described above for the nDGP (left panel) and Kmouflage (right panel) models where the benchmark case is V10 (black solid line). We find a permille agreement between all the different schemes, and different numbers of cycles used to solve the PDEs, on almost all scales. However, the running time is larger when using more cycles or iterations, i.e., the slowest simulations are those using V10. The F-cycles and W-cycles are more efficient in reducing the residual, which is not surprising given that they walk more times across the fine and coarse multigrid levels. However, they are also slower than V2. As a compromise between accuracy and cost, we have therefore decided to always use V2 in our cosmological runs. It is actually incredible to reach convergence with just two V-cycles (and two Gauss-Seidel passings of the entire mesh in each cycle), for nonlinear equations in the DGP and Kmouflage models. 

\subsection{Scaling tests}
\label{sec:scaling}
\begin{figure*}
    \centering
    \includegraphics[width=0.497\textwidth]{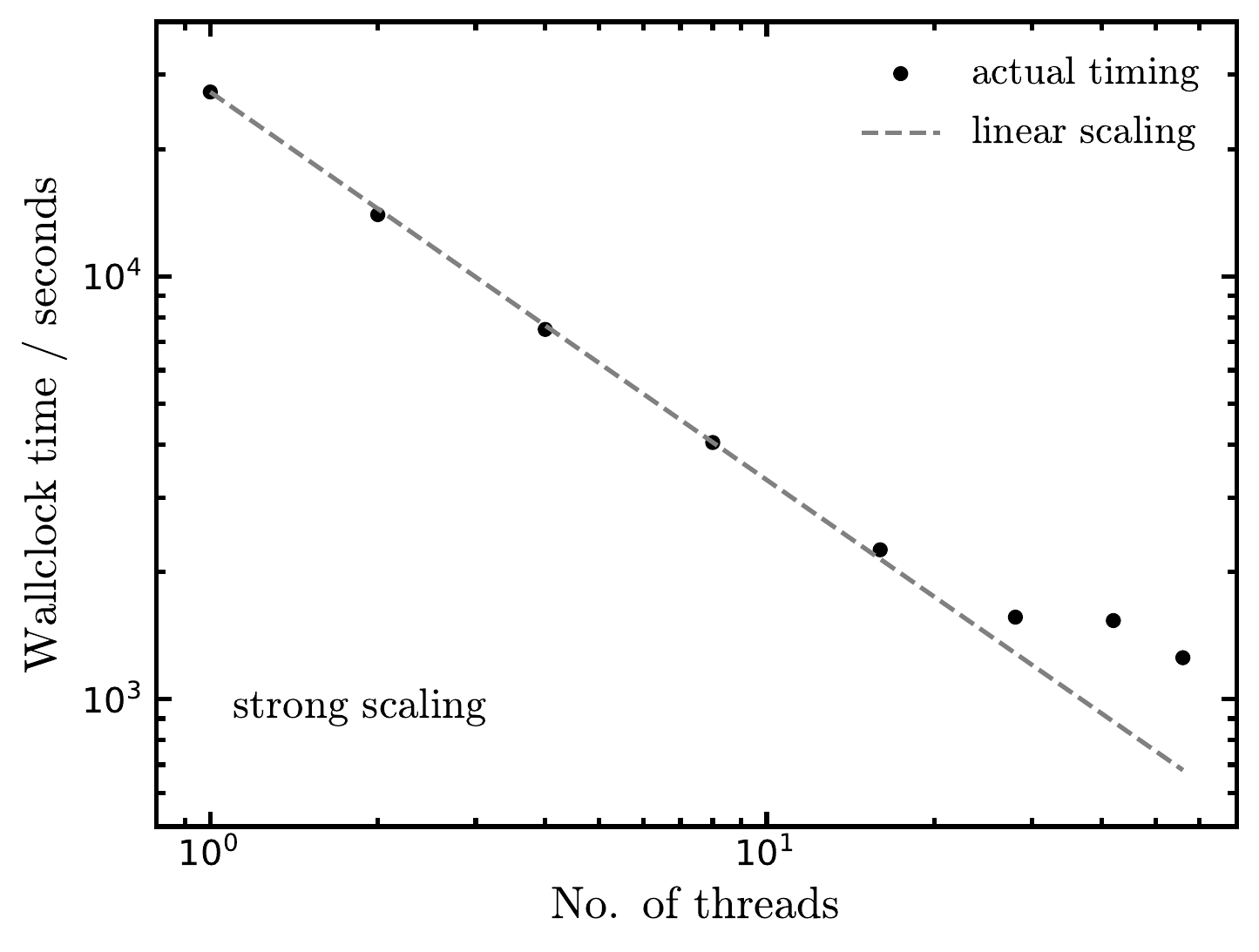}
    \includegraphics[width=0.497\textwidth]{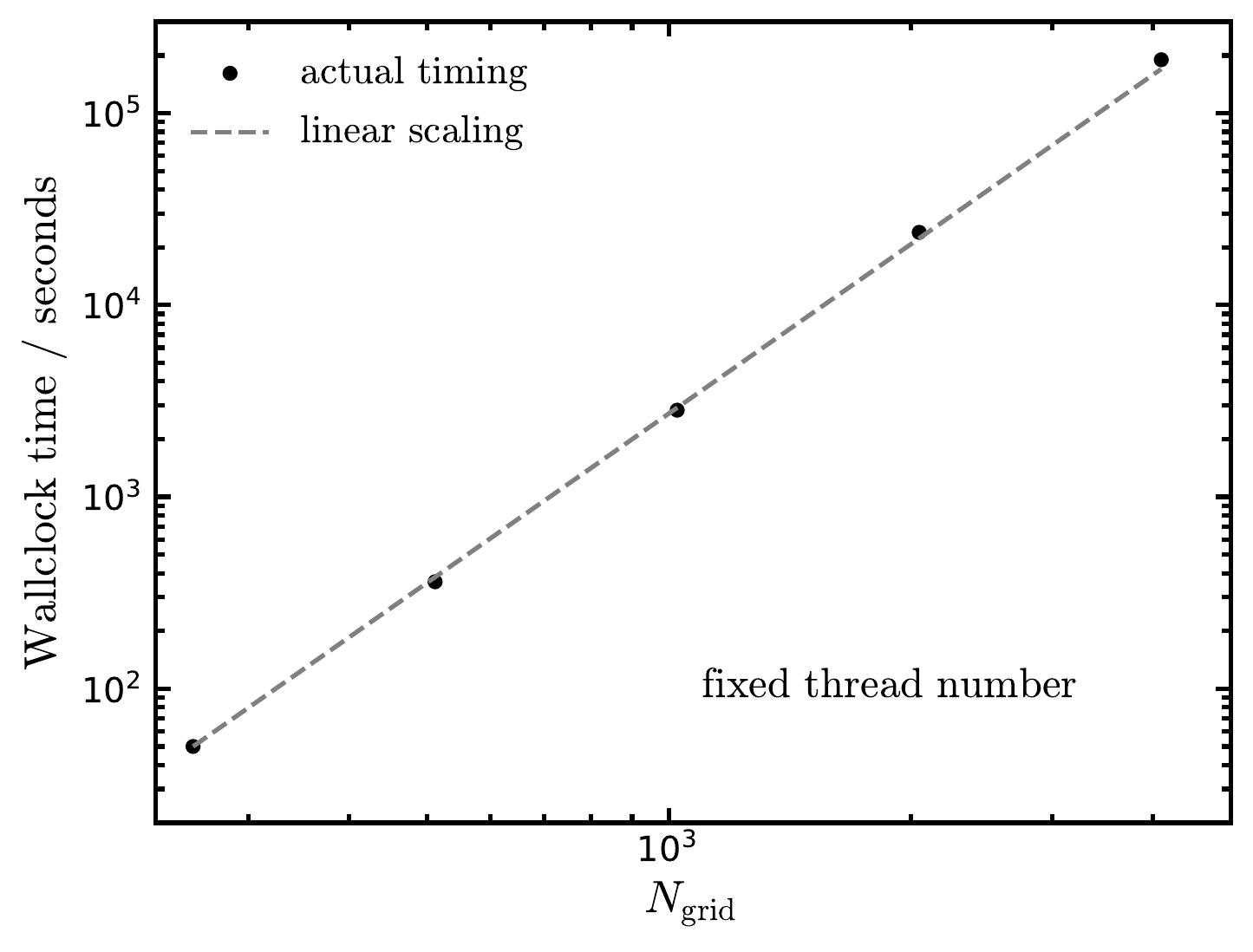}
    \caption{Results of the strong scaling ({\it left panel}) and fixed-thread-number ({\it right panel}) tests of {\sc mg-glam}. The black dots correspond to the wallclock running time of a series of cosmological simulations of the nDGP model with $H_0r_c=1$, while the grey dashed lines show the ideal linear scaling relation.}
    \label{fig:scaling}
\end{figure*}

To test the parallelisation performance and scalability of {\sc mg-glam}, we have run a series of simulations for the nDGP model with $H_0r_c=1$, with varying sizes and/or resolutions. The strong scaling is shown in the left panel of Fig.~\ref{fig:scaling}, where we test the speed-up of the code when varying the number of {\sc openmp} threads while fixing the size of the simulation. The test simulations follow the evolution of $N_{\rm p}^3 = 256^3$ particles in a box of size $L=128\Mpch$ with $512^3$ grids. We vary the number of threads from 1 to 56 (symbols) and found a nearly perfect agreement with the ideal linear scaling relation (dashed line) when using up to 16 threads. The code also shows good scalability when using up to 56 threads, and the deviation from ideal scaling is likely caused by the fact that the test run has a small size so that the overhead becomes a significant fraction of the total time when using too many threads.

The right panel of Fig.~\ref{fig:scaling} displays the result of the 
tests with fixed number of \textsc{openmp} threads (56), but varying the simulation size. For this test we run five simulations with different number of grid points and DM particles, $N_{\rm g} = 256,\,512,\,1024,\,2048$ and $4096$ (symbols) with $N_{\rm p} = N_{\rm g}/2$ and $L=512h^{-1}\mathrm{Mpc}$ in all cases. Again we find a nearly perfect agreement with an ideal linear scaling (dashed line).

These tests suggest that \textsc{mg-glam} has excellent scalability, and the running times for the simulations performed in this work can be used to reliably predict the requirement for even larger runs.
  
\subsection{Resolution tests}
\label{sec:res}
\begin{figure*}
    \centering
    \includegraphics[width=0.49\textwidth]{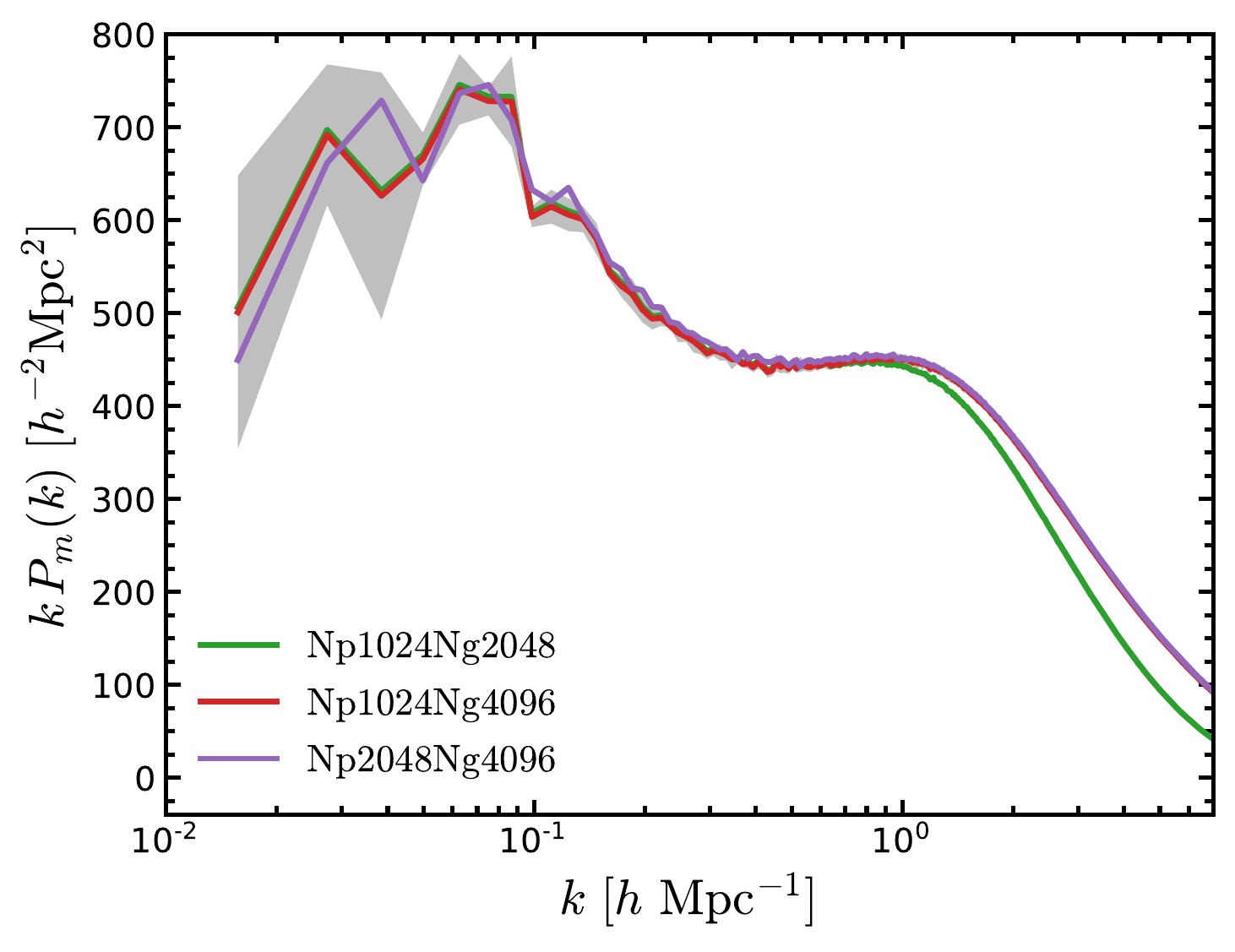}
    \includegraphics[width=0.50\textwidth]{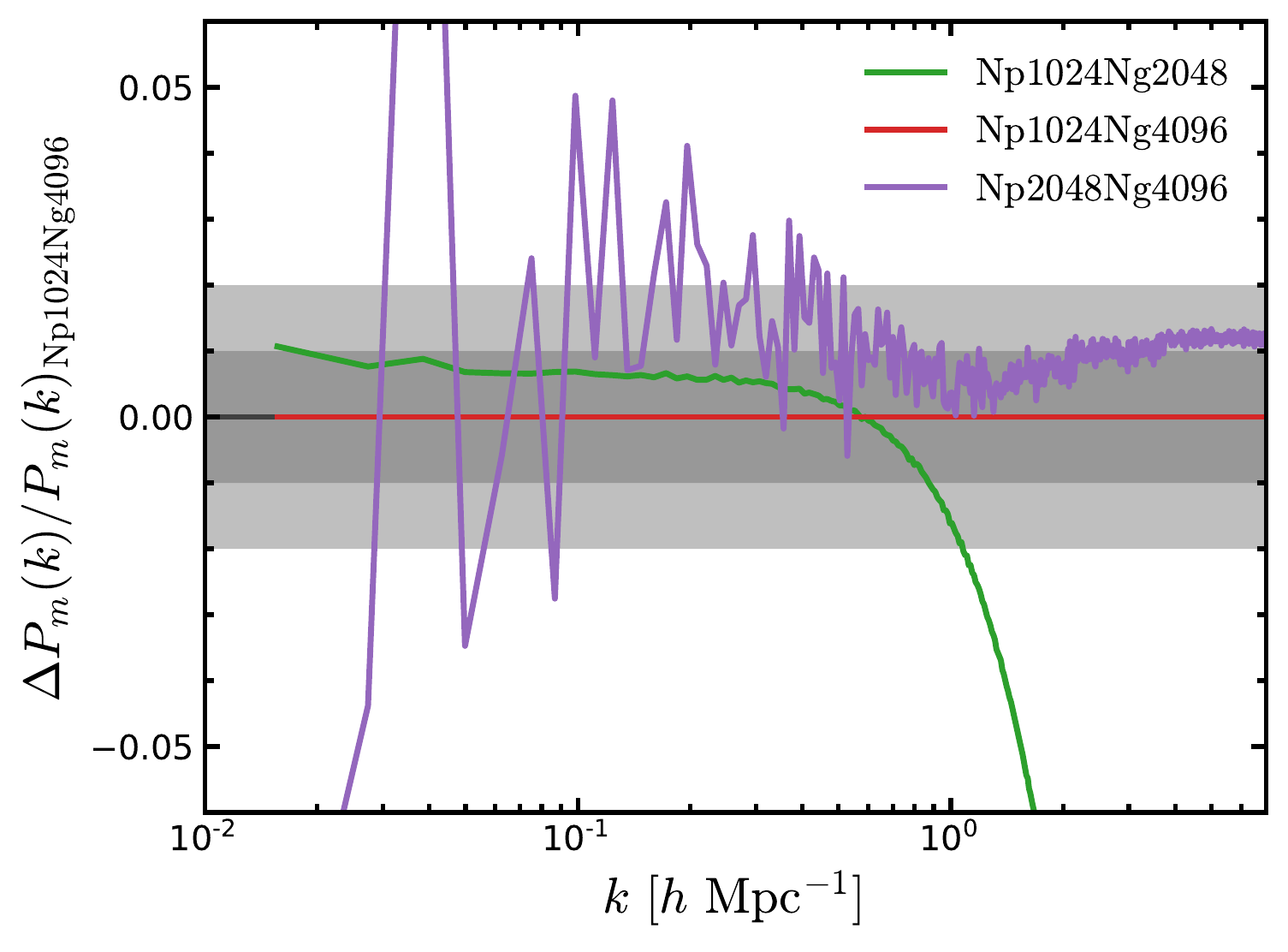}
    \caption{{\it Left panel:} Comparison of the measured nonlinear power spectra from {\sc mg-glam} simulations of the nDGP $H_0r_c=1$ model at $z=0$ with different force and mass resolutions: $(N_{\rm p}, N_{\rm g}) = (1024, 2048)$ (green line), $(1024, 4096)$ (red) and $(2048, 4096)$ (purple). The shaded region corresponds to the $1\sigma$ error bar over five independent realisations of the Np1024Ng4096 simulations. {\it Right panel:} Relative difference between the different simulations with respect to the Np1024Ng4096 case. The light and dark grey shaded regions show the two and one per cent deviations.}
    \label{fig:Pk_resolution}
\end{figure*}

\begin{figure*}
    \centering
    \includegraphics[width=0.49\textwidth]{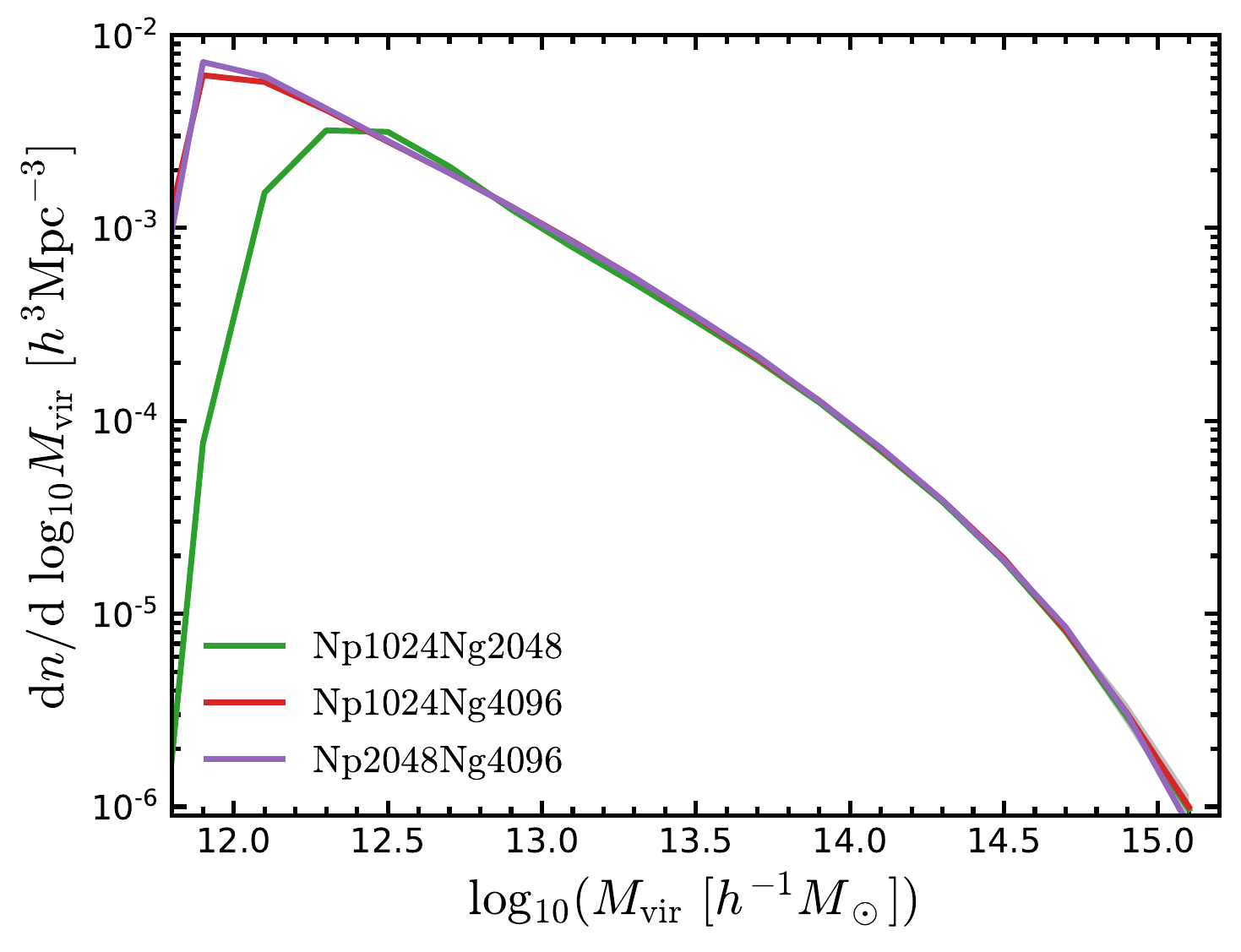}
    \includegraphics[width=0.50\textwidth]{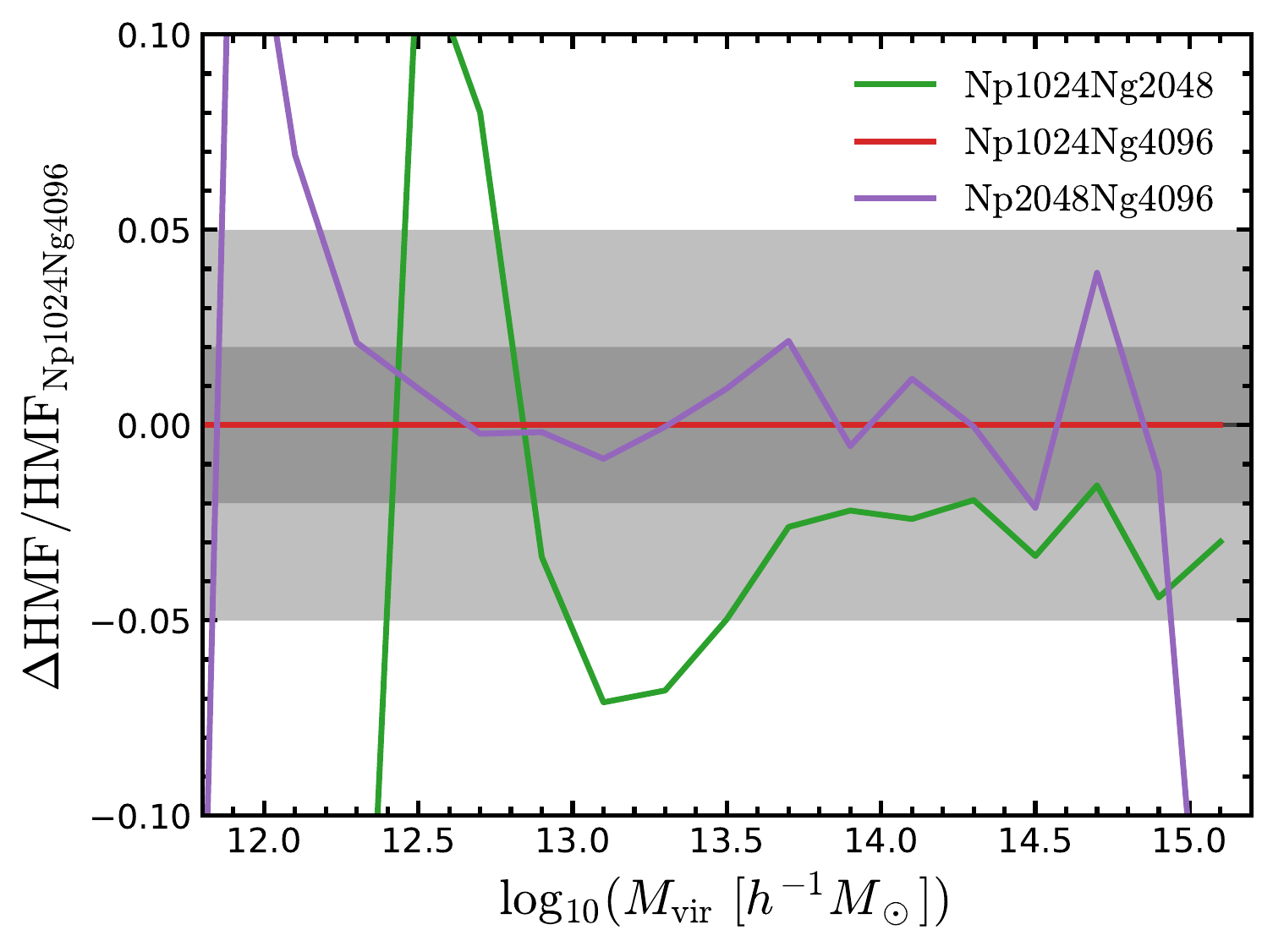}
    \caption{The same as Fig.~\ref{fig:Pk_resolution} but for the differential halo mass function. Note that the light and dark grey shaded regions in the {\it right panel} denote the five and two per cent deviations.}
    \label{fig:HMF_resolution}
\end{figure*}

We performed a series of mass and force resolution tests for the nDGP model with $H_0r_c=1$. To do so, we ran three sets of five independent simulations with fixed box size, $L=512\Mpch$, and varying grid size and number of particles: $(N_{\rm g}, N_{\rm p}) = (2048, 1024)$, $(4096, 1024)$ and $(4096, 2048)$. The setup of each simulation leads to the following mass and force resolution: $(m_{\rm p}, \Delta x) = (1.07\times 10^{10}\Msh,\,0.25\Mpch)$, $(1.07\times 10^{10}\Msh,\,0.125\Mpch)$ and $(1.34\times 10^{9}\Msh,\,0.125\Mpch)$, respectively. We have lower resolution runs than these, such as those used in the right panel of Fig.~\ref{fig:scaling}, but these are not used in this comparison.

The measured nonlinear power spectra at $z=0$ are shown in the left panel of Fig.~\ref{fig:Pk_resolution}, where we have multiplied $P_m(k)$ by the wavenumber $(k)$ to enhance any difference on large-scales. We find a good agreement on large-scales, where the measurements of the Np1024Ng2048 and Np2048Np4096 simulations are well within the error bars of the Np1024Ng4096 case. In the right panel of Fig.~\ref{fig:Pk_resolution} we confirm a one per cent agreement between all simulations on scales $k\lesssim 1\hMpc$. It also shows that for $N_{\rm g}=4096$, increasing $N_{\rm p}$ from $1024$ to $2048$ does not make a big difference.

The effects of mass and force resolution on the halo mass function (HMF) are shown in Fig.~\ref{fig:HMF_resolution}. First, we observe an improvement of the completeness of the HMF down to $M_{\rm vir} \sim 10^{12} \Msh$ for the highest force resolution simulations, i.e., those configurations with $\Delta x = 0.125\Mpch$ or $N_{\rm g} = 4096$ (see left panel of Fig.~\ref{fig:HMF_resolution}). In addition, the right panel of Fig.~\ref{fig:HMF_resolution} shows the level of agreement between the different configurations. We found that the $N_{\rm g} = 4096$ cases have a $2\%$ agreement over a large range of masses, $10^{12.3}\Msh< M_{\rm vir} < 10^{15}\Msh$, while the $N_{\rm g} = 2048$ simulations show good convergence (better than $5\%$ agreement) for haloes with mass $M_{\rm vir} > 10^{12.3} \Msh$. To have complete halo catalogues down to $10^{12.5}\Msh$, the resolution of L512Np1024Ng2048 seems to be fine, while to have haloes down to $10^{12}\Msh$ we need the resolution of L512Np2048Ng4096.

\subsection{Comparisons with previous simulations}
\label{sec:comparisons}
\begin{figure*}
    \centering
    \includegraphics[width=0.49\textwidth]{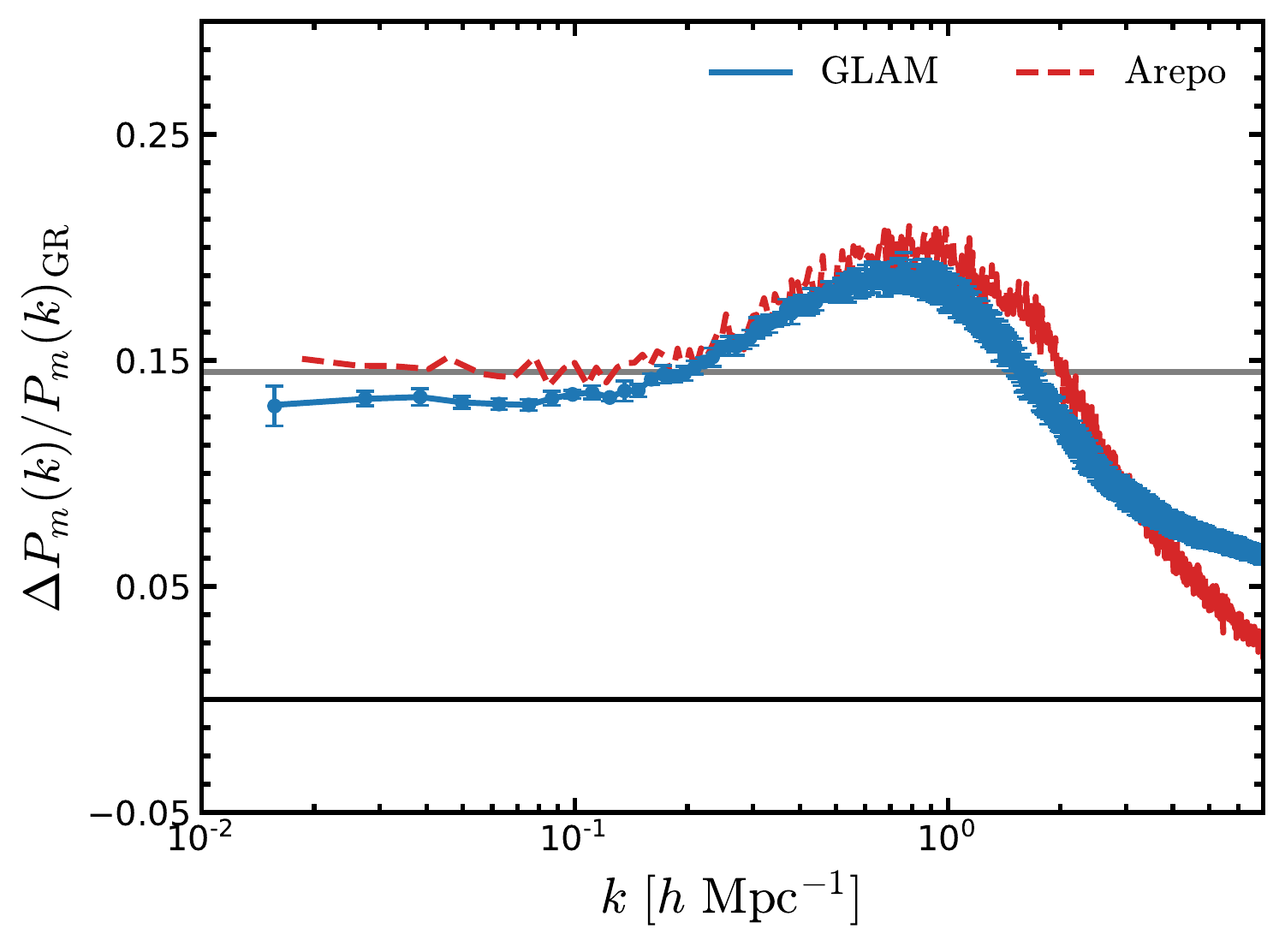}
    \includegraphics[width=0.49\textwidth]{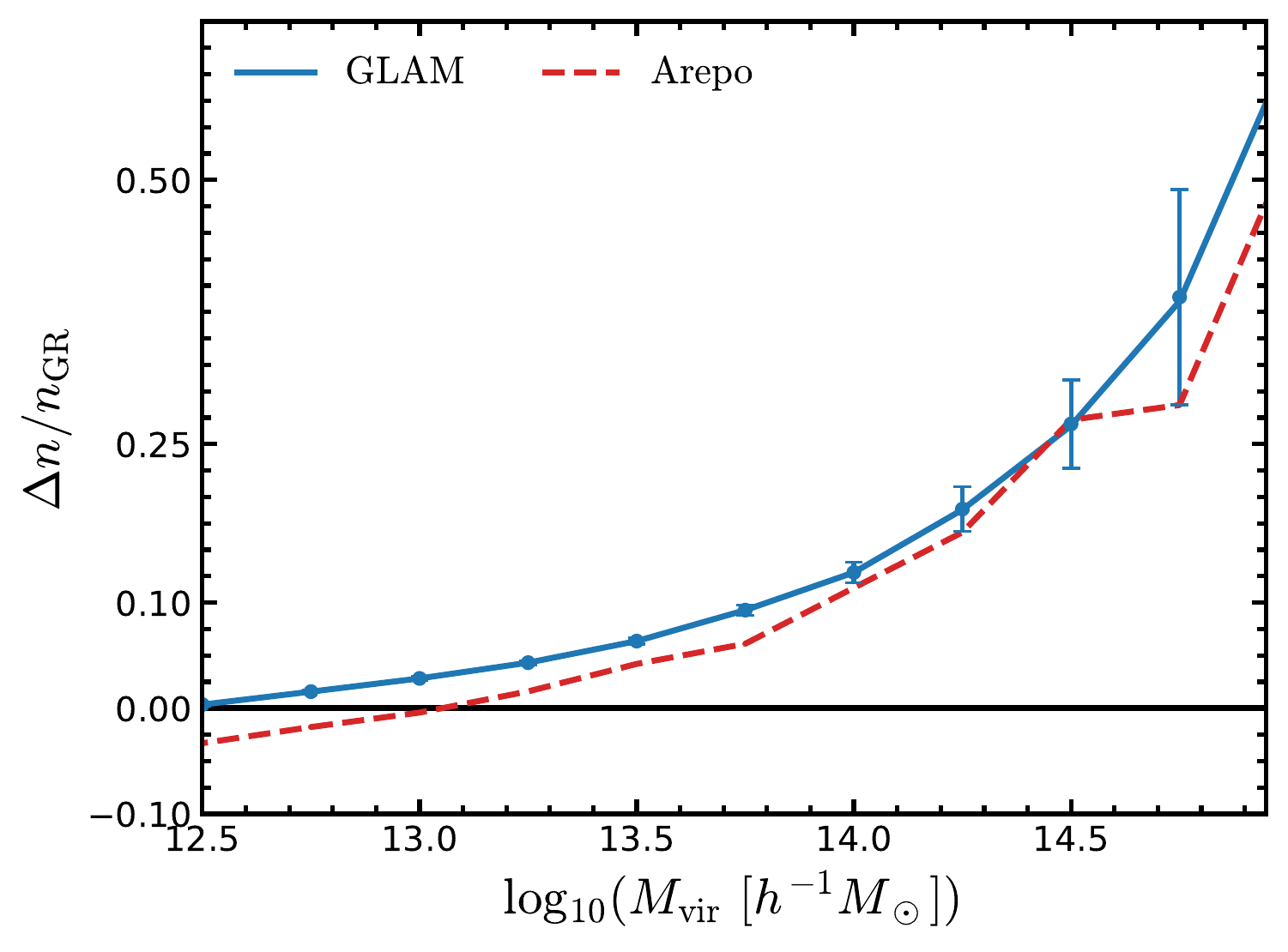}
    \caption{Comparison of the measured matter power spectrum ({\it left panel}) and halo mass function ({\it right panel}) enhancement from the simulations of the nDGP model with $H_0r_{\rm c}=1$ performed with the {\sc mg-glam} (the blue line with error bars correspond to the mean and standard deviation over 10 independent realisations) and the {\sc mg-arepo} (red dashed lines) codes.}
    \label{fig:codes_test}
\end{figure*}

Finally, we compare the dark matter power spectrum and the abundance of dark matter haloes of the nDGP ($H_0r_{\rm c}=1$) model at the present time measured from our {\sc mg-glam} simulations with those from the $L=500\Mpch$ simulations presented in \citep{Mitchell:2021aex} ran with the {\sc mg-arepo} code \citep{Hernandez-Aguayo:2020_MGAREPO_code_paper}.

The {\sc mg-arepo} simulation follows the evolution of one realisation of $1024^3$ particles in a box of size $500\Mpch$, with a force resolution $0.01\Mpch$ and mass resolution $m_{\rm p} = 9.98\times10^9\Msh$. 
We take advantage of the performance of {\sc mg-glam} to run 10 independent realisations of the same nDGP model, using the same linear theory power spectrum as for the {\sc mg-arepo} runs.
For the {\sc mg-glam} simulations we use a box of size $512\Mpch$ and a mesh with $N^3_{\rm g} = 2048^3$ grid points, giving a force resolution and particle mass of $\Delta x=0.25\Mpch$ and $m_{\rm p} = 1.06\times10^{10}\Msh$, respectively.

The left panel of Fig.~\ref{fig:codes_test} shows the comparison of the power spectrum enhancement predicted from the mean over 10 realisations of {\sc mg-glam} (solid blue line) and {\sc mg-arepo} (dashed red line). We find a good agreement between the measurements of both codes on scales $k\lesssim3\hMpc$ (with the smaller-scale discrepancy due to the lower force resolution of the {\sc mg-glam} runs), and the $P(k)$ enhancement approaches to the linear theory prediction (solid horizontal grey line) on large scales. \textsc{mg-glam} slightly under-predicts the power spectrum enhancement at large, linear scales, and this effect appears to be systematic, which is independent of the simulation box size or resolution. However, we have performed checks by running simulations of the same nDGP model using the \textsc{ecosmog} code, and found the same behaviour, which to a less extent also exists in \textsc{mg-arepo} simulations (the red dashed line here is a particular realisation). In any case, the agreement between these two codes is consistent with that between \textsc{ecosmog} and \textsc{mg-arepo}, cf.~Fig.~A1 of \cite{Hernandez-Aguayo:2020_MGAREPO_code_paper}.

The comparison of the cumulative halo mass function enhancement measured from {\sc mg-glam} (solid blue line with error bars) and {\sc mg-arepo} (dashed red line) is presented in the right panel of Fig.~\ref{fig:codes_test}. For the latter we have run the halo finder with the same virial mass overdensity halo definition as adopted for \textsc{mg-glam}, to be consistent. We again find a good, percent-level, agreement between the results of both codes, especially for high-mass haloes where the {\sc mg-arepo} measurement is well within the {\sc mg-glam} error bars (standard deviation of the 10 realisations). The \textsc{mg-arepo} prediction appears to be slightly but consistently lower than that of \textsc{mg-glam}. Indeed, while in \textsc{mg-arepo} the nDGP model enhances the abundance of large haloes and reduces it for small haloes, for \textsc{mg-glam} the abundance is always enhanced; the latter behaviour is seen in all the \textsc{ecosmog} simulations, e.g., Fig.~2 of \cite{Alam:2020jdv} of the nDGP model. This is unlikely due to the different halo finding algorithms, since \cite{Alam:2020jdv} does not use the BDM halo finder and yet finds the same behaviour. Rather, we suspect that this small discrepancy between \textsc{mg-glam} and \textsc{mg-arepo} is caused by differences in other code details, such as force calculation.

All in all, we conclude that the \textsc{mg-glam} code has passed various tests, and is ready for massive productions of simulations and mock catalogues. We will demonstrate a small-scale---in terms of the very low cost compared to \textsc{mg-arepo} and \textsc{ecosmog} simulations---application in the next section.

\section{Cosmological simulations}
\label{sec:cosmo_sims}
As a taster of the \textsc{mg-glam} code, we have conducted a large suite of dark-matter only simulations of the nDGP model and a few Kmouflage simulations, to have a quick look at the nonlinear matter power spectrum and the halo mass function in these classes of models. For the former, we have run $30$ nDGP models with $H_0r_{\rm c}$ logarithmically spaced between $0.25$ and $10$, and for the latter we have simulated 3 Kmouflage models with $(n=3,\, K_0 = 1)$, $(n=2,\,K_0 = 1)$ and $(n=2,\,K_0 = 0.5)$, all using $\beta_{\rm Kmo}=0.2$; for each Kmouflage model, we also run a `linearised' counterpart using Eq.~\eqref{eq:linearised_Kmo}, which is obtained by linearising the full field equation of motion by dropping all nonlinear terms. All the simulations have a box size of $L=512\Mpch$, $N^3_{\rm g}=2048^3$ grid cells and contain $1024^3$ dark matter particles, giving a mass resolution of $m_{\rm p} = 1.06\times10^{10}\Msh$. 

For all simulations, we use the same $\Lambda$CDM linear perturbation theory power spectrum to generate the initial conditions at $z_{\rm ini}=100$ using the on-the-fly algorithm of {\sc mg-glam}. The cosmological parameters are chosen from those reported by the Planck collaboration \citep{Ade:2015xua}: 
$$\{\Omega_{\rm b}, \Omega_{\rm m}, h, n_s, \sigma_8\} = \{0.0486,0.3089,0.6774,0.9667,0.8159\}.$$ The linear matter power spectrum is generated using the \href{https://camb.info/}{\textsc{camb}} code. The reason we can use the same initial condition for all simulations is that the effect of the scalar field is very weak at $z>100$; we have checked that even the strongest Kmouflage model studied in this work only differs from $\Lambda$CDM by $\mathcal{O}\left(0.1\%\right)$ in the linear matter power spectrum at $z=100$.

\subsection{Matter power spectrum}
\label{sec:Pkm}
\begin{figure*}
    \centering
    \includegraphics[width=0.49\textwidth]{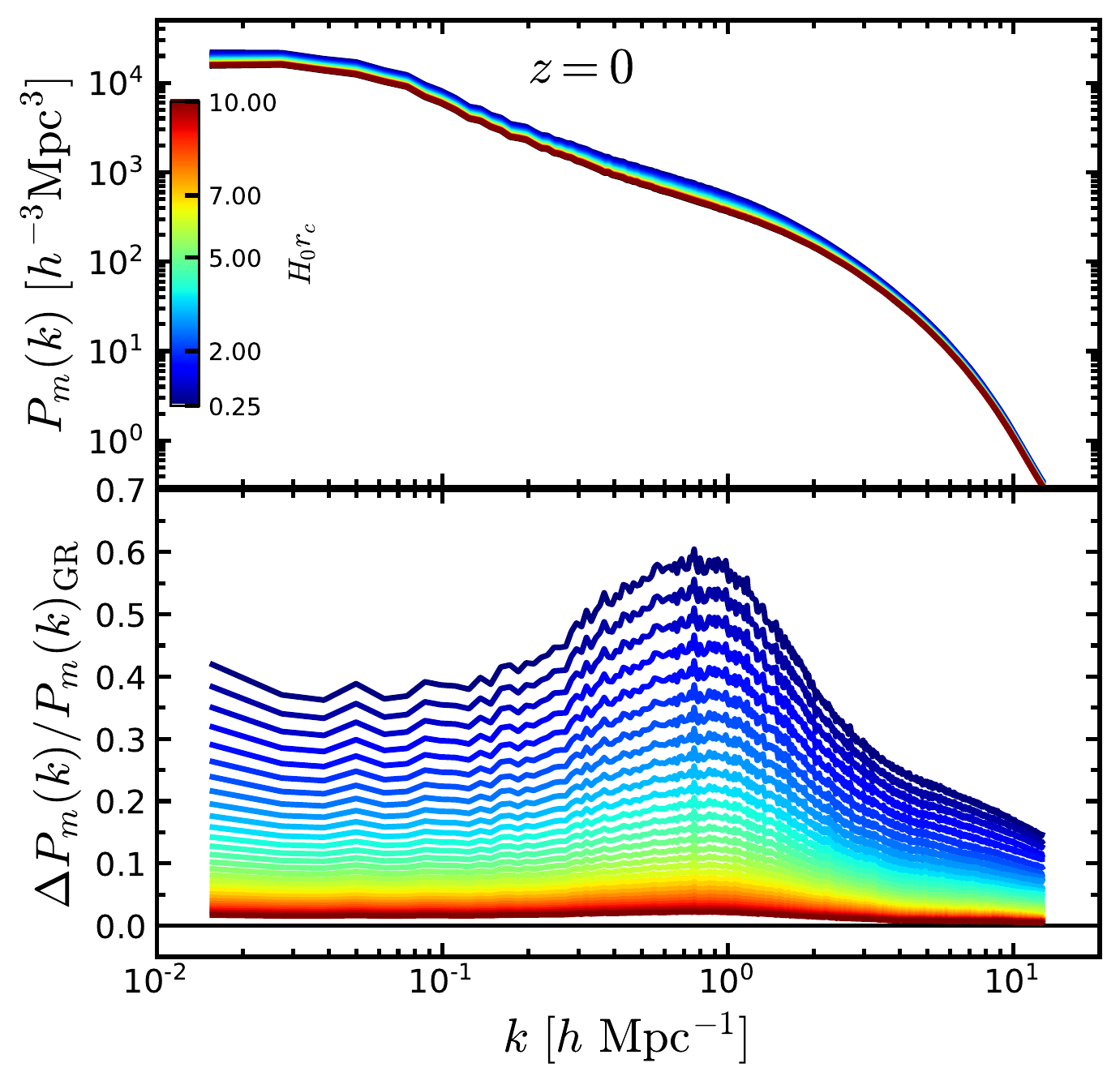}
    \includegraphics[width=0.49\textwidth]{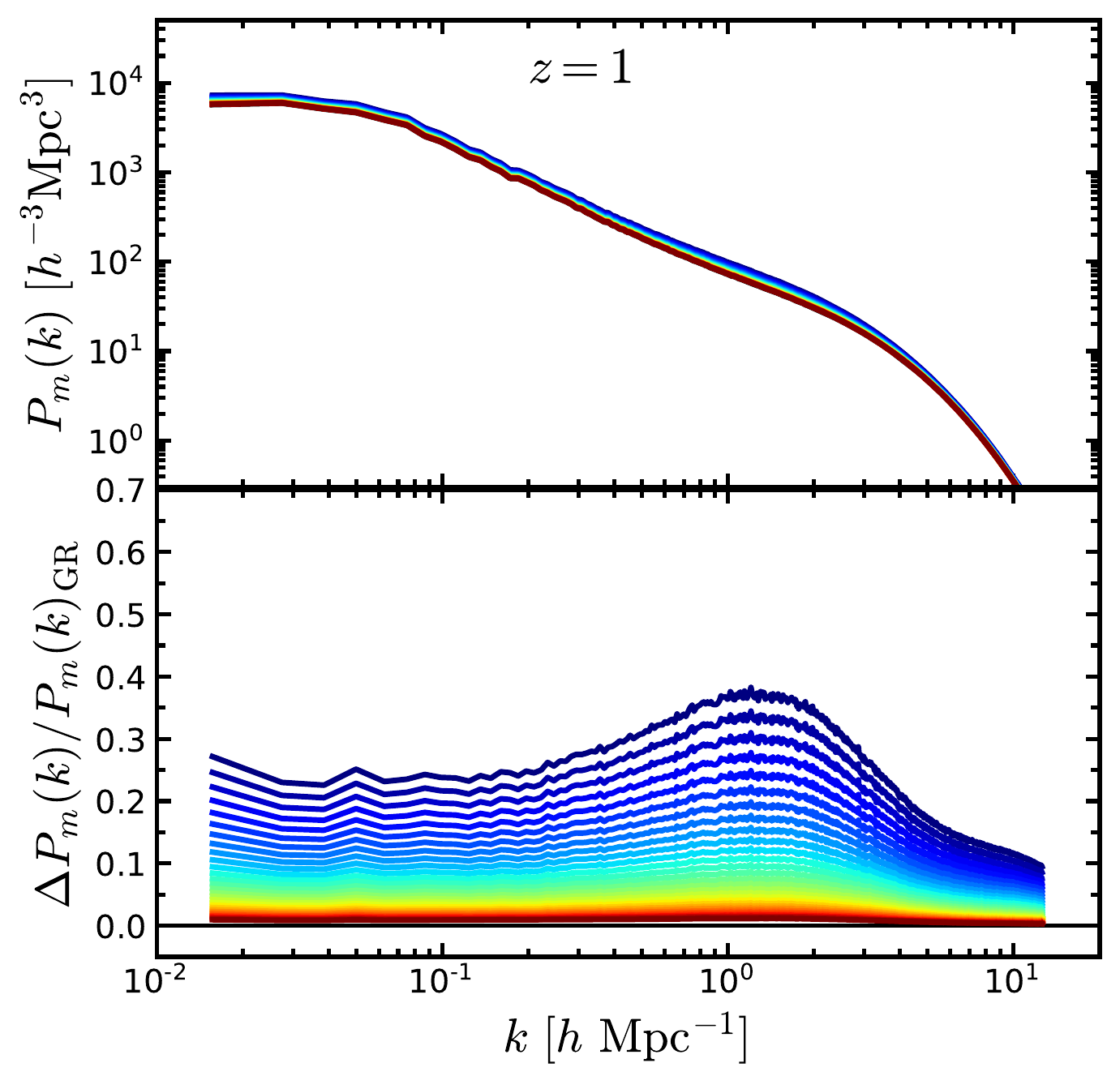}
    \caption{Measured non-linear matter power spectrum from {\sc mg-glam} simulations for 30 nDGP models with $H_0r_{\rm c}$ logarithmically spaced between $0.25$ and $10$ (indicated by the colour bar in the upper left panel) at $z=0$ ({\it left panel}) and $z=1$ ({\it right panel}). The {\it lower subpanels} show the relative differences with respect to a $\Lambda$CDM model with the same cosmological parameters and simulation specifications. We have used $L=512h^{-1}$Mpc, $N_{\rm p}=1024$ and $N_{\rm g}=2048$.}
    \label{fig:Pk_nDGP}
\end{figure*}

\begin{figure*}
    \centering
    \includegraphics[width=0.49\textwidth]{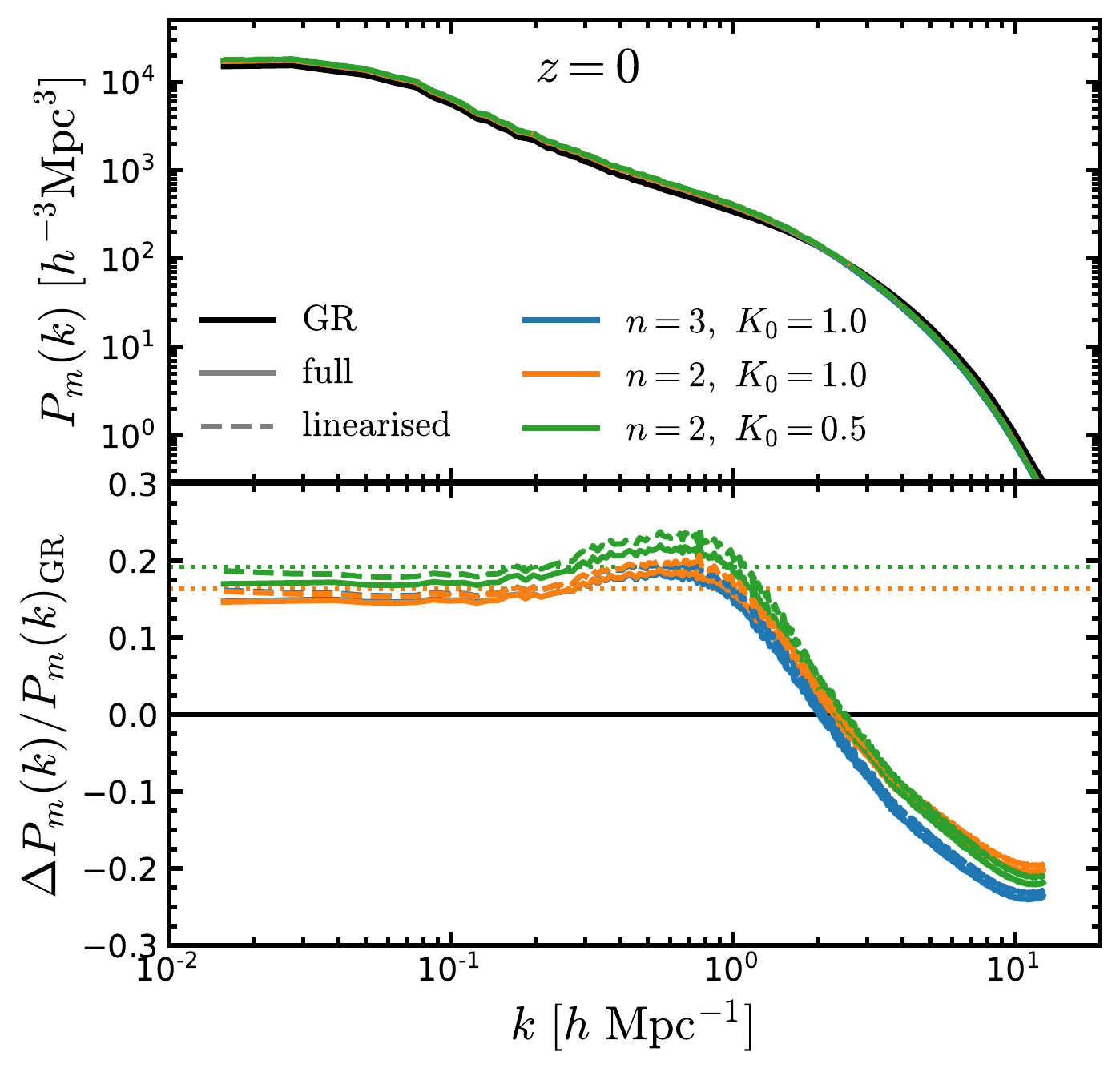}
    \includegraphics[width=0.49\textwidth]{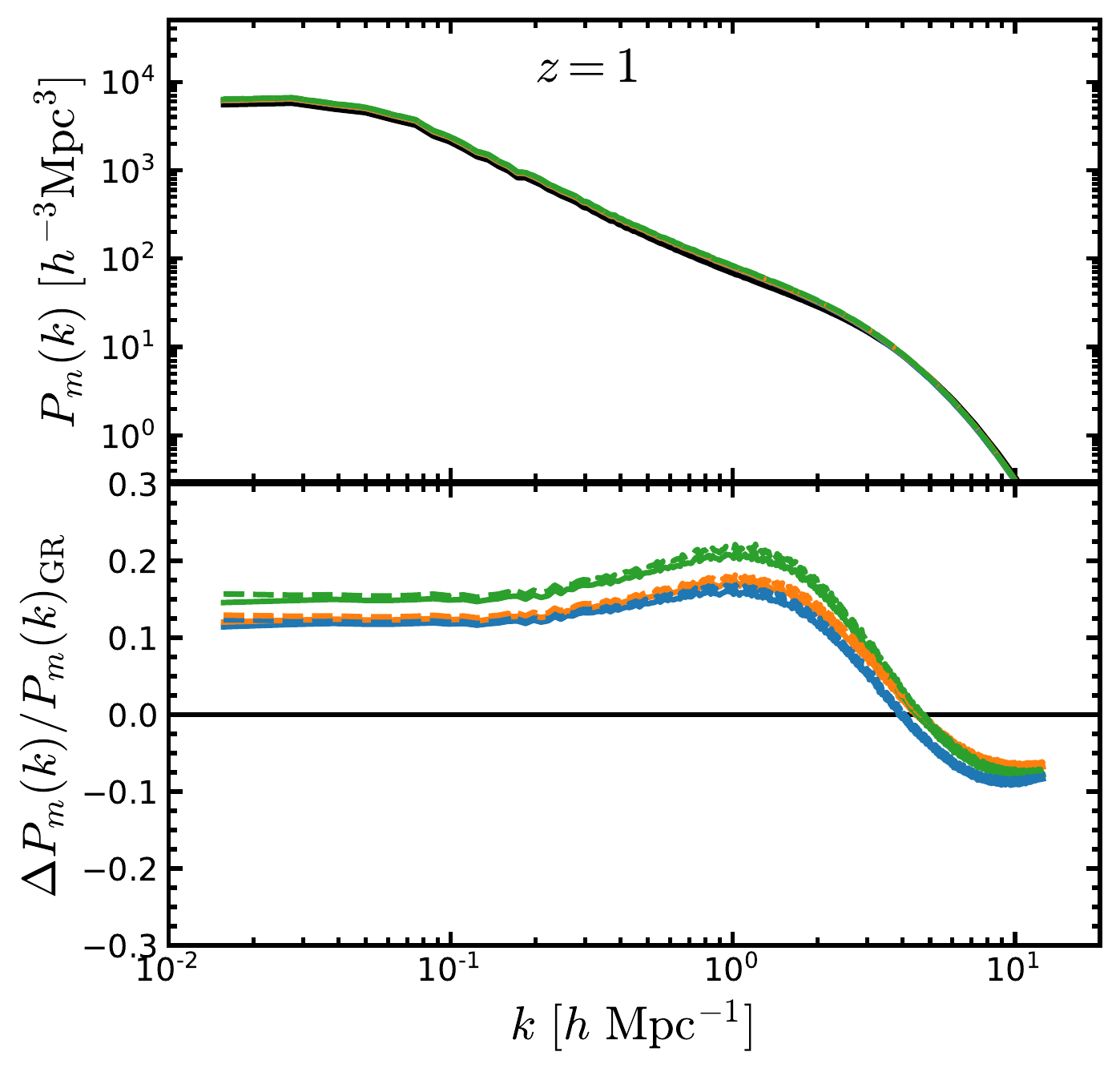}
    \caption{Measured nonlinear matter power spectra from {\sc mg-glam} simulations for three linearised (dashed lines) and fully nonlinear (solid lines) Kmouflage simulations, with $n=3$, $K_0 = 1$ (blue lines), $n=2$, $K_0 = 1$ (orange lines) and $n=2$, $K_0 = 0.5$ (green lines), at $z=0$ ({\it left panel}) and $z=1$ ({\it right panel}). The {\it lower subpanels} show the relative difference with respect to a $\Lambda$CDM model with the same cosmological parameters and simulation specifications. All simulations use $L=512h^{-1}$Mpc, $N_{\rm p}=1024$ and $N_{\rm g}=2048$.}
    \label{fig:Pk_Kmo}
\end{figure*}

The measured power spectra for all 30 nDGP models are displayed in Fig.~\ref{fig:Pk_nDGP} at $z=0$ (left panel) and $z=1$ (right panel). The colorbar displays the values of $H_0r_{\rm c}$ from the strongest ($H_0r_{\rm c} = 0.25$; bluest solid line) to the weakest models ($H_0r_{\rm c} =10$, reddest solid line). From the lower subpanels, we see that we can cover a wide range of enhancement amplitudes of the power spectrum, with the relative differences between the nDGP and GR models spanning from $\approx1\%$ to $40\%$ on large scales at $z=0$. At earlier times ($z=1$; right panel), the behaviour is qualitatively similar, but the enhancement is generally smaller ($\approx0.5\%$--$25\%$ on large scales) as the fifth force has had less time to take effect. 

The effect of the Vainshtein screening mechanism is reflected by the decay of the power spectrum enhancement towards $0$ at small scales (large $k$). However, notice that at this resolution, we can only trust the result at $k\lesssim3h$/Mpc, as shown by the comparison between \textsc{mg-glam} and \textsc{mg-arepo} in \S~\ref{sec:comparisons}. Should the simulations be run at a higher resolution, we expect the decay to $0$ to happen faster at $k>3h$/Mpc. This decay is because, according to the halo model \cite{Cooray:2002dia} of structure formation, the small-scale matter power spectrum is determined by the one-halo term, which in turn depends on the inner density profiles of dark matter haloes; the Vainshtein screening mechanism can effectively suppress the relative strength of the fifth force, cf.~Eq.~\eqref{eq:V_screening}, inside and near massive bodies such as haloes \cite{Falck:2015rsa}, so that in Vainshtein-type models the halo density profile is close to $\Lambda$CDM \cite{Barreira:2014zza,Becker:2020yim,Mitchell:2021aex}.  

In Fig.~\ref{fig:Pk_Kmo}, we show the nonlinear matter power spectra from our three pairs of linearised (dashed lines) and fully non-linear (solid lines) Kmouflage simulations with $n=3,\, K_0 = 1$ (blue lines), $n=2,\,K_0 = 1$ (orange), and $n=2,\,K_0 = 0.5$ (green) at $z=0$ (left panel) and $z=1$ (right). To perform the linearised simulations we solved the linearised Kmouflage equation of motion, Eq.~\eqref{eq:linearised_Kmo}, equivalent to removing the screening effect. 

The lower subpanels of Fig.~\ref{fig:Pk_Kmo} display the relative difference between the measured power spectra of the Kmouflage models and GR. In addition to the results of the full and linearised simulations, we also show in dotted lines the linear-theory predictions at $z=0$ (left panel), obtained using the modified version of the \href{https://camb.info/}{\sc camb} code developed in \citep{Barreira:2014gwa}. In general, we find that the linearised simulations give similar results to those of their full nonlinear counterparts; also, all measurements approach to the linear theory predictions on large scales. This shows that the Kmouflage screening mechanism is not efficient \cite{Brax:2015lra} in suppressing the effect of the fifth force in cosmic structure formation. This is related to the way in which screening works in this class of models, which requires $|\boldsymbol{\nabla}\varphi|\gg|\dot{\bar{\varphi}}|\sim H_0$, a condition that is likely to be satisfied only on small (e.g., sub-galactic) scales. A corollary from this is that, in cosmological simulations, solving the fully nonlinear Kmouflage equation of motion may not be as important as for the other models such as nDGP and $f(R)$ gravity \cite{Ruan:2021_twin_paper}.

Since this is the first time that cosmological simulations for the Kmouflage model are conducted, let us comment on the qualitative behaviour shown in the lower subpanels of Fig.~\ref{fig:Pk_Kmo}. Overall, the power spectrum enhancement in this model looks very similar to that in the nDGP model, cf.~Fig.~\ref{fig:Pk_nDGP}, but there is a critical difference: here the enhancement becomes negative at small scales, $k\gtrsim2h$/Mpc. We have already seen that this can not be due to the Kmouflage screening mechanism --- actually, it is due to the \textit{lack of} screening. Unlike in nDGP, here even inside dark matter haloes particles still feel a strong fifth force which has a nearly constant ratio with the strength of Newtonian gravity, and on top of this the direction-dependent force discussed below Eq.~\eqref{eq:kmo_lin_growth} can also speed up the particles; the result of these two forces is that particles gain a higher kinetic energy, tend to move into or stay in the outer regions of dark matter haloes and thus reduce the clustering on small scales as compared to $\Lambda$CDM. Such distinct behaviours between the nDGP and Kmouflage matter power spectra may offer a potential way to distinguish between them observationally, although that is beyond the scope of this paper.

\subsection{Halo mass functions}
\label{sec:haloes}
\begin{figure*}
    \centering
    \includegraphics[width=0.49\textwidth]{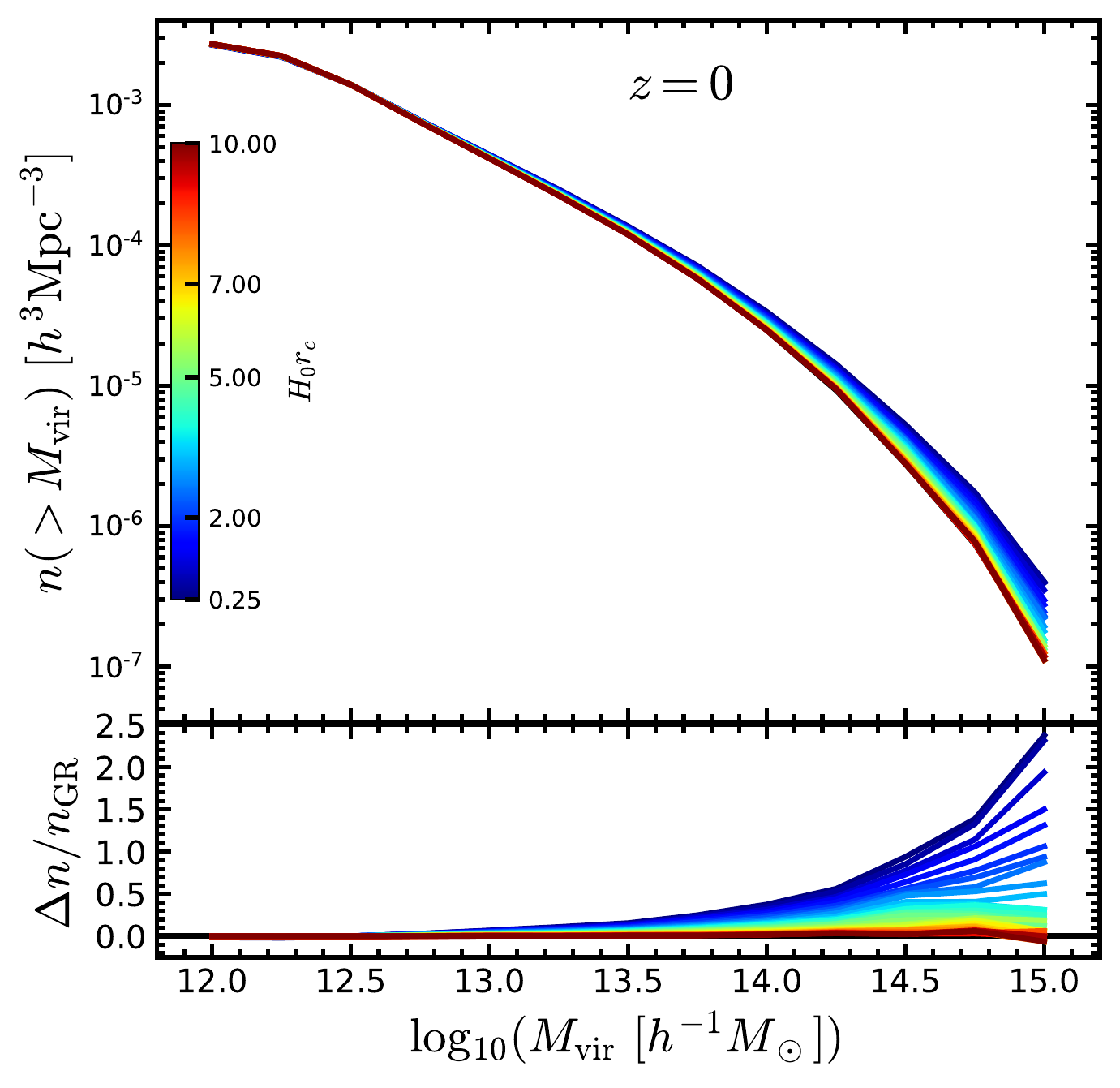}
    \includegraphics[width=0.49\textwidth]{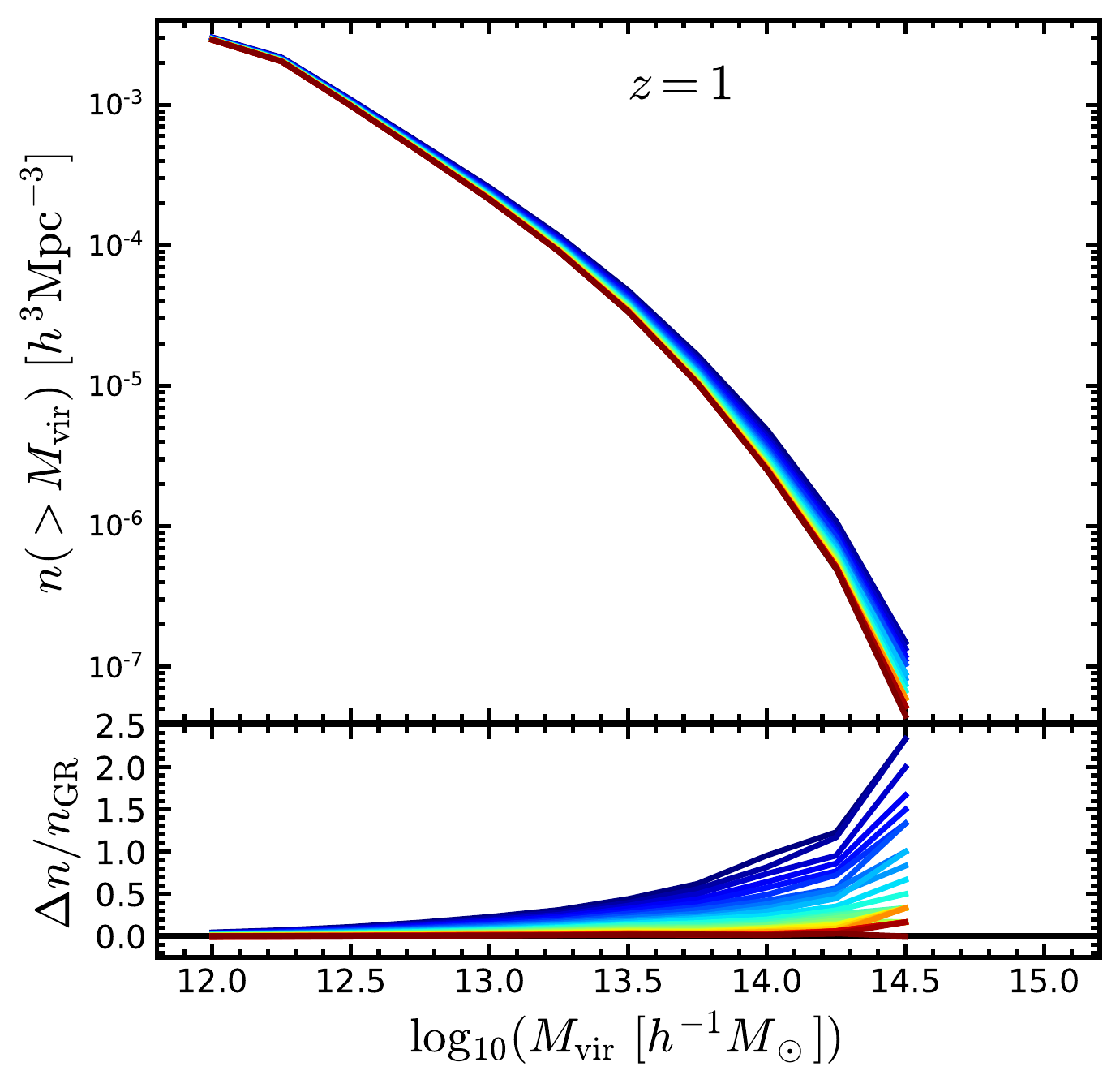}
    \caption{Cumulative halo mass functions from {\sc mg-glam} simulations for 30 nDGP models with $H_0r_{\rm c}$ logarithmically spaced between $0.25$ and $10$ (indicated by the colour bar in the top left panel) at $z=0$ ({\it left panel}) and $z=1$ ({\it right panel}). The {\it lower subpanels} show the relative difference with respect to a $\Lambda$CDM model with the same cosmological parameters and simulation specifications. All these simulations use $L=512h^{-1}$Mpc, $N_{\rm p}=1024$ and $N_{\rm g}=2048$.}
    \label{fig:cHMF_nDGP}
\end{figure*}

\begin{figure*}
    \centering
    \includegraphics[width=0.497\textwidth]{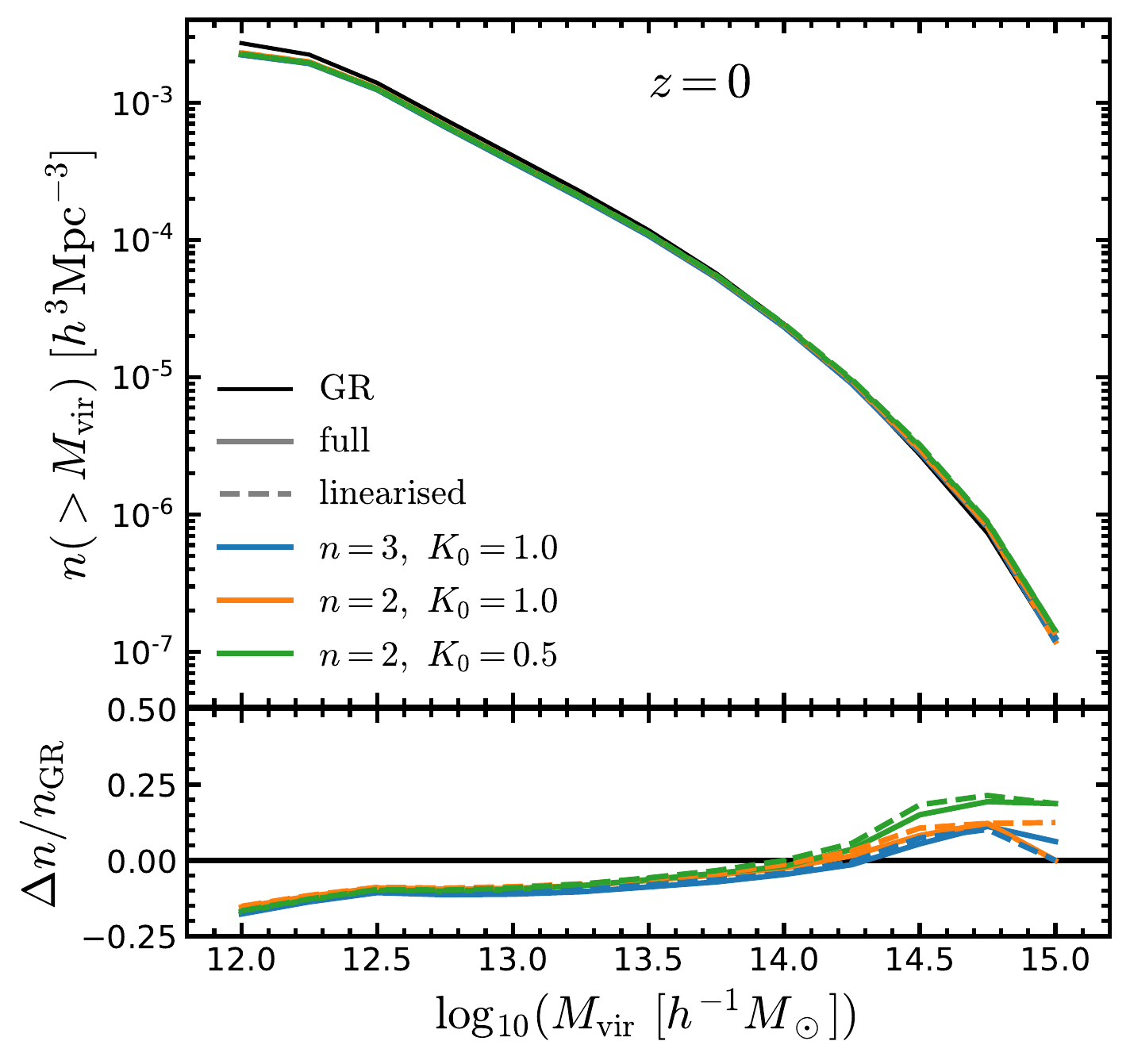}
    \includegraphics[width=0.497\textwidth]{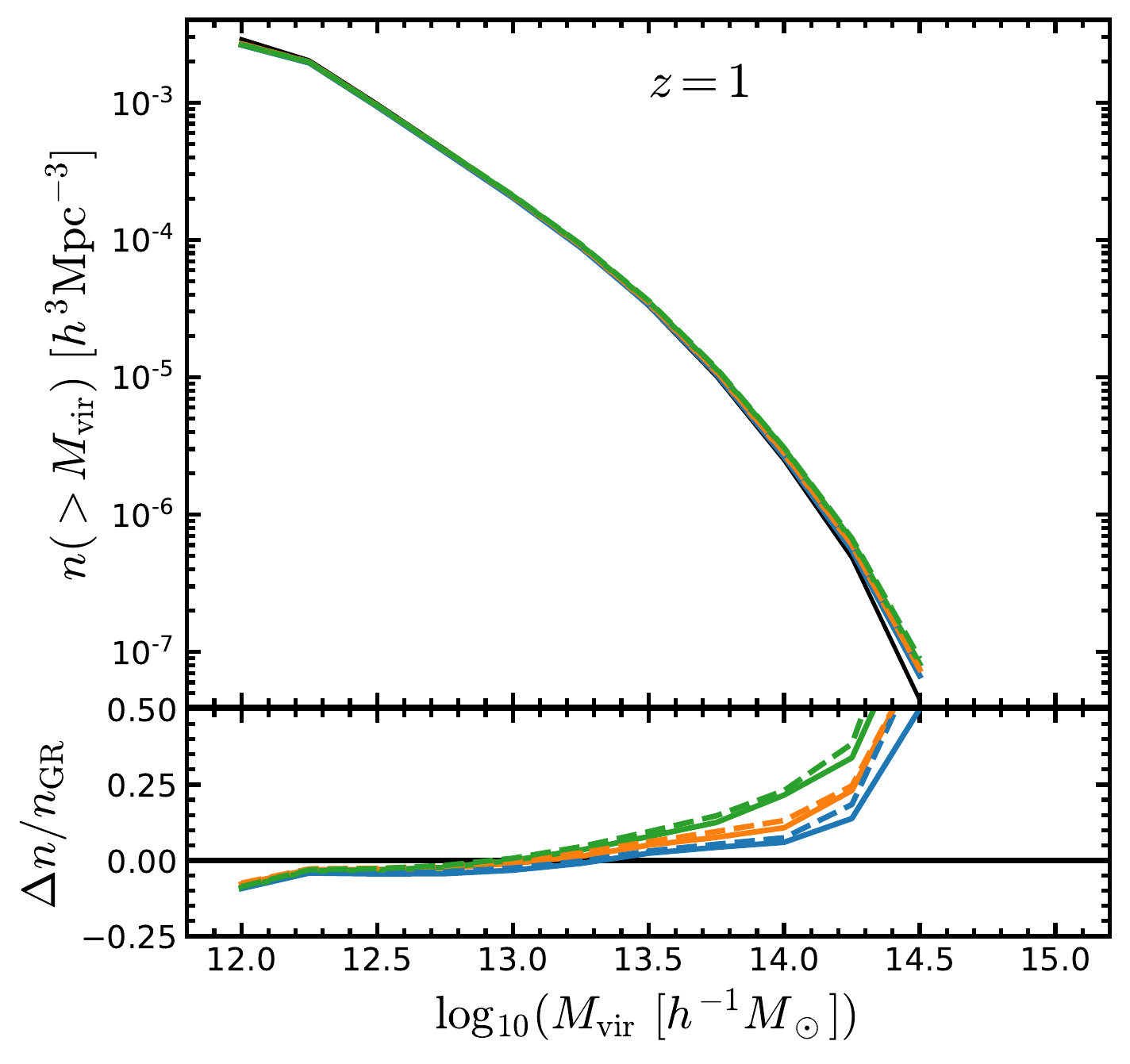}
    \caption{Cumulative halo mass functions from {\sc mg-glam} simulations for 3 linearised (dashed lines) and fully nonlinear (solid lines) Kmouflage simulations with $n=3$, $K_0 = 1$ (blue lines), $n=2$, $K_0 = 1$ (orange lines) and $n=2$, $K_0 = 0.5$ (green lines) at $z=0$ ({\it left panel}) and $z=1$ ({\it right panel}). The {\it lower subpanels} show the relative difference with respect to a $\Lambda$CDM model with the same cosmological parameters and simulation specifications. All simulations use $L=512h^{-1}$Mpc, $N_{\rm p}=1024$ and $N_{\rm g}=2048$.}
    \label{fig:cHMF_Kmo}
\end{figure*}

Modified gravity and screening mechanism effects can also be studied by exploring dark matter halo populations. In Figs.~\ref{fig:cHMF_nDGP} and \ref{fig:cHMF_Kmo} we show the cumulative halo mass function (cHMF), which defines the number density of dark matter haloes more massive that a given halo mass $M_{\rm vir}$, measured from our BDM halo catalogues at $z=0$ (left panels) and $z=1$ (right panels). For nDGP the $30$ models are colour-coded in the same away as in Fig.~\ref{fig:Pk_nDGP}.

From the lower subpanels of Fig.~\ref{fig:cHMF_nDGP}, we see that the abundance of haloes is enhanced by the fifth force, especially at low redshifts and for high-mass haloes. The same behaviour has been found and discussed in previous works, e.g., \cite{Alam:2020jdv,Mitchell:2021aex,Winther:2015wla}. We also notice that the enhancement over $\Lambda$CDM is positive for the whole halo mass range, not just for massive haloes, as already discussed in \S~\ref{sec:comparisons}. 
The abundance of haloes is enhanced from $\approx1$ to $250$ percent for the different nDGP models. The large increase of high-mass haloes in the less efficiently screened nDGP models (models with $H_0r_{\rm c} < 5$) is due to the accretion of surrounding matter around these massive objects thanks to the enhanced gravity force: these objects, often being the dominating object within some large surrounding region, can attract matter from the whole region, including the accretion of smaller haloes to them, and so the fifth force can strongly boost their masses; on the other hand, smaller objects, while also experiencing the fifth force \cite{Hernandez-Aguayo:2020_MGAREPO_code_paper}, are more likely to meet competitors and so their masses grow less.

On the other hand, the lower subpanels of Fig.~\ref{fig:cHMF_Kmo} show the relative difference of the cHMFs between the Kmouflage models and $\Lambda$CDM. In the same figure we compare the predictions from the linearised Kmouflage simulations (dashed lines) with their fully nonlinear counterparts (solid lines). Each pair of Kmouflage simulations produce roughly the same abundances of dark matter haloes, as evident from the overlap between dashed and solid lines in the entire mass range used to measure the cHMFs, confirming that the effects of Kmoulfage screening are marginal. The abundance of massive haloes is enhanced by $\approx 50$ percent at $z=1$ and $\approx 20$ percent at $z=0$, consistent with the redshift evolution of the matter power spectrum shown in the lower panels of Fig.~\ref{fig:Pk_Kmo}. 

Also, we find that the Kmouflage model produces fewer low-mass haloes than GR, especially at lower redshifts, and we believe this is the consequence of the competition between the four effects of the Kmouflage model, discussed below Eq.~\eqref{eq:K_Xbar}. As we have demonstrated in \S~\ref{sec:bkg_tests} for a few cases of fixed $n$ and $\beta_{\rm Kmo}$, this competition can be complicated and not analytically predictable. As a result, to disentangle the four effects and to rank their relative importance, we need to switch them on and off individually to observe the impact on cosmological observables. While this is apparently an interesting and important thing to do, it is beyond the scope of this paper and so we will leave such a study to future works.

Finally, before concluding this section, it is worthwhile to mention that, at the simulation resolution used here, we can already get the HMF complete down to $10^{12.5}h^{-1}M_\odot$, as shown in \cite{Ruan:2021_twin_paper,Hernandez-Aguayo:2020oiw}.

\subsection{Discussion}
\label{subsect:cosmo_runs_summary}
In this section we have had an initial taste of the \textsc{mg-glam} code, by running a large suite of simulations covering all three classes of models studied in this paper.

One particularly relevant aspect of the \textsc{mg-glam} code is its fast speed (cf.~\S~\ref{sec:scaling}). The 30 nDGP simulations described in this section have been run using 56 threads with \textsc{openmp} parallelisation, and we find that the run time for the majority of them is $\sim23,000$ seconds, or equivalently $\simeq357$ CPU hours, roughly $105$ times faster than \textsc{mg-arepo}, and $300$ times faster than \textsc{ecosmog}, for the same simulation specifications. With such a high efficiency, we can easily ramp up the simulation programme to include many more models and parameter choices, and increase the size and/or resolution of the runs, e.g., using boxes of at least $1h^{-1}\mathrm{Gpc}$. The Kmouflage simulations, while having a different screening mechanism, take about $22,000$ seconds each, similar to the nDGP runs. This is not unexpected given that in both models we use the same number of V-cycles and 157 time-steps. As part of the resolution tests in \S~\ref{sec:res}, we have also run a few even larger simulations for $\Lambda$CDM and N1, e.g., with $L = 512 \, h^{-1}\mathrm{Mpc}$, $N_{\rm p}=2048$ and $N_{\rm g}=4096$. These runs took around $40,000$ seconds for $\Lambda$CDM and $116,000$ seconds (wallclock time) for N1, using $128$ threads on the SKUN8@IAA supercomputer at the IAA-CSIC in Spain, suggesting that a single run of specification L1000Np2048Ng4096, which would be useful for cosmological (e.g., galaxy clustering and galaxy clusters) analyses should take at most 1.3 days to complete and is therefore easily affordable with existing computing resources.

On the other hand, efficiency should not be achieved at the cost of a significant loss of accuracy. For the runs used here, we have used a mesh resolution of $0.25h^{-1}\mathrm{Mpc}$, which is sufficient to achieve percent-level accuracy of the matter power spectrum at $k\lesssim1h\mathrm{Mpc}^{-1}$ \cite{Klypin:2017iwu}, matter power spectrum enhancement at $k\lesssim3h\mathrm{Mpc}^{-1}$, and (main) halo mass function down to $\sim10^{12.5}h^{-1}M_\odot$ \cite{Ruan:2021_twin_paper}. The particle number, $N_{\rm p}^3$, in \textsc{glam} simulations is normally set according to $N_{\rm p}=N_{\rm g}/2$, so that in the simulations here we have used $1024^3$ particles. However, we have checked that increasing the particle number to $2048^3$ has little impact on the halo mass function (cf. \S~\ref{sec:res}). We notice that the completeness level of the HMFs here is similar to \textsc{ecosmog} runs with the same simulation specifications, suggesting that \textsc{mg-glam} is capable of striking an optimal balance between cost and accuracy. 

\section{Summary and conclusions}
\label{sec:conclusions}
In this paper, along with a companion paper \citep{Ruan:2021_twin_paper}, we have presented the {\sc mg-glam} code, which is an extension of the {\sc glam} pipeline \citep{Klypin:2017iwu} that enables very efficient and accurate production of \textit{full} $N$-body simulations in a large variety of modified gravity models, with the ultimate objective of covering all such models of interest. We have focused on the description and numerical implementation of models with derivative coupling terms, while our twin paper \citep{Ruan:2021_twin_paper} explores the conformally coupled scalar field models, including thin-shell screening models such as $f(R)$ gravity and symmetrons, as well as the usual coupled scalar field models.

We studied two classes of derivative coupling models, the Vainshtein-type and the Kmouflage-type gravity models, which employ the {\it Vainshtein} and {\it Kmouflage} screening mechanism, respectively. As an example of Vainshtein-type models, we considered the nDGP braneworld model, which serves as a prototype for other classes of models such as Galileons, vector Galileons, generalised Galileons and kinetic-gravity braiding models. The Kmouflage models are comparatively new in the context of cosmological simulations, and we have proposed a new numerical algorithm to solve their equations of motion in this work. This algorithm, and its implementation in \textsc{mg-glam}, can be easily generalise to simulate other classes of interesting models such as k-essence, MOND, and the scalar \cite{Brax:2019fzb} or vector \cite{Jimenez:2020bgw} dark matter models with non-canonical kinetic terms of the k-essence type and possibly a generic interaction potential.

To implement these models into the parent code {\sc glam}, we have added subroutines to solve the nonlinear partial differential equations that govern the formation of cosmological structures in such models (cf.~\S~\ref{subsect:extradof}). These nonlinear PDEs are solved using the multigrid Gauss-Seidel relaxation technique, which uses one of three different arrangements of the multigrid solver (V-cycles, F-cycles and W-cycles). In addition, we have included some background cosmology solvers for the Kmouflage model (cf.~\S~\ref{sec:BG_solvers}). For both classes of models, we have designed the relaxation algorithm to avoid the Newton-Gauss-Seidel iteration commonly used for nonlinear PDEs, which generally slows down the convergence and is sometimes unstable. This is a key to the performance of \textsc{mg-glam}, which we find to be $100$--$300$ times faster than earlier modified gravity codes such as \textsc{mg-arepo} and \textsc{ecosmog} for the same mass resolution; the force resolution is lower as \textsc{mg-glam} uses a fixed mesh resolution, while the other codes use adaptive mesh refinements; but even with the resolution used in this work, \textsc{mg-glam} is able to accurately predict the halo mass function down to $\approx10^{12.5}h^{-1}M_\odot$ (comparable to the performance of \textsc{ecosmog}) and the power spectrum enhancement down to $k\approx3\hMpc$. 

We have performed a series of tests to check that our implementation of the multigrid solvers works correctly, using different density configurations for which we can obtain analytical expressions of the scalar field solution (cf.~\S~\ref{sec:code_tests}), and found that the {\sc mg-glam} numerical solutions agree very well with the analytical expectations. We have shown that using only 2 V-cycles, we can reach convergence for the nonlinear equations in the nDGP and Kmouflage models.  Also, we have compared the solutions of the background scalar field and the modified expansion rate in the Kmouflage model obtained with {\sc mg-glam} and {\sc camb} \citep{Barreira:2014gwa}, finding excellent agreement between both codes. Finally, we have compared the power spectrum enhancement and the abundance of dark matter haloes for one nDGP model ($H_0r_{\rm c}=1.0$) predicted by {\sc mg-glam} and the {\sc mg-arepo} code \citep{Hernandez-Aguayo:2020_MGAREPO_code_paper}. To do so, we ran 10 independent {\sc mg-glam} realisations (to reduce cosmic variance) and use the L500-N1 simulation presented in \citep{Mitchell:2021aex}. In general, {\sc mg-glam} is able to reproduce the power spectrum enhancement and the abundance of dark matter haloes from those high-resolution simulations with high accuracy.

For the first time, we have been able to run a large suite of nDGP simulations, for 30 models with $H_0r_{\rm c}$ logarithmically spaced between $0.25$ and $10$, and carried out the first fully nonlinear $N$-body simulations for three Kmouflage models with $\beta_{\rm Kmo} = 0.2$ and $(n=3,\,K_0=1)$, $(n=2,\,K_0=1)$ and $(n=2,\,K_0=0.5)$. In addition, we have run linearised simulations for each of the Kmouflage models mentioned above. With this large suite of MG simulations we are able to study in great detail the interplay between modified gravity effects and screening mechanism on structure formation, as we have shown in the nonlinear matter power spectra and cumulative halo mass function predictions, Figs.~\ref{fig:Pk_nDGP}--\ref{fig:cHMF_Kmo}. Our nDGP simulations clearly demonstrate the effect of Vainshtein screening in the matter power spectrum, and how that evolves with time and depends on $H_0r_{\rm c}$. The Kmofulage simulations, on the other hand, indicates that the Kmouflage screening mechanism is is much less efficient in the cosmological regime, as the fully nonlinear and linearised simulations give similar predictions of the matter power spectrum and halo mass function; this agrees with expectations. 

The development of {\sc mg-glam} will help in the construction of a large number of galaxy mock catalogues in MG theories for Stage-IV galaxy surveys, such as DESI and Euclid. Owing to its high efficiency and accuracy, this code can be used to perform $>\mathcal{O}(100)$ large ($L>1.0h^{-1}$Gpc at least) and high-resolution ($m_{\rm p}<10^{10}h^{-1}M_\odot$) simulations for each modified gravity model, with minimal computational cost. These will allow for variations of not only the gravitational but also cosmological parameters, and subsequently the construction of accurate emulators for various physical quantities in different gravity models. This will open up a wide range of possibilities for future works to test gravity using cosmological observations. The prescriptions to populate dark matter haloes with galaxies will be explored in an upcoming paper, as well as a more detailed study of halo properties, including halo clustering, will be left in future works.


\acknowledgments
We thank Phil Brax, Jose Beltran Jimenez and Dario Bettoni for helpful discussions on the Kmouflage model. 
CH-A acknowledges support from the Excellence Cluster ORIGINS which is funded by the Deutsche Forschungsgemeinschaft (DFG, German Research Foundation) under Germany's Excellence Strategy - EXC-2094-390783311. 
C-ZR, BL and CA are supported by the European Research Council through ERC Starting Grant ERC-StG-716532-PUNCA. 
BL and CMB acknowledge support from the Science Technology Facilities Council (STFC) through ST/T000244/1 and ST/P000541/1. 
AK and FP thank the support of the Spanish Ministry of Science and Innovation funding grant PGC2018- 101931-B-I00.
This work used the DiRAC@Durham facility managed by the Institute for Computational Cosmology on behalf of the STFC DiRAC HPC Facility (\url{www.dirac.ac.uk}). The equipment was funded by BEIS capital funding via STFC capital grants ST/K00042X/1, ST/P002293/1, ST/R002371/1 and ST/S002502/1, Durham University and STFC operations grant ST/R000832/1. DiRAC is part of the National e-Infrastructure. 
This work used the skun6@IAA facility (\url{www.skiesanduniverses.org}) managed by the Instituto de Astrof\'{i}sica de Andaluc\'{i}a (CSIC). The equipment was funded by the Spanish Ministry of Science EU-FEDER infrastructure grants EQC2018-004366-P and EQC2019-006089-P.








\bibliographystyle{utphys}
\bibliography{refs.bib}

\end{document}